\pdfoutput=1
\providecommand*{\reserveinserts}[1]{}
\documentclass[AMA,STIX1COL]{WileyNJD-v2}

\articletype{Article Type}%

\usepackage{graphicx}
\usepackage{amsmath,amssymb}
\usepackage{multirow}
\usepackage{multicol}
\usepackage{colortbl}
\usepackage{xcolor}
\usepackage{booktabs}
\usepackage{mathtools}
\usepackage{rotating}
\newcommand{\rot}[1]{\rotatebox{90}{#1}}

\received{25 October 2021}
\revised{ }
\accepted{ }

\makeatletter
\@ifundefined{theorem}{\newtheorem{theorem}{Theorem}}{}
\@ifundefined{lemma}{\newtheorem{lemma}{Lemma}}{}
\@ifundefined{proposition}{\newtheorem{proposition}{Proposition}}{}
\@ifundefined{corollary}{\newtheorem{corollary}{Corollary}}{}
\@ifundefined{definition}{\newtheorem{definition}{Definition}}{}
\@ifundefined{assumption}{\newtheorem{assumption}{Assumption}}{}
\@ifundefined{remark}{\newtheorem{remark}{Remark}}{}
\@ifundefined{example}{\newtheorem{example}{Example}}{}
\@ifundefined{proof}{\newenvironment{proof}{\par\noindent\textit{Proof.}\ }{\hfill$\square$\par\medskip}}{}
\makeatother

\begin{document}

\title{Sensor Monitoring Towards Industry 4.0: Bridging Concepts on the $C_n$ Multivariate Control Chart}

\author[neoma]{Diego C. Nascimento}

\author[ufba]{Ana C. S. Batista}

\author[icmc]{Francisco Louzada}

\author[ufba]{Paulo H. Ferreira*}

\authormark{NASCIMENTO ET AL.}

\address[neoma]{\orgname{NEOMA Business School}, \orgaddress{\state{Mont-Saint-Aignan}, \country{France}}}

\address[ufba]{\orgdiv{Department of Statistics}, \orgname{Federal University of Bahia}, \orgaddress{\state{Salvador-BA}, \country{Brazil}}}

\address[icmc]{\orgdiv{Institute of Mathematics and Computer Sciences}, \orgname{University of São Paulo}, \orgaddress{\state{S\~ao Carlos-SP}, \country{Brazil}}}

\corres{*Corresponding author name: Paulo H. Ferreira. \email{paulohenri@ufba.br}}

%\presentaddress{This is sample for present address text this is sample for present address text}

\abstract[Summary]{The main challenge in monitoring multiple sensors is the high dimensionality that a process can easily reach. From a decision-making perspective, converting these multivariate dynamics into an informative visualization is also not trivial. Statistical Process Control (SPC) techniques detect changes in quality characteristics, but they are difficult to apply beyond the normality (Gaussian) assumption. For instance, the most well-known SPC multivariate control chart is based on Hotelling's $T^2$ statistic, which assumes that the process data follow a multivariate normal distribution (which rarely occurs in practice) and that the observations are serially independent. This study focuses on the multivariate case (dimension $d \geq 2$), regardless of the data dependence or marginal asymmetry. We propose a new control chart for change detection in copula-based multivariate processes, called the $C_n$-chart, which is based on a density level-set tolerance region built from a copula function, and we derive its run-length properties. The methodology also quantifies the (not necessarily linear) dependence among the variables. Two case studies are presented, one on water quality control and the other on monitoring brass-steel thermostats, complemented by a simulation study of run-length performance and computational cost. The results show that the proposed methodology captures both the association among variables and the anomalies, and that it accommodates variable selection through conditional (regression) modeling of the margins.
}


\jnlcitation{\cname{
\author{D.C. Nascimento}, 
\author{A.C.S. Batista}, 
\author{F. Louzada}, and
\author{P.H. Ferreira}} (\cyear{2026}),
 \ctitle{Sensor Monitoring Towards Industry 4.0: Bridging Concepts on the $C_n$ Multivariate Control Chart}, \cvol{00:1--26}.
 }

\maketitle

%\footnotetext{\textbf{Abbreviations:} ANA, anti-nuclear antibodies; APC, antigen-presenting cells; IRF, interferon regulatory factor}

\section{Introduction} \label{sec:intro}

Industry 4.0 represents a transformative wave of technological advancement that is reshaping the manufacturing landscape (Javaid et al. 2022). Central to this revolution are sensors, which play a pivotal role in collecting vast amounts of data from various points within the production process (De Felice 2021). These sensors connect the physical and digital worlds, providing real-time information on everything from machine performance to product quality. However, as Industry 4.0 progresses, the challenge of high dimensionality arises. With an ever-increasing number of sensors generating data, industries are faced with the task of managing and making sense of high-dimensional data sets. \nocite{javaid2022understanding,de2021innovation}

%% QUALITY IN INDUSTRY 4.0
Regardless of the industrial sector (product or service), quality is a key source of competitive advantage. This is where Industry 4.0 sensor-driven capabilities can provide a higher-resolution understanding of the process, bridging industry standards with customers' expectations, especially in increasingly competitive markets. In statistical terms, consistency and precision are expressed through process variability. According to Montgomery (2013), the traditional definition of quality states that products and services must comply with the specifications required by their users. Louzada et al. (2013) added that quality is a concept of paramount importance to any company because it is through its consideration, improvement, and continuous application that the level of excellence can be achieved in any of the objectives set and activities performed. \nocite{montgomery2013,louzada2013controle}

% SPC IN INDUSTRY 4.0
Statistical Process Control (SPC) is a powerful set of problem-solving tools used to reduce variability and achieve stable performance of production processes or services (Montgomery 2013). One of the most used SPC tools is the $\bar{X}$ (mean or average) control chart, adopted for process monitoring and control in the univariate case. However, in practice, monitoring and control usually involve several (related) variables. Therefore, studies such as Bersimis et al. (2007) and Epprecht et al. (2013), amongst others, suggest, as an alternative to the Shewhart $\bar{X}$ univariate control chart, using Hotelling's $T^2$ control chart (Hotelling 1947), which is a multivariate approach that considers jointly varying variables and assumes that they are multivariate normally distributed and serially uncorrelated.
As a proposal to circumvent the autocorrelation problem among the data, Boone (2010) studied the $z$ control chart, which, although presenting some good results, outperformed the traditional Hotelling's $T^2$ control chart only when there was a shift in the mean vector while multivariate normality was maintained. As an alternative to assuming multivariate normality, %\textcolor{blue}{nonparametric control charts and mixture of models were proposed, which showed better results than the Hotelling's $T^2$ method when there was no certainty of the data generative distribution.} 
which is rarely observed in practice, one may instead specify a distinct probability distribution for each marginal and model their joint behavior using a copula function (Nelsen, 2006). This approach also enables the modeling of nonlinear dependence structures among the variables, which are common in industrial data. For some examples of the application of copulas in industrial contexts, particularly in multivariate process monitoring, see, for instance, Krupskii et al. (2020), Xu et al. (2020), and Ahmad et al. (2023).
\nocite{montgomery2013,hotelling1947multivariate,boone2010contributions,nelsen2006introduction, krupskii2020copula,xu2020fault,ahmad2023monitoring,epprecht2013reducing,bersimis2007multivariate}

Transforming variables/data (e.g., through the multivariate Box-Cox transformation; Velilla 1993) can be helpful in satisfying the underlying assumption of multivariate normality required by Hotelling’s $T^2$ control chart. However, such transformations may compromise important aspects, particularly interpretability, as the resulting control statistics no longer correspond to the original measurement scales (Montgomery 2013).
\nocite{VELILLA1993259}

To detect changes in the distribution of sequentially observed independent random vectors, Baíllo and Cuevas (2006) proposed tolerance regions defined in terms of density level sets estimated in a plug-in fashion, and compared a parametric (Gaussian mixture) estimator with a nonparametric (kernel) estimator. Later, Verdier (2013) estimated the density level sets with copula models (Nelsen 2006; Joe 2014), which are more flexible than Gaussian mixtures and avoid the curse of dimensionality faced by nonparametric estimators. \nocite{baillo2006parametric,verdier2013application,nelsen2006introduction,joe2014dependence}

Building on this line of work, this study proposes a flexible, distribution-based alternative to the traditional control tolerance region for the multidimensional setting ($d \geq 2$). The joint distribution is estimated by linking the fitted marginal distributions of the process through a copula function. Relative to Verdier (2013), the contributions are: (i) the $C_n$ multivariate control chart (or simply, $C_n$-chart), which plots the estimated joint density of each observation against its index, so that the result is always a two-dimensional (2D) chart, whatever the value of $d$; (ii) margins that may include a regression structure on covariates; (iii) exact and asymptotic run-length properties of the chart, including its relation to univariate charts and to Hotelling's $T^2$ chart; and (iv) a simulation study of run-length performance and computational cost up to $d=20$.

This article is structured in four sections and two supplementary materials. Section \ref{sec:meth} discusses the copula-based multivariate SPC control chart tool and its run-length extension (exact and asymptotic properties). Then, Section \ref{sec:verify} illustrates two case studies (bivariate and trivariate, respectively) with the proposed $C_n$-chart, followed by a simulation study of its run-length performance and computational cost. Finally, Section \ref{sec:conc} presents the conclusions, limitations, and future works. Additionally, two supplementary materials are also presented: Supplementary Material A provides a brief review of the parametric copula theory and the most used models in the literature, and Supplementary Material B presents a simulation study comparing copula-based tolerance regions with the Hotelling's $T^2$ region under several marginal and dependence scenarios.

%%%%%%%%%%% METODOLOGIA

\section{METHODS}
\label{sec:meth}

The control chart was developed in the 1920s by Walter A. Shewhart (Shewhart 1931), a physicist at the Bell Telephone Laboratories, and is a widely used process monitoring technique that aims to detect (as soon as possible) the occurrence of special causes leading to a change (or drift) in the process behavior. If the process is under statistical control, we expect the collected data to vary around a fixed mean at random; that is, the observations are assumed to be independent (in particular, not autocorrelated). When dealing with a quality characteristic expressed on a numerical scale, it is usually of interest to monitor both its mean value and variability using control charts. \nocite{shewhart1931}
\begin{figure}[h!]
    \centering
    \includegraphics[scale=0.5]{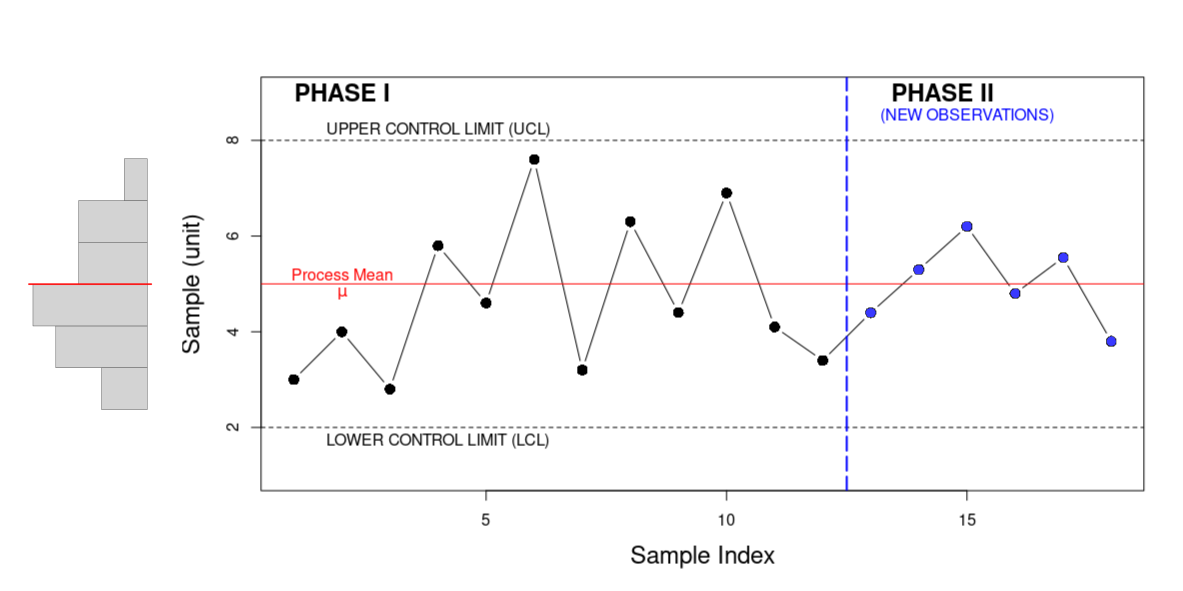}
    \caption{An illustration of the Shewhart X-bar univariate control chart. The left-hand panel shows the histogram of the variable, and the right-hand panel shows the control chart, with each observation plotted over time, the process mean as the center line, and the lower and upper control limits.}
    \label{fig:Control}
\end{figure}

A typical control chart contains a center line (CL), which represents the average value for the quality characteristic of interest (the process mean), and two other lines (or boundaries) called the lower control limit (LCL) and upper control limit (UCL), which are usually set at three standard deviations from the center line (Montgomery 2013). These boundaries are defined so that the process can be said to be under statistical control if all plotted sample points are between them (illustrated by Figure \ref{fig:Control}). The control chart strategy can be used in stages (or phases) I and II of the SPC analysis. Phase I consists of a retrospective analysis as an estimation step, which defines the ``reliable'' control limits (i.e., when the process is under statistical control) to be used in the next stage. Phase II involves process monitoring, comparing the estimates obtained from periodically sampled data with these control limits. \nocite{montgomery2013}

To monitor a process, one needs to understand how likely some events are to be observed. Hence, parametric models are extremely useful for statistical inference, enabling generalizations and supporting decision-making with low computational costs. Nevertheless, for many years, normality has been assumed in SPC techniques, which is often not the case (especially in multivariate settings). It is therefore more practical to relax the normality assumption and let the model adapt to the shape of the data. Therefore, this study proposes combining marginal estimation with copula theory (Sklar's theorem; Sklar 1959) to draw the joint distribution relationship across the multivariate response. This allows a flexible delimitation of the tolerance region, based on the estimated joint density, which is useful for anomaly detection and also quantifies the dependence among the responses. \nocite{sklar1959fonctions}

Let us suppose a process (production or service) presents $d$ response variables (i.e., sensors), in which their marginal densities $f_j$ can be estimated, for $j=1,2,\ldots,d$. Therefore, their cumulative distributions $F_j$ can be estimated straightforwardly, for $j=1,2,\ldots,d$. By Sklar's theorem, there exists a copula density $c$ for the probability integral transformed sample (for further details, see Supplementary Material A), which connects these marginals and yields the joint density $h$ of the response variables ($X_1,X_2,\ldots,X_d$). That is,
\begin{equation}
h\left(X_1,X_2,\ldots,X_d\right)=c\left(F_1(X_1),F_2(X_2),\ldots,F_d(X_d)\right)\prod_{j=1}^{d}f_j(X_j).
\end{equation}

Following Nelsen (2006), a copula $C$ is a $d$-dimensional distribution function $C: [0,1]^d \rightarrow [0,1]$, such that its univariate margins have uniform distributions on the interval $[0,1]$, i.e., $U_{j} \sim \text{U}(0,1)$, for $j=1,2,\ldots,d$. That is, writing $H$ for the joint distribution function of $(X_1,X_2,\ldots,X_d)$, \nocite{nelsen2006introduction}
%Uma c\'opula $C$ \'e uma fun\c{c}\~ao de distribui\c{c}\~ao $d$-dimensional $C$: [0,1]$^d \rightarrow$ [0,1], de modo que suas marginais univariadas tenham distribui\c{c}\~ao uniforme no intervalo $[0,1]$, i.e. $U_{j} \sim \text{Uniforme}(0,1)$, para $j=1,2,\ldots,d$. Uma c\'opula multivariada d-dimensional \'e uma fun\c{c}\~ao de distribui\c{c}\~ao $C$, como segue: 
\begin{align}
C\left(u_1,u_2, \ldots,u_d\right) &= \mathbb{P}\left(U_1 \le u_1,U_2 \le u_2,\ldots,U_d \le u_d\right) \nonumber \\
 &=H\left(F_1^{-1}(u_1),F_2^{-1}(u_2),\ldots,F_d^{-1}(u_d)\right),
\end{align}

\noindent where, writing $u_j=F_j(x_j)$ with $u_j\in[0,1]$, we have $x_j=F_j^{-1}(u_j)$. Thus, we can calculate the density $c$ of a copula $C$ for the transformed sample as follows:
%Assim, podemos calcular a densidade de uma c\'opula  $c(u_1,u_2,\ldots,u_d)$ da seguinte maneira:
\begin{align}\label{eq_dcopula}
c\left(u_1,u_2,\ldots,u_d\right)&=\frac{\partial^{d}}{\partial u_1 \partial u_2 \cdots \partial u_d }C\left(u_1,u_2,\ldots,u_d\right) \nonumber \\ &=\frac{\partial^{d}}{\partial u_1 \partial u_2  \cdots \partial u_d }H\left(F_1^{-1}(u_1),F_2^{-1}(u_2),\ldots,F_d^{-1}(u_d)\right) \nonumber \\
&=\frac{h\left(F_1^{-1}(u_1),F_2^{-1}(u_2),\ldots,F_d^{-1}(u_d)\right)}{\prod_{j=1}^{d}f_j(F_j^{-1}(u_j))}.
\end{align}

%% ADD FIG. - The transformation estimator
\begin{figure}[!ht]
    \centering
    \includegraphics[width=\textwidth]{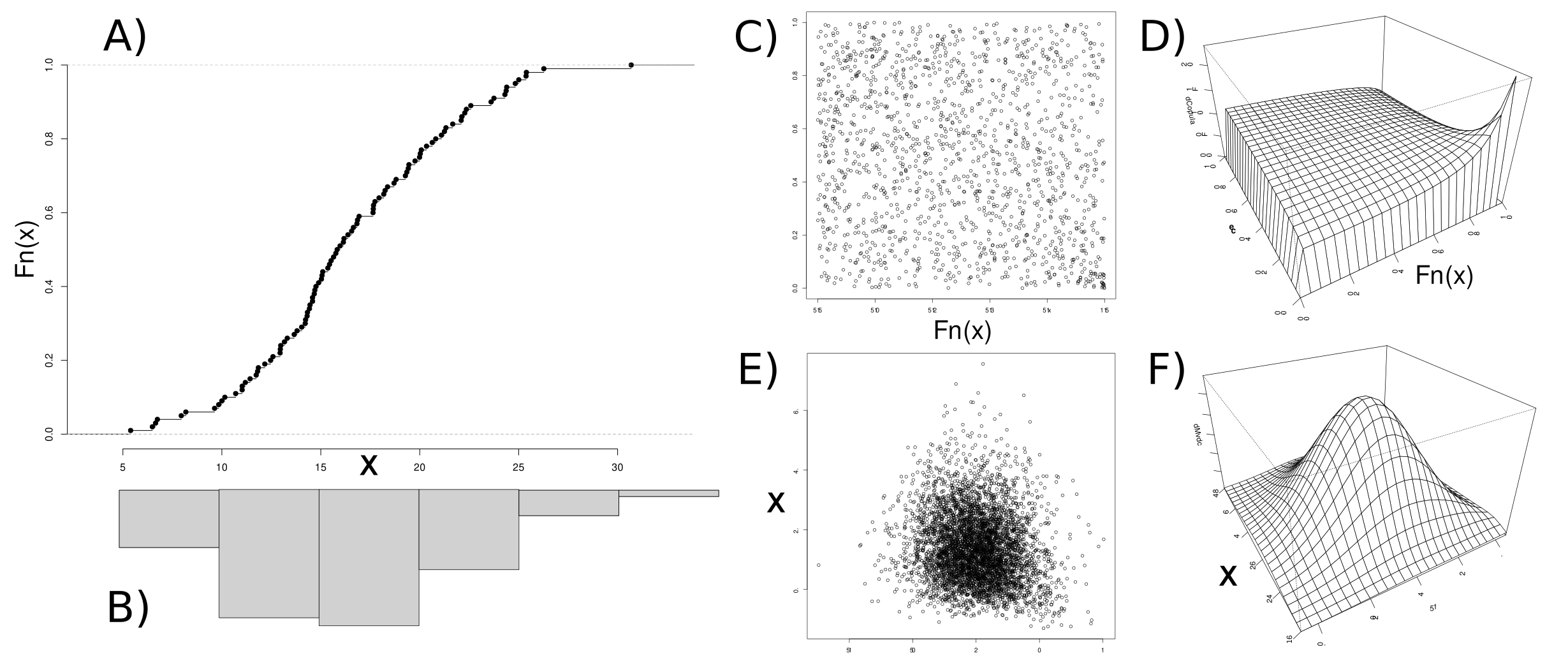}
    \caption{Illustration of the transformation estimator: A) variable $X$ - empirical cumulative distribution function (CDF); B) variable $X$ - histogram; C) copula sample; D) kernel estimate of the copula density; E) transformed sample $(X,Y)$; and F) kernel density estimate for the transformed sample. Notice that the empirical CDF, $F_n(x)$, produces the copula (pseudo-)sample $(u_X,u_Y)$ from which the copula density $c(u_X,u_Y)$ is estimated, and the marginal densities (e.g., $f_X$) map it back to the joint density of the original variables, $h(X,Y)$.}
    \label{fig:VISUALcopula}
\end{figure}

Figure \ref{fig:VISUALcopula} illustrates the main elements of the copula theory presented in the previous equations. Once the copula has been estimated from the transformed sample, the joint density is fully specified (a review of popular parametric copula families can be found in Supplementary Material A). The degree of dependence between the random variables is summarized by a measure that depends on the copula family adopted:
%Uma vez definida a relação entre a construção da densidade conjunta, o grau de associação entre variáveis aleatórias dependerá do tipo de cópula adotado, já incluida na sua estrutura:

\begin{itemize}
    \item Elliptical copulas (Gaussian and Student-t) $\rightarrow$ Pearson correlation coefficient ($\rho$);
    \item Archimedean copulas $\rightarrow$ Kendall's $\tau$;
    \item Nonparametric copulas $\rightarrow$ Kendall's $\tau$ or Spearman's $\rho_{\text{S}}$.
\end{itemize}

Given the critical role that the choice of copula plays in modeling dependence, one of the major challenges in copula-based methods is selecting an appropriate copula function. In this study, we address this issue by employing the Akaike information criterion (AIC; Akaike 1977) for model selection. This is followed by a goodness-of-fit (GOF) test tailored for copulas (for further details, see Genest et al. 2009). The procedure is implemented using the \textit{gofCopula}($\cdot$) function from the  \texttt{copula} package in R (Yan 2007). \nocite{akaike1977,genest2009goodness,yan2007enjoy}

In the following subsection, we discuss how the copula-based joint density defines the probabilistic bounds of the tolerance region.

\subsection{Tolerance Region}\label{sec:tolreg}

Baíllo and Cuevas (2006) addressed the detection of changes in the distribution of sequentially observed independent random vectors, a major challenge for SPC, especially when the vector's multivariate distribution, prior to the change, is not normal. To overcome this problem, the authors used tolerance regions as a detection tool, which are defined in terms of density level sets. These, in turn, are estimated in a plug-in fashion. This approach was previously used in the Gaussian case (Fuchs and Kenett 1998), and in the nonparametric setting (Baíllo et al. 2001). Polansky (2001) also used this idea to assess the capability of a manufacturing process. \nocite{baillo2006parametric,fuchs1998multivariate,polansky2001smooth}

%If a problem persists in this problem, the authors will use tolerance regions as an action detection tool, such as what are the categories of use in terms of sets of price variables, which, in turn, are estimated to be. from form plug-in. Such an approach was employed in the previous mind for the Gaussian case in Fuchs and Kenett (1998), and not parametric in Ba Illoetal. (2001). Ja Polansky (2001) used this idea to evaluate the ability of a fabricated process.

Baíllo and Cuevas (2006) also compared, through a simulation study, based on the proposed approach, a parametric density estimator (Gaussian mixture) and a nonparametric density estimator (kernel). Cadre et al. (2013) revealed the importance of this approach: ``density level sets are used in a variety of scientific areas, including statistics and machine learning, medical imaging, computer vision or remote sensing, and with applications for unsupervised classification/clustering, pattern recognition, anomalies or detection of new characteristics, for example''. Verdier (2013) constructed a tolerance region obtained from a density level estimate based on the plug-in approach, where density is estimated using copula modeling, claiming that it has the advantage of being more flexible than Gaussian mixture modeling, and avoids the dimensionality problem encountered with nonparametric methods. \nocite{baillo2006parametric,cadre2013estimation,verdier2013application}

Let $\boldsymbol{X}=(X_1,X_2,\ldots,X_d)^{\top}$ be a random vector on ${\mathbb{R}}^d$ with  (univariate) continuous CDFs $F_j$, for $j=1,2,\ldots,d$. Also, let $h$ be the (joint) density function of $\boldsymbol{X}$. Our goal is to decide whether a new observation, $\boldsymbol{X}_{n+k}$, for $k\geq 1$ (Phase II data), was generated from $h$. This decision will be based on a preliminary random sample from $h$, that is, from a set of independent and identically distributed (IID) observations, $\boldsymbol{X}_1,\ldots,\boldsymbol{X}_n$, collected while the process is in control (Phase I data). Once the tolerance region is defined, if $\boldsymbol{X}_{n+k}$ falls outside it, then the process is declared out of control.

\subsection{Density Level Set} \label{den}

Consider, as the tolerance region, the level set of the density $h$ with prespecified probability content $1-\alpha$, $\left\{\boldsymbol{x} \in \mathbb{R}^{d} \mid h(\boldsymbol{x}) \geq c_{\alpha}\right\}$ (hereafter, simply $\left\{h(\boldsymbol{x}) \geq c_{\alpha}\right\}$). Since $h$ is unknown, it is replaced by a plug-in estimator $\hat{h}_n$. Therefore, an ``out-of-control'' alarm is triggered whenever $\boldsymbol{X}_{n+k} \notin \left\{\hat{h}_n(\boldsymbol{x}) \geq \hat{c}_n\right\}$, for $k\geq 1$, so that, while the process is in control, a false alarm occurs with probability approximately $\alpha$, where $\hat{c}_n$ is the estimated value of $c_{\alpha}$ satisfying (Verdier 2013): \nocite{verdier2013application}
%\textcolor{blue}{\begin{equation}
%    \int_{\left\{\hat{h}_n(\boldsymbol{x}) \geq \hat{c}_n\right\}} \hat{h}_n(\boldsymbol{x}) d\boldsymbol{x} =
%   Q_{\boldsymbol{X}}(\omega)
%    =\inf \left\{\boldsymbol{x}:c_{\boldsymbol{X}}(\boldsymbol{x})\geq \omega \right\} = 1- \alpha.
%\end{equation}}
\begin{equation}
    \int_{\left\{\hat{h}_n(\boldsymbol{x}) \geq \hat{c}_n\right\}} \hat{h}_n(\boldsymbol{x}) d\boldsymbol{x} =
    1- \alpha.
\end{equation}
%In other words, \textcolor{blue}{$c_n$ is the $\alpha$-th quantile of $\hat{h}_n(\boldsymbol{x})$}, \textcolor{blue}{$Q_{\boldsymbol{X}}$ is the quantile function of the random vector $\boldsymbol{X}$}, and $0 \leq \omega \leq 1$ is the quantile value. 
In other words, $\hat{c}_n$ is the $\alpha$-th quantile of the random variable $\hat{h}_n(\boldsymbol{X})$, with $\boldsymbol{X}\sim\hat{h}_n$.

In practice, Baíllo and Cuevas (2006) suggested obtaining $\hat{c}_n$ via a Monte Carlo procedure, in which an auxiliary sample $\boldsymbol{X}_1^*,\ldots,\boldsymbol{X}_N^ *$ is generated from $\hat{h}_n$. Thus, $\hat{c}_n$ will be given by the $\alpha$-th quantile of $\hat{h}_n\left(\boldsymbol{X}_1^*\right),\ldots,\hat{h}_n\left(\boldsymbol{X}_N^*\right)$. The authors further proved that, under certain conditions, the false alarm probability obtained by this approach converges to $\alpha$ as $N$ tends to infinity. For more details, see Baíllo and Cuevas (2006), Cadre et al. (2013), and Verdier (2013). 

In this study, the density $h$ will be estimated through copula modeling, that is,
\begin{align}
\hat{h}_n\left(X_1,X_2,\ldots,X_d\right)&=\hat{c}\left(\hat{F}_1(X_1),\hat{F}_2(X_2),\ldots,\hat{F}_d(X_d)\right)\prod_{j=1}^{d}\hat{f}_j(X_j),
\end{align}
where $\hat{c}$ denotes the estimated copula density; and $\hat{f}_j$ and $\hat{F}_j$ represent, respectively, the estimated marginal densities and cumulative distributions, for $j=1,2,\ldots,d$. In the parametric case, which is the focus of this paper, $f_j=f_j(\cdot\mid \boldsymbol{\theta}_j)$ and $F_j=F_j(\cdot\mid \boldsymbol{\theta}_j)$, where $\boldsymbol{\theta}_j$ denotes the parameter vector of margin $j$, for $j=1,2,\ldots,d$. For some parametric families of copulas (i.e., when $c=c_{\boldsymbol{\psi}}$, with $\boldsymbol{\psi}$ being the copula parameter vector), see Supplementary Material A. Maximum likelihood (ML) estimation is commonly adopted to estimate the marginal and copula parameters simultaneously in the parametric method. An alternative and also popular method for estimating copula-based models is the Inference Function for Margins (IFM) method, also referred to as the two-stage estimation method, proposed by Joe and Xu (1996). More details on these estimation approaches can be found in Supplementary Material A. \nocite{joe1996estimation}

\begin{figure}[!ht]
    \centering
    \includegraphics[width=\textwidth]{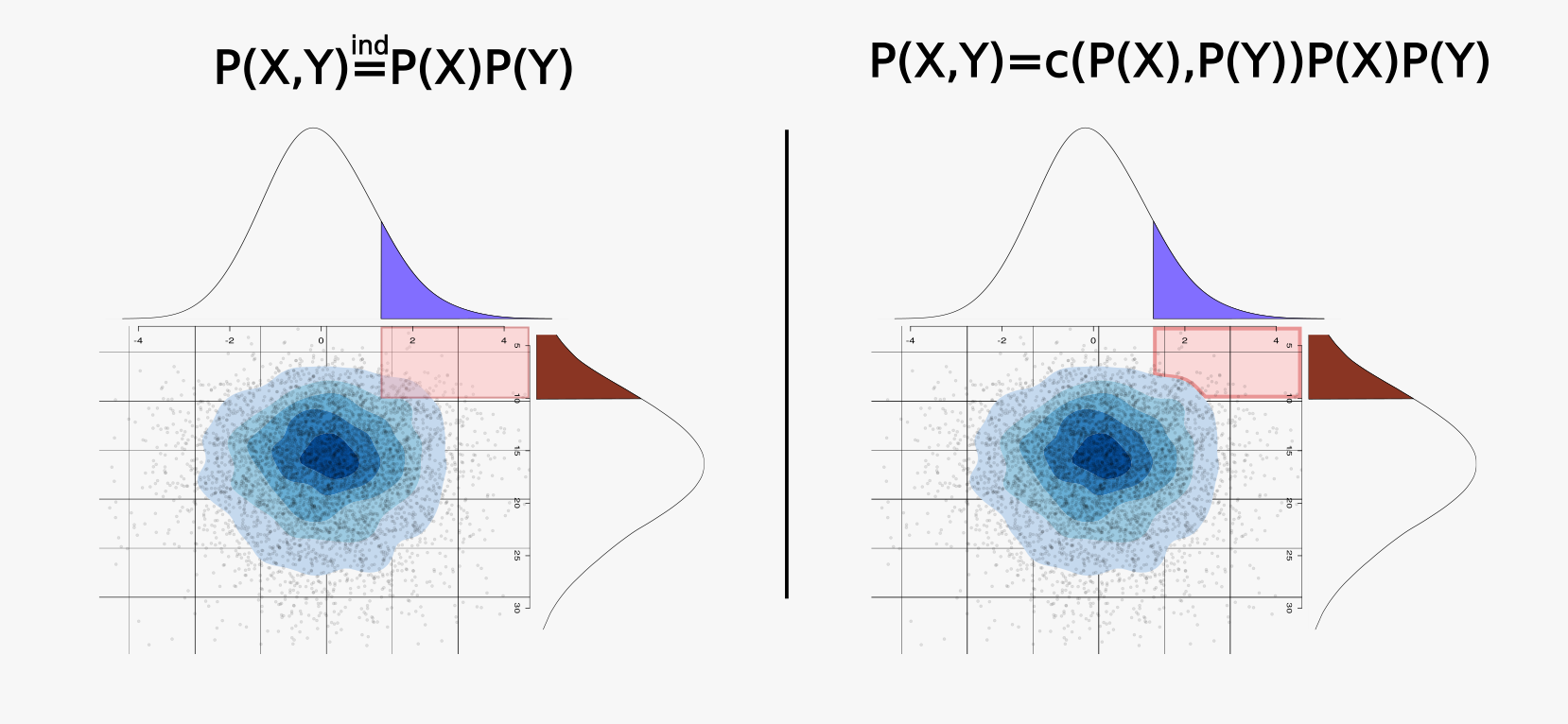}
    \caption{Joint distribution under the assumption of independence (left-hand panel) and with dependence modeled by a copula (right-hand panel) between the two variables. The one-sided tolerance region, with $\alpha=0.1$, is colored and expressed in light blue and light red for the marginals, and in pink for the joint distribution (bounded by the selected limits).}
    \label{fig:IND-COPULA}
\end{figure}

Figure \ref{fig:IND-COPULA} illustrates a situation where a one-sided tolerance region is adopted, comparing the joint distribution under independence with that obtained with a copula function. \nocite{baillo2006parametric, cadre2013estimation, verdier2013application} 

%Neste trabalho, utilizamos fun\c{c}\~oes c\'opula para estimar $h_n$ nos casos param\'etricos e semi-param\'etrico, e estimadores Kernel para o caso n\~ao-param\'etrico.
 
The proposed copula-based tolerance region is not restricted to a rectangular box, which is the joint acceptance region implied by a set of univariate charts (e.g., Figure 7 of Fonseca et al. (2021) treats the jointly distributed variables as independent, so that the bivariate control region is a rectangle). In the bivariate case with two-sided marginal limits, the area outside the rectangular box splits into eight regions: an exceedance of only the upper (or only the lower) limit of $X$; of only the upper (or only the lower) limit of $Y$; joint upper exceedances; joint lower exceedances; and mixed exceedances (the upper limit of $X$ with the lower limit of $Y$, or the reverse). Because the $C_n$-chart flags every observation outside the copula-based tolerance region, whatever its position relative to the marginal limits, it covers all these regions at once. It also flags observations that lie inside all marginal limits but are unusual given the dependence structure, and it accepts marginal exceedances that agree with that structure; these regions are formalized in Subsection~\ref{sec:arl-exact}. \nocite{foncesa2021}

\subsection{The $C_n$-Chart and Its Run-Length Properties}\label{sec:cnchart}

In this study, we develop a new and flexible multivariate control chart, called $C_n$-chart, aiming to visually detect changes occurring, e.g., in the process mean vector, especially in high dimensions ($d\gg 2$), as the information about the process condition (in or out of statistical control) is presented in a 2D plot, regardless of the data dimensionality. %since the tolerance region presented here only covers the bivariate case.
 
The proposed multivariate control chart is built on the density level set of Subsection~\ref{den}: in Phase I, $\hat{c}_n$ is estimated at a significance level $\alpha$. The $C_n$-chart replaces the lower and upper limits of several univariate control charts by a single lower limit, $\hat{c}_n$, applied to the density value $\hat{h}_i=\hat{h}_n(\boldsymbol{x}_i)$ of each observation $\boldsymbol{x}_i=(x_{1i},x_{2i},\ldots,x_{di})^{\top}$, for $i=1,2,\ldots,n$, which is calculated from the estimated copula model and then plotted. In Phase II, the densities of new observations are monitored based on the (one-sided or LCL) $C_n$-chart obtained in Phase I. If $\hat{h}_k < \hat{c}_n$, for some $k=n+1,n+2,\ldots$, an alarm is generated and its cause must be investigated.

The performance of the proposed SPC tool is illustrated in Section~\ref{sec:verify}, with bivariate (Subsection \ref{sec:ap_biv}) and trivariate (Subsection \ref{sec:ap_tri}) examples.
%%%%%%%%%%%%%%%%%%%%%%%%%%%%%%%%%%%%%%%%%%%%%%%%%%%%%
The following subsubsections derive exact and asymptotic run-length properties of the $C_n$-chart, which signals whenever the joint density of a new observation falls below the level $c_\alpha$, as in Verdier (2013). The $C_n$-chart is compared with two other charts: the set of univariate (marginal) control charts, whose joint acceptance region is a rectangular box, and a \emph{gated} variant of the $C_n$-chart, which signals only when an observation also exceeds at least one marginal control limit and serves as a theoretical bridge between the two.

\subsubsection{Setting and assumptions}\label{sec:arl-setting}

Let $\boldsymbol{X}_i=(X_{1i},X_{2i},\ldots,X_{di})^{\top}$ be the $i$th observation of the $d$ response variables and $\boldsymbol{Z}_i\in\mathbb{R}^q$ the corresponding vector of covariates, possibly empty. Each margin $j$, for $j=1,2,\ldots,d$, follows a regression model
\begin{equation}\label{arl:reg}
X_{ji}=r_j\bigl(\boldsymbol{Z}_i,\varepsilon_{ji};\boldsymbol{\beta}_j\bigr),
\end{equation}
where $r_j$ is continuous and strictly increasing in its error argument, $\boldsymbol{\beta}_j$ is the vector of regression coefficients, and the error $\varepsilon_{ji}$ has a continuous CDF $F_j(\cdot\mid\boldsymbol{\theta}_j)$ with density $f_j(\cdot\mid\boldsymbol{\theta}_j)$. The linear model \eqref{red-mod} of Subsection~\ref{sec:ap_biv} corresponds to $r_j(\boldsymbol{z},e;\boldsymbol{\beta}_j)=\beta_{0}^{*}+\beta_{2}^{*}z+e$. A margin without covariates has $r_j(\boldsymbol{z},e;\boldsymbol{\beta}_j)=e$, so that $\varepsilon_{ji}=X_{ji}$. Write $\varepsilon_{ji}=r_j^{-1}(X_{ji};\boldsymbol{Z}_i,\boldsymbol{\beta}_j)$ for the inverse in the error argument, $\boldsymbol{\varepsilon}_i=(\varepsilon_{1i},\ldots,\varepsilon_{di})^{\top}$ for the monitored vector, and
\[
\boldsymbol{\vartheta}=\bigl(\boldsymbol{\beta}_1^{\top},\ldots,\boldsymbol{\beta}_d^{\top},\boldsymbol{\theta}_1^{\top},\ldots,\boldsymbol{\theta}_d^{\top},\boldsymbol{\psi}^{\top}\bigr)^{\top}\in\mathbb{R}^{p}
\]
for the full parameter vector, where $\boldsymbol{\psi}$ is the copula parameter vector. Its in-control value is $\boldsymbol{\vartheta}_0$.

\begin{assumption}[In-control model]\label{ass:ic}
The pairs $(\boldsymbol{X}_i,\boldsymbol{Z}_i)$, $i\ge1$, are IID, and for every $\boldsymbol{z}$, conditionally on $\boldsymbol{Z}_i=\boldsymbol{z}$, the vector $\boldsymbol{\varepsilon}_i$ has distribution function
\[
H(\boldsymbol{e})=C_{\boldsymbol{\psi}}\bigl(F_1(e_1\mid\boldsymbol{\theta}_1),\ldots,F_d(e_d\mid\boldsymbol{\theta}_d)\bigr),
\]
where $C_{\boldsymbol{\psi}}$ is a copula with density $c_{\boldsymbol{\psi}}>0$ on $(0,1)^d$. In particular, neither the copula nor the error distributions depend on the covariate values.
\end{assumption}

\begin{lemma}[Distribution of the monitored vector]\label{lem:iid}
Under Assumption~\ref{ass:ic}, the vectors $\boldsymbol{\varepsilon}_i$, $i\ge1$, are IID, each $\boldsymbol{\varepsilon}_i$ is independent of $\boldsymbol{Z}_i$, and $\boldsymbol{\varepsilon}_i$ has joint density
\begin{equation}\label{arl:h}
h(\boldsymbol{e})=c_{\boldsymbol{\psi}}\bigl(F_1(e_1\mid\boldsymbol{\theta}_1),\ldots,F_d(e_d\mid\boldsymbol{\theta}_d)\bigr)\prod_{j=1}^{d}f_j(e_j\mid\boldsymbol{\theta}_j).
\end{equation}
Moreover, $\boldsymbol{U}_i=\bigl(F_1(\varepsilon_{1i}\mid\boldsymbol{\theta}_1),\ldots,F_d(\varepsilon_{di}\mid\boldsymbol{\theta}_d)\bigr)^{\top}$ has distribution function $C_{\boldsymbol{\psi}}$.
\end{lemma}

\begin{proof}
For fixed $\boldsymbol{\beta}$, $\boldsymbol{\varepsilon}_i=g(\boldsymbol{X}_i,\boldsymbol{Z}_i)$ for one measurable map $g$, so the $\boldsymbol{\varepsilon}_i$ are IID because the pairs $(\boldsymbol{X}_i,\boldsymbol{Z}_i)$ are. By Assumption~\ref{ass:ic}, $\mathbb{P}(\boldsymbol{\varepsilon}_i\le\boldsymbol{e}\mid\boldsymbol{Z}_i)=H(\boldsymbol{e})$ almost surely, so for every Borel set $B\subseteq\mathbb{R}^q$,
\[
\mathbb{P}(\boldsymbol{\varepsilon}_i\le\boldsymbol{e},\ \boldsymbol{Z}_i\in B)=\mathbb{E}\bigl[\mathbf{1}\{\boldsymbol{Z}_i\in B\}\,\mathbb{P}(\boldsymbol{\varepsilon}_i\le\boldsymbol{e}\mid\boldsymbol{Z}_i)\bigr]=H(\boldsymbol{e})\,\mathbb{P}(\boldsymbol{Z}_i\in B).
\]
The sets $\{\boldsymbol{\varepsilon}_i\le\boldsymbol{e}\}$ form a $\pi$-system generating the $\sigma$-field of $\boldsymbol{\varepsilon}_i$, so by the $\pi$--$\lambda$ theorem $\boldsymbol{\varepsilon}_i$ is independent of $\boldsymbol{Z}_i$ and has distribution function $H$. Since each $F_j$ is continuous with density $f_j$ and $C_{\boldsymbol{\psi}}$ has density $c_{\boldsymbol{\psi}}$, differentiating $H$ gives \eqref{arl:h}, as in \eqref{eq_dcopula}. Finally, let $q_j(u)=\sup\{e:F_j(e\mid\boldsymbol{\theta}_j)\le u\}$ for $u\in(0,1)$. Continuity of $F_j$ gives $F_j(q_j(u)\mid\boldsymbol{\theta}_j)=u$ and $\{F_j(\varepsilon_{ji}\mid\boldsymbol{\theta}_j)\le u\}=\{\varepsilon_{ji}\le q_j(u)\}$, hence $\mathbb{P}(\boldsymbol{U}_i\le\boldsymbol{u})=H(q_1(u_1),\ldots,q_d(u_d))=C_{\boldsymbol{\psi}}(\boldsymbol{u})$.
\end{proof}

Fix a significance level $\alpha\in(0,1)$, and let $\mathbb{P}_0$ denote probability under the in-control model.

\begin{assumption}[No flat regions]\label{ass:cont}
Under $\mathbb{P}_0$, the random variable $h(\boldsymbol{\varepsilon})$ has a continuous distribution.
\end{assumption}

Under Assumption~\ref{ass:cont} there is $c_\alpha>0$ with $\mathbb{P}_0\{h(\boldsymbol{\varepsilon})<c_\alpha\}=\alpha$; that is, $c_\alpha$ is the $\alpha$-th quantile of $h(\boldsymbol{\varepsilon})$, whose plug-in estimate is $\hat{c}_n$ (Subsection~\ref{den}).

\begin{definition}[Charts]\label{def:charts}
\begin{enumerate}
\item The \emph{tolerance region} is $\mathcal{R}=\{\boldsymbol{e}:h(\boldsymbol{e})\ge c_\alpha\}$, so that $\mathbb{P}_0(\boldsymbol{\varepsilon}\notin\mathcal{R})=\alpha$.
\item The $C_n$-chart (with known parameters) signals at the Phase II observation $n+k$ if and only if $h(\boldsymbol{\varepsilon}_{n+k})<c_\alpha$. Its signal set is $\mathcal{S}=\mathcal{R}^c$.
\item The marginal charts use the box $\mathcal{M}=\prod_{j=1}^{d}[\mathrm{LCL}_j,\mathrm{UCL}_j]$ and signal when $\boldsymbol{\varepsilon}_{n+k}\in\mathcal{S}_B=\mathcal{M}^c$, that is, when at least one margin is out of control. Their per-margin false-alarm probabilities are $a_j=\mathbb{P}_0(\varepsilon_j\notin[\mathrm{LCL}_j,\mathrm{UCL}_j])$, and $\alpha_{\mathcal{M}}=\mathbb{P}_0(\boldsymbol{\varepsilon}\notin\mathcal{M})$.
\item The gated $C_n$-chart has signal set $\mathcal{S}_G=\mathcal{M}^c\cap\mathcal{R}^c$.
\item For a signal set $\mathcal{A}$, the run length is $\mathrm{RL}_{\mathcal{A}}=\inf\{k\ge1:\boldsymbol{\varepsilon}_{n+k}\in\mathcal{A}\}$, with $\inf\emptyset=\infty$. We write $\mathrm{RL}=\mathrm{RL}_{\mathcal{S}}$, $\mathrm{RL}_B=\mathrm{RL}_{\mathcal{S}_B}$ and $\mathrm{RL}_G=\mathrm{RL}_{\mathcal{S}_G}$, and similarly for the average run lengths (ARL).
\item The in-control \emph{joint-only} mass is $\delta_0=\mathbb{P}_0(\boldsymbol{\varepsilon}\in\mathcal{M}\cap\mathcal{R}^c)$: alarms of the $C_n$-chart while every margin is within its limits. The in-control \emph{absorbed} mass is $\eta_0=\mathbb{P}_0(\boldsymbol{\varepsilon}\in\mathcal{M}^c\cap\mathcal{R})$: marginal alarms that agree with the dependence structure and are accepted by the $C_n$-chart.
\end{enumerate}
\end{definition}

\begin{assumption}[Sustained shift]\label{ass:oc}
From $k=1$ onwards, the monitored vectors $\boldsymbol{\varepsilon}_{n+k}$, computed with the in-control parameters $\boldsymbol{\vartheta}_0$, are IID with a law $\mathbb{P}_1$. The in-control case is $\mathbb{P}_1=\mathbb{P}_0$.
\end{assumption}

Assumption~\ref{ass:oc} covers, for example, a sustained change in the regression coefficients, in the error distributions or in the copula, provided the covariates remain IID.

\begin{lemma}[Geometric run length]\label{lem:geom}
Under Assumption~\ref{ass:oc}, let $\mathcal{A}$ be a Borel set and $\pi=\mathbb{P}_1(\mathcal{A})$. If $\pi>0$, then $\mathbb{P}(\mathrm{RL}_{\mathcal{A}}=\ell)=(1-\pi)^{\ell-1}\pi$ for $\ell\ge1$, $\mathrm{ARL}_{\mathcal{A}}=1/\pi$ and $\mathrm{SDRL}_{\mathcal{A}}=\sqrt{1-\pi}/\pi$. If $\pi=0$, then $\mathrm{RL}_{\mathcal{A}}=\infty$ almost surely.
\end{lemma}

\begin{proof}
By independence, $\mathbb{P}(\mathrm{RL}_{\mathcal{A}}>\ell)=\mathbb{P}(\boldsymbol{\varepsilon}_{n+1}\notin\mathcal{A},\ldots,\boldsymbol{\varepsilon}_{n+\ell}\notin\mathcal{A})=(1-\pi)^{\ell}$. If $\pi=0$, this equals one for every $\ell$, so $\mathrm{RL}_{\mathcal{A}}=\infty$ almost surely. If $\pi>0$, then $\mathbb{P}(\mathrm{RL}_{\mathcal{A}}=\ell)=(1-\pi)^{\ell-1}-(1-\pi)^{\ell}=(1-\pi)^{\ell-1}\pi$, the geometric law with success probability $\pi$, whose mean is $1/\pi$ and variance $(1-\pi)/\pi^2$.
\end{proof}

\subsubsection{Exact run-length results}\label{sec:arl-exact}

\begin{theorem}[Exact ARL and comparison with the marginal charts]\label{thm:decomp}
Under Assumptions~\ref{ass:ic}--\ref{ass:oc}, with the convention $1/0=\infty$:
\begin{enumerate}
\item[(i)] The in-control run length of the $C_n$-chart is geometric with $\mathrm{ARL}_0=1/\alpha$ and $\mathrm{SDRL}_0=\sqrt{1-\alpha}/\alpha$, whatever the copula family, the margins, the dimension $d$ and the covariates.
\item[(ii)] For every $\mathbb{P}_1$,
\[
\mathrm{ARL}(\mathbb{P}_1)^{-1}-\mathrm{ARL}_B(\mathbb{P}_1)^{-1}=\mathbb{P}_1(\mathcal{M}\cap\mathcal{R}^c)-\mathbb{P}_1(\mathcal{M}^c\cap\mathcal{R}).
\]
Hence $\mathrm{ARL}(\mathbb{P}_1)\le\mathrm{ARL}_B(\mathbb{P}_1)$ if and only if the joint-only probability is at least the absorbed probability, $\mathbb{P}_1(\mathcal{M}\cap\mathcal{R}^c)\ge\mathbb{P}_1(\mathcal{M}^c\cap\mathcal{R})$. In control, $\alpha=\alpha_{\mathcal{M}}+\delta_0-\eta_0$.
\item[(iii)] The gated $C_n$-chart satisfies
\[
\mathrm{ARL}_G(\mathbb{P}_1)=\frac{1}{\mathbb{P}_1(\mathcal{R}^c)-\mathbb{P}_1(\mathcal{M}\cap\mathcal{R}^c)}\;\ge\;\max\{\mathrm{ARL}(\mathbb{P}_1),\,\mathrm{ARL}_B(\mathbb{P}_1)\},
\]
and in control
\[
\mathrm{ARL}_{G,0}=\frac{1}{\alpha-\delta_0}=\frac{1}{\alpha_{\mathcal{M}}-\eta_0}.
\]
\item[(iv)] Pathwise, $\mathrm{RL}_G\ge\max\{\mathrm{RL},\mathrm{RL}_B\}$.
\end{enumerate}
\end{theorem}

\begin{proof}
(i) Apply Lemma~\ref{lem:geom} with $\mathbb{P}_1=\mathbb{P}_0$ and $\mathcal{A}=\mathcal{R}^c$, for which $\mathbb{P}_0(\mathcal{R}^c)=\alpha$ by Assumption~\ref{ass:cont}. Nothing in this argument depends on the copula, the margins, $d$ or the covariates.

(ii) The sets $\mathcal{R}^c$ and $\mathcal{M}^c$ are the disjoint unions $\mathcal{R}^c=(\mathcal{M}^c\cap\mathcal{R}^c)\cup(\mathcal{M}\cap\mathcal{R}^c)$ and $\mathcal{M}^c=(\mathcal{M}^c\cap\mathcal{R}^c)\cup(\mathcal{M}^c\cap\mathcal{R})$. Subtracting,
\[
\mathbb{P}_1(\mathcal{R}^c)-\mathbb{P}_1(\mathcal{M}^c)=\mathbb{P}_1(\mathcal{M}\cap\mathcal{R}^c)-\mathbb{P}_1(\mathcal{M}^c\cap\mathcal{R}).
\]
By Lemma~\ref{lem:geom}, $\mathrm{ARL}(\mathbb{P}_1)^{-1}=\mathbb{P}_1(\mathcal{R}^c)$ and $\mathrm{ARL}_B(\mathbb{P}_1)^{-1}=\mathbb{P}_1(\mathcal{M}^c)$, including the case of a zero signal probability. The in-control identity follows with $\mathbb{P}_1=\mathbb{P}_0$.

(iii) From the first decomposition in (ii), $\mathbb{P}_1(\mathcal{S}_G)=\mathbb{P}_1(\mathcal{R}^c)-\mathbb{P}_1(\mathcal{M}\cap\mathcal{R}^c)$, and from the second, $\mathbb{P}_1(\mathcal{S}_G)=\mathbb{P}_1(\mathcal{M}^c)-\mathbb{P}_1(\mathcal{M}^c\cap\mathcal{R})$. Thus $\mathbb{P}_1(\mathcal{S}_G)\le\min\{\mathbb{P}_1(\mathcal{R}^c),\mathbb{P}_1(\mathcal{M}^c)\}$, and the inequality follows from Lemma~\ref{lem:geom} because $x\mapsto1/x$ is decreasing. The in-control values follow with $\mathbb{P}_1=\mathbb{P}_0$.

(iv) Since $\mathcal{S}_G\subseteq\mathcal{R}^c$ and $\mathcal{S}_G\subseteq\mathcal{M}^c$, a signal of the gated chart is also a signal of the $C_n$-chart and of the marginal charts, so its first signal cannot occur earlier than theirs.
\end{proof}

\begin{remark}
Theorem~\ref{thm:decomp}(ii) formalizes the regions discussed in Subsection~\ref{sec:tolreg}. Relative to the marginal charts, the $C_n$-chart gains the joint-only alarms in $\mathcal{M}\cap\mathcal{R}^c$ and gives up the absorbed alarms in $\mathcal{M}^c\cap\mathcal{R}$; observations in $\mathcal{M}^c\cap\mathcal{R}^c$ are signaled by both. 
%In the first water quality example, samples \#2, \#12 and \#54 lie in $\mathcal{M}^c\cap\mathcal{R}^c$, whereas sample \#5 lies in $\mathcal{M}^c\cap\mathcal{R}$: it is out of control for the marginal chart of $X$ but is absorbed by the $C_n$-chart (Figure~\ref{fig:MARGs}). 
Note also that if each marginal chart runs at level $\alpha$, then $\alpha\le\alpha_{\mathcal{M}}\le d\alpha$, and typically $\alpha_{\mathcal{M}}>\alpha$, whereas the false-alarm probability of the $C_n$-chart is exactly $\alpha$.
\end{remark}

\begin{corollary}[Shifts in the dependence only]\label{cor:dep}
Under Assumptions~\ref{ass:ic}--\ref{ass:oc}, suppose that the univariate margins of $\mathbb{P}_1$ coincide with those of $\mathbb{P}_0$, so that only the dependence changes. Then
\[
\mathrm{ARL}_G(\mathbb{P}_1)\ge\mathrm{ARL}_B(\mathbb{P}_1)\ge\Bigl(\sum_{j=1}^{d}a_j\Bigr)^{-1}.
\]
No such bound holds for the $C_n$-chart (Example~\ref{ex:counter}).
\end{corollary}

\begin{proof}
By the union bound,
\[
\begin{aligned}
\mathbb{P}_1(\mathcal{M}^c)&=\mathbb{P}_1\Bigl(\textstyle\bigcup_{j=1}^{d}\{\varepsilon_j\notin[\mathrm{LCL}_j,\mathrm{UCL}_j]\}\Bigr)\\
&\le\sum_{j=1}^{d}\mathbb{P}_1\bigl(\varepsilon_j\notin[\mathrm{LCL}_j,\mathrm{UCL}_j]\bigr)=\sum_{j=1}^{d}a_j,
\end{aligned}
\]
where the last equality uses that the $j$th margin of $\mathbb{P}_1$ equals that of $\mathbb{P}_0$. Lemma~\ref{lem:geom} and Theorem~\ref{thm:decomp}(iii) complete the proof.
\end{proof}

\begin{proposition}[Calibration of the gated chart]\label{prop:calib}
Under Assumptions~\ref{ass:ic}--\ref{ass:cont}, let $\pi_G(t)=\mathbb{P}_0\{\boldsymbol{\varepsilon}\in\mathcal{M}^c,\ h(\boldsymbol{\varepsilon})<t\}$ for $t>0$. Then $\pi_G$ is continuous and nondecreasing, with $\lim_{t\downarrow0}\pi_G(t)=0$ and $\lim_{t\to\infty}\pi_G(t)=\alpha_{\mathcal{M}}$. Hence, if $\alpha<\alpha_{\mathcal{M}}$, there exists $t^*$ such that the gated chart with signal set $\mathcal{M}^c\cap\{h<t^*\}$ has in-control ARL exactly $1/\alpha$. If, in addition, $\delta_0>0$, every such $t^*$ satisfies $t^*>c_\alpha$.
\end{proposition}

\begin{proof}
Write $Y=h(\boldsymbol{\varepsilon})$. If $t<t'$, then $\{Y<t\}\subseteq\{Y<t'\}$, so $\pi_G$ is nondecreasing and $0\le\pi_G(t')-\pi_G(t)\le\mathbb{P}_0(t\le Y<t')$. By Assumption~\ref{ass:cont}, the distribution function of $Y$ is continuous, so the right-hand side tends to zero as $t'-t\to0$, and $\pi_G$ is continuous. By continuity of probability, $\lim_{t\downarrow0}\pi_G(t)\le\mathbb{P}_0(Y\le0)=\int_{\{h=0\}}h=0$ and $\lim_{t\to\infty}\pi_G(t)=\mathbb{P}_0(\boldsymbol{\varepsilon}\in\mathcal{M}^c)=\alpha_{\mathcal{M}}$. The intermediate value theorem gives $t^*$ with $\pi_G(t^*)=\alpha$, and Lemma~\ref{lem:geom} gives the in-control ARL $1/\alpha$. Finally, Theorem~\ref{thm:decomp}(iii) gives $\pi_G(c_\alpha)=\alpha-\delta_0<\alpha$ when $\delta_0>0$; if some $t^*\le c_\alpha$ satisfied $\pi_G(t^*)=\alpha$, monotonicity would give $\alpha\le\pi_G(c_\alpha)<\alpha$, a contradiction.
\end{proof}

\subsubsection{Coincidence and convergence}\label{sec:arl-conv}

\begin{theorem}[Coincidence and convergence]\label{thm:conv}
Under Assumptions~\ref{ass:ic}--\ref{ass:oc}:
\begin{enumerate}
\item[(a)] If $\mathcal{M}\subseteq\mathcal{R}$, then $\mathcal{S}_G=\mathcal{S}\subseteq\mathcal{S}_B$, so that $\mathrm{RL}_B\le\mathrm{RL}=\mathrm{RL}_G$ pathwise. If $\mathcal{R}\subseteq\mathcal{M}$, then $\mathcal{S}_G=\mathcal{S}_B\subseteq\mathcal{S}$, so that $\mathrm{RL}\le\mathrm{RL}_B=\mathrm{RL}_G$ pathwise.
\item[(b)] If $\mathbb{P}_1(\mathcal{R}^c)>0$, the probability that the first alarm of the $C_n$-chart is a joint-only alarm is
\[
\mathbb{P}(\mathrm{RL}_G\neq\mathrm{RL})=\frac{\mathbb{P}_1(\mathcal{M}\cap\mathcal{R}^c)}{\mathbb{P}_1(\mathcal{R}^c)}.
\]
If $\mathbb{P}_1(\mathcal{M}^c)>0$, the probability that the first alarm of the marginal charts is absorbed by the $C_n$-chart is $\mathbb{P}(\mathrm{RL}_G\neq\mathrm{RL}_B)=\mathbb{P}_1(\mathcal{M}^c\cap\mathcal{R})/\mathbb{P}_1(\mathcal{M}^c)$. In control, these probabilities equal $\delta_0/\alpha$ and $\eta_0/\alpha_{\mathcal{M}}$.
\item[(c)] Let $(\mathcal{M}_m)_{m\ge1}$ be marginal boxes with Lebesgue measure $\lambda(\mathcal{M}_m)\to0$, let $\mathrm{RL}_{G,m}$ be the run length of the gated chart with box $\mathcal{M}_m$, keep $\mathcal{R}$ fixed, and suppose $\mathbb{P}_1$ has a density $g_1$. Then $\mathrm{ARL}_{G,m}(\mathbb{P}_1)\to\mathrm{ARL}(\mathbb{P}_1)$ and, if $\mathbb{P}_1(\mathcal{R}^c)>0$, $\mathbb{P}(\mathrm{RL}_{G,m}\neq\mathrm{RL})\to0$. If $g_1\le L$, then
\[
\bigl|\mathrm{ARL}_{G,m}(\mathbb{P}_1)^{-1}-\mathrm{ARL}(\mathbb{P}_1)^{-1}\bigr|\le L\,\lambda(\mathcal{M}_m).
\]
\end{enumerate}
\end{theorem}

\begin{proof}
(a) If $\mathcal{M}\subseteq\mathcal{R}$, then $\mathcal{R}^c\subseteq\mathcal{M}^c$, so $\mathcal{S}_G=\mathcal{M}^c\cap\mathcal{R}^c=\mathcal{R}^c=\mathcal{S}\subseteq\mathcal{M}^c=\mathcal{S}_B$. A smaller signal set is entered no earlier, which gives the pathwise relations. The second statement follows by exchanging the roles of $\mathcal{M}$ and $\mathcal{R}$.

(b) By Theorem~\ref{thm:decomp}(iv), $\mathrm{RL}_G\ge\mathrm{RL}$. At time $\mathrm{RL}<\infty$ the observation lies in $\mathcal{R}^c$, so the gated chart also signals then if and only if it lies in $\mathcal{M}^c$. Hence $\{\mathrm{RL}_G\neq\mathrm{RL}\}=\{\mathrm{RL}<\infty,\ \boldsymbol{\varepsilon}_{n+\mathrm{RL}}\in\mathcal{M}\cap\mathcal{R}^c\}$. Since $\mathbb{P}_1(\mathcal{R}^c)>0$ implies $\mathrm{RL}<\infty$ almost surely (Lemma~\ref{lem:geom}),
\[
\begin{aligned}
\mathbb{P}(\mathrm{RL}_G\neq\mathrm{RL})&=\sum_{\ell\ge1}\mathbb{P}\bigl(\boldsymbol{\varepsilon}_{n+1}\in\mathcal{R},\ldots,\boldsymbol{\varepsilon}_{n+\ell-1}\in\mathcal{R},\ \boldsymbol{\varepsilon}_{n+\ell}\in\mathcal{M}\cap\mathcal{R}^c\bigr)\\
&=\sum_{\ell\ge1}\{1-\mathbb{P}_1(\mathcal{R}^c)\}^{\ell-1}\,\mathbb{P}_1(\mathcal{M}\cap\mathcal{R}^c)=\frac{\mathbb{P}_1(\mathcal{M}\cap\mathcal{R}^c)}{\mathbb{P}_1(\mathcal{R}^c)}.
\end{aligned}
\]
The statement for $\mathrm{RL}_B$ follows in the same way with $\mathcal{M}$ and $\mathcal{R}$ exchanged, and the in-control values follow from Definition~\ref{def:charts}.

(c) By Theorem~\ref{thm:decomp}(iii), $0\le\mathrm{ARL}(\mathbb{P}_1)^{-1}-\mathrm{ARL}_{G,m}(\mathbb{P}_1)^{-1}=\mathbb{P}_1(\mathcal{M}_m\cap\mathcal{R}^c)\le\int_{\mathcal{M}_m}g_1\,d\lambda$. By the absolute continuity of the Lebesgue integral, for every $\epsilon>0$ there is $\eta>0$ such that $\lambda(A)<\eta$ implies $\int_A g_1\,d\lambda<\epsilon$, so the right-hand side tends to zero. If $\mathbb{P}_1(\mathcal{R}^c)>0$, continuity of $x\mapsto1/x$ gives $\mathrm{ARL}_{G,m}(\mathbb{P}_1)\to\mathrm{ARL}(\mathbb{P}_1)$, and part (b) gives $\mathbb{P}(\mathrm{RL}_{G,m}\neq\mathrm{RL})\le\int_{\mathcal{M}_m}g_1\,d\lambda/\mathbb{P}_1(\mathcal{R}^c)\to0$. If $\mathbb{P}_1(\mathcal{R}^c)=0$, both ARLs are infinite. If $g_1\le L$, then $\int_{\mathcal{M}_m}g_1\,d\lambda\le L\lambda(\mathcal{M}_m)$.
\end{proof}

\begin{remark}
Part (a) shows that the $C_n$-chart and the marginal charts are the two extremes of the gated construction: the gated chart reduces to the $C_n$-chart when the box lies inside the tolerance region, and to the marginal charts when the tolerance region lies inside the box. Part (b) measures, for any fixed box, how often the first alarms of the charts differ, and part (c) shows that the gated chart recovers the $C_n$-chart as the marginal limits tighten.
\end{remark}

\subsubsection{Effect of Phase I estimation}\label{sec:arl-phase1}

In practice $\boldsymbol{\vartheta}_0$ is replaced by an estimator $\hat{\boldsymbol{\vartheta}}_n$ computed from the Phase I sample $(\boldsymbol{X}_1,\boldsymbol{Z}_1),\ldots,(\boldsymbol{X}_n,\boldsymbol{Z}_n)$, for instance by ML or IFM (Joe and Xu 1996). For a parameter value $\boldsymbol{\vartheta}$, the Phase II monitored vector is $\boldsymbol{\varepsilon}_{n+k}(\boldsymbol{\beta})=\bigl(r_1^{-1}(X_{1,n+k};\boldsymbol{Z}_{n+k},\boldsymbol{\beta}_1),\ldots,r_d^{-1}(X_{d,n+k};\boldsymbol{Z}_{n+k},\boldsymbol{\beta}_d)\bigr)^{\top}$, the density $h(\cdot\mid\boldsymbol{\theta},\boldsymbol{\psi})$ is \eqref{arl:h} at $(\boldsymbol{\theta},\boldsymbol{\psi})$, and $c_\alpha(\boldsymbol{\theta},\boldsymbol{\psi})$ is its $\alpha$-th quantile. The estimated $C_n$-chart signals when $\boldsymbol{\varepsilon}_{n+k}(\hat{\boldsymbol{\beta}})$ falls in
\[
\mathcal{S}(\boldsymbol{\theta},\boldsymbol{\psi})=\bigl\{\boldsymbol{e}:\ h(\boldsymbol{e}\mid\boldsymbol{\theta},\boldsymbol{\psi})<c_\alpha(\boldsymbol{\theta},\boldsymbol{\psi})\bigr\}
\]
evaluated at $(\hat{\boldsymbol{\theta}},\hat{\boldsymbol{\psi}})$; with an infinite Monte Carlo sample, $\hat{c}_n=c_\alpha(\hat{\boldsymbol{\theta}},\hat{\boldsymbol{\psi}})$ and $\hat{h}_n=h(\cdot\mid\hat{\boldsymbol{\theta}},\hat{\boldsymbol{\psi}})$.

\begin{assumption}[Estimation]\label{ass:est}
\begin{enumerate}
\item[(a)] The Phase II data are independent of the Phase I data, and $\sqrt{n}(\hat{\boldsymbol{\vartheta}}_n-\boldsymbol{\vartheta}_0)\xrightarrow{d}N_p(\boldsymbol{0},\boldsymbol{\Omega})$.
\item[(b)] The in-control signal probability $\varrho(\boldsymbol{\vartheta})=\mathbb{P}_0\{\boldsymbol{\varepsilon}_{n+k}(\boldsymbol{\beta})\in\mathcal{S}(\boldsymbol{\theta},\boldsymbol{\psi})\}$ is differentiable at $\boldsymbol{\vartheta}_0$ with gradient $\nabla\varrho_0$.
\end{enumerate}
\end{assumption}

Note that $\varrho(\boldsymbol{\vartheta}_0)=\alpha$. Let $\mathrm{CARL}_n=\mathbb{E}(\mathrm{RL}\mid\text{Phase I data})$ denote the conditional in-control ARL of the estimated $C_n$-chart, and let $\sigma_\varrho^2=\nabla\varrho_0^{\top}\boldsymbol{\Omega}\,\nabla\varrho_0$.

\begin{theorem}[Conditional in-control ARL]\label{thm:phase1}
Under Assumptions~\ref{ass:ic}, \ref{ass:cont} and \ref{ass:est}:
\begin{enumerate}
\item[(i)] Conditionally on the Phase I data, the run length is geometric with parameter $\varrho(\hat{\boldsymbol{\vartheta}}_n)$, so $\mathrm{CARL}_n=1/\varrho(\hat{\boldsymbol{\vartheta}}_n)$.
\item[(ii)] $\mathrm{CARL}_n\xrightarrow{\mathbb{P}}1/\alpha$.
\item[(iii)] $\sqrt{n}\,(\mathrm{CARL}_n-1/\alpha)\xrightarrow{d}N\bigl(0,\ \sigma_\varrho^2/\alpha^{4}\bigr)$.
\item[(iv)] If $\sigma_\varrho>0$, then for every $b\ge0$,
\[
\mathbb{P}\Bigl\{\mathrm{CARL}_n\le\frac{1}{\alpha}\Bigl(1-\frac{b}{\sqrt{n}}\Bigr)\Bigr\}\longrightarrow\Phi\Bigl(-\frac{b\,\alpha}{\sigma_\varrho}\Bigr).
\]
\end{enumerate}
\end{theorem}

\begin{proof}
(i) For a fixed $\boldsymbol{\vartheta}$, $\boldsymbol{\varepsilon}_{n+k}(\boldsymbol{\beta})$ is one measurable function of $(\boldsymbol{X}_{n+k},\boldsymbol{Z}_{n+k})$, so under Assumption~\ref{ass:ic} the sequence $\{\boldsymbol{\varepsilon}_{n+k}(\boldsymbol{\beta})\}_{k\ge1}$ is IID with signal probability $\varrho(\boldsymbol{\vartheta})$ for the set $\mathcal{S}(\boldsymbol{\theta},\boldsymbol{\psi})$. By the independence in Assumption~\ref{ass:est}(a), conditioning on the Phase I data amounts to fixing $\boldsymbol{\vartheta}=\hat{\boldsymbol{\vartheta}}_n$, and Lemma~\ref{lem:geom} gives a geometric run length with mean $1/\varrho(\hat{\boldsymbol{\vartheta}}_n)$.

(ii) Since $\sqrt{n}(\hat{\boldsymbol{\vartheta}}_n-\boldsymbol{\vartheta}_0)$ is tight, $\hat{\boldsymbol{\vartheta}}_n\xrightarrow{\mathbb{P}}\boldsymbol{\vartheta}_0$. Differentiability at $\boldsymbol{\vartheta}_0$ implies continuity there, so $\varrho(\hat{\boldsymbol{\vartheta}}_n)\xrightarrow{\mathbb{P}}\alpha>0$ by the continuous mapping theorem, and then $1/\varrho(\hat{\boldsymbol{\vartheta}}_n)\xrightarrow{\mathbb{P}}1/\alpha$.

(iii) Differentiability at $\boldsymbol{\vartheta}_0$ means $\varrho(\boldsymbol{\vartheta})=\alpha+\nabla\varrho_0^{\top}(\boldsymbol{\vartheta}-\boldsymbol{\vartheta}_0)+\epsilon_\varrho(\boldsymbol{\vartheta})$ with $\epsilon_\varrho(\boldsymbol{\vartheta})=o(\|\boldsymbol{\vartheta}-\boldsymbol{\vartheta}_0\|)$. Since $\|\hat{\boldsymbol{\vartheta}}_n-\boldsymbol{\vartheta}_0\|=O_{\mathbb{P}}(n^{-1/2})$, $\sqrt{n}\,\epsilon_\varrho(\hat{\boldsymbol{\vartheta}}_n)=o_{\mathbb{P}}(1)$, and Slutsky's lemma gives
\[
\sqrt{n}\{\varrho(\hat{\boldsymbol{\vartheta}}_n)-\alpha\}=\nabla\varrho_0^{\top}\sqrt{n}(\hat{\boldsymbol{\vartheta}}_n-\boldsymbol{\vartheta}_0)+o_{\mathbb{P}}(1)\xrightarrow{d}N(0,\sigma_\varrho^2).
\]
The map $x\mapsto1/x$ is differentiable at $\alpha$ with derivative $-1/\alpha^{2}$, so the delta method yields (iii).

(iv) Let $Y_n=\sqrt{n}\{\varrho(\hat{\boldsymbol{\vartheta}}_n)-\alpha\}$. For $n>b^2$, and since $\mathrm{CARL}_n=\infty$ when $\varrho(\hat{\boldsymbol{\vartheta}}_n)=0$,
\[
\mathrm{CARL}_n\le\frac{1}{\alpha}\Bigl(1-\frac{b}{\sqrt{n}}\Bigr)
\iff\varrho(\hat{\boldsymbol{\vartheta}}_n)\ge\frac{\alpha}{1-b/\sqrt{n}}
\iff Y_n-\frac{b\,\alpha}{1-b/\sqrt{n}}\ge0.
\]
By (iii) and Slutsky's lemma, $Y_n-b\alpha/(1-b/\sqrt{n})\xrightarrow{d}N(-b\alpha,\sigma_\varrho^2)$, a continuous law, so the probability converges to $1-\Phi(b\alpha/\sigma_\varrho)=\Phi(-b\alpha/\sigma_\varrho)$.
\end{proof}

\begin{remark}
(1) In this regression setting the IID property used in part (i) follows from Assumption~\ref{ass:ic} for every parameter value, so no further condition on the Phase II residuals is needed.
(2) Theorem~\ref{thm:phase1} describes the chart with an infinite Monte Carlo sample. A finite auxiliary sample of size $N$ adds a further error in $\hat{c}_n$ that vanishes as $N\to\infty$ (Baíllo and Cuevas 2006).
(3) Differentiability in Assumption~\ref{ass:est}(b) is a regularity condition on the boundary of $\mathcal{S}(\boldsymbol{\theta},\boldsymbol{\psi})$; it holds, for example, when $h$ is continuously differentiable in $(\boldsymbol{e},\boldsymbol{\vartheta})$ and $\nabla_{\boldsymbol{e}}h\neq\boldsymbol{0}$ on the level set $\{h=c_\alpha\}$. It is assumed here rather than proved.
(4) The relative standard error of $\mathrm{CARL}_n$ is approximately $\sigma_\varrho/(\alpha\sqrt{n})$, so small Phase I samples (such as $n=30$ and $n=28$ in Section~\ref{sec:verify})
can give conditional in-control ARLs noticeably different from $1/\alpha$. The same theorem holds for the gated chart with $\alpha$ replaced by $\alpha-\delta_0$.
\end{remark}
\nocite{joe1996estimation,baillo2006parametric}

\subsubsection{Hotelling's $T^2$ versus the Gaussian copula with normal margins}\label{sec:arl-gauss}

\begin{lemma}[Hotelling ellipsoid]\label{lem:hotelling}
Suppose that $C_{\boldsymbol{\psi}}$ is the Gaussian copula with positive definite correlation matrix $\Sigma$ and that $F_j$ is the $\text{Normal}(\mu_j,\sigma_j^2)$ CDF, for $j=1,\ldots,d$. Let $\mathbf{D}_\sigma=\operatorname{diag}(\sigma_1,\ldots,\sigma_d)$, $\xi_j=(e_j-\mu_j)/\sigma_j$ and $Q(\boldsymbol{e})=\boldsymbol{\xi}^{\top}\Sigma^{-1}\boldsymbol{\xi}$. Then $h$ is the $N_d(\boldsymbol{\mu},\mathbf{D}_\sigma\Sigma\mathbf{D}_\sigma)$ density, Assumption~\ref{ass:cont} holds, and
\[
\mathcal{R}=\bigl\{\boldsymbol{e}:\ Q(\boldsymbol{e})\le\chi^2_{d,1-\alpha}\bigr\},
\]
where $\chi^2_{d,1-\alpha}$ is the $(1-\alpha)$-quantile of the $\chi^2_d$ distribution. Hence the $C_n$-chart coincides with Hotelling's $T^2$ chart with known parameters. For $d=2$, $\chi^2_{2,1-\alpha}=2\ln(1/\alpha)$.
\end{lemma}

\begin{proof}
By Lemma~\ref{lem:iid}, $\boldsymbol{U}=(F_1(\varepsilon_1),\ldots,F_d(\varepsilon_d))$ has distribution function $C_{\boldsymbol{\psi}}(\boldsymbol{u})=\Phi_d(\Phi^{-1}(u_1),\ldots,\Phi^{-1}(u_d);\Sigma)$, so $\boldsymbol{\Xi}=(\Phi^{-1}(U_1),\ldots,\Phi^{-1}(U_d))\sim N_d(\boldsymbol{0},\Sigma)$. Since $\Phi^{-1}(F_j(e))=(e-\mu_j)/\sigma_j$, we have $\boldsymbol{\varepsilon}=\boldsymbol{\mu}+\mathbf{D}_\sigma\boldsymbol{\Xi}\sim N_d(\boldsymbol{\mu},\mathbf{D}_\sigma\Sigma\mathbf{D}_\sigma)$, whose density is $h(\boldsymbol{e})=\kappa_0\exp\{-Q(\boldsymbol{e})/2\}$ with $\kappa_0=(2\pi)^{-d/2}|\mathbf{D}_\sigma\Sigma\mathbf{D}_\sigma|^{-1/2}$. Under $\mathbb{P}_0$, $Q(\boldsymbol{\varepsilon})=\boldsymbol{\Xi}^{\top}\Sigma^{-1}\boldsymbol{\Xi}\sim\chi^2_d$, a continuous law, so $h(\boldsymbol{\varepsilon})$, a strictly decreasing function of $Q(\boldsymbol{\varepsilon})$, is continuous. For $t\in(0,\kappa_0)$, $\{h\ge t\}=\{Q\le-2\ln(t/\kappa_0)\}$, and $\mathbb{P}_0\{Q(\boldsymbol{\varepsilon})>x\}=\alpha$ if and only if $x=\chi^2_{d,1-\alpha}$. For $d=2$, $\chi^2_2$ is exponential with mean $2$, so $\chi^2_{2,1-\alpha}=2\ln(1/\alpha)$.
\end{proof}

\begin{example}[Dependence shift]\label{ex:counter}
Take $d=2$, $\rho=0.9$, $\alpha=0.01$ and marginal limits $\mu_j\pm3\sigma_j$, so that $a_1=a_2=0.0027$. Let $\mathbb{P}_1$ reverse the dependence, $\xi_2=-\xi_1$ with $\xi_1\sim N(0,1)$, which leaves both margins unchanged. Under $\mathbb{P}_1$, $Q(\boldsymbol{\varepsilon})=2\xi_1^2/(1-\rho)=20\xi_1^2$, and by Lemma~\ref{lem:hotelling} the $C_n$-chart signals when $20\xi_1^2>2\ln100\approx9.210$, that is, when $|\xi_1|>0.679$. Hence $\mathbb{P}_1(\mathcal{R}^c)\approx0.497$ and $\mathrm{ARL}(\mathbb{P}_1)\approx2.0$. The marginal charts signal only when $|\xi_1|>3$, so $\mathrm{ARL}_B(\mathbb{P}_1)\approx370$, and the gated chart has the same ARL because $|\xi_1|>3$ implies $20\xi_1^2>9.210$. The $C_n$-chart therefore detects almost at once a change in the dependence that the marginal charts essentially cannot detect, in agreement with Corollary~\ref{cor:dep}.
\end{example}

For a sign vector $\boldsymbol{\varsigma}\in\{-1,1\}^d$, let $\mathbf{D}_\varsigma=\operatorname{diag}(\boldsymbol{\varsigma})$ and, for $s>0$,
\[
\mathcal{O}_{\boldsymbol{\varsigma}}(s)=\{\boldsymbol{e}:\ \varsigma_j\xi_j>s,\ j=1,\ldots,d\}
\]
be the set of joint exceedances with sign pattern $\boldsymbol{\varsigma}$, in the standardized coordinates $\xi_j=(e_j-\mu_j)/\sigma_j$. With symmetric limits $\mathrm{LCL}_j=\mu_j-s\sigma_j$ and $\mathrm{UCL}_j=\mu_j+s\sigma_j$, every point of $\mathcal{O}_{\boldsymbol{\varsigma}}(s)$ exceeds all marginal limits, and $a_j=a=2\{1-\Phi(s)\}$.

\begin{theorem}[Joint exceedances are never absorbed]\label{thm:orthant}
In the setting of Lemma~\ref{lem:hotelling} with symmetric limits $\mu_j\pm s\sigma_j$, suppose that
\begin{enumerate}
\item[(a)] $\mathbf{D}_\varsigma\Sigma^{-1}\mathbf{D}_\varsigma\mathbf{1}\ge\boldsymbol{0}$ componentwise, and
\item[(b)] $s^2\gamma_{\boldsymbol{\varsigma}}>\chi^2_{d,1-\alpha}$, where $\gamma_{\boldsymbol{\varsigma}}=\boldsymbol{\varsigma}^{\top}\Sigma^{-1}\boldsymbol{\varsigma}$.
\end{enumerate}
Then $\mathcal{O}_{\boldsymbol{\varsigma}}(s)\subseteq\mathcal{S}_G\subseteq\mathcal{S}$. Consequently, for every law $\mathbb{P}_1$, every observation in $\mathcal{O}_{\boldsymbol{\varsigma}}(s)$ triggers a signal of the $C_n$-chart and of the gated chart, and $\mathbb{P}_0\{\boldsymbol{\varepsilon}\in\mathcal{O}_{\boldsymbol{\varsigma}}(s)\cap\mathcal{R}\}=0$: the $C_n$-chart never absorbs an in-control joint exceedance with pattern $\boldsymbol{\varsigma}$.
\end{theorem}

\begin{proof}
Put $\boldsymbol{y}=\mathbf{D}_\varsigma\boldsymbol{\xi}$ and $\mathbf{A}=\mathbf{D}_\varsigma\Sigma^{-1}\mathbf{D}_\varsigma$, which is positive definite because $\mathbf{D}_\varsigma$ is orthogonal. Then $Q(\boldsymbol{e})=\tilde Q(\boldsymbol{y})=\boldsymbol{y}^{\top}\mathbf{A}\boldsymbol{y}$, and $\mathcal{O}_{\boldsymbol{\varsigma}}(s)$ corresponds to $\{\boldsymbol{y}:y_j>s\ \forall j\}$, a subset of the closed convex set $K=\{\boldsymbol{y}:\boldsymbol{y}\ge s\mathbf{1}\}$. The function $\tilde Q$ is convex and differentiable, so for $\boldsymbol{y}\in K$ and $\boldsymbol{y}^*=s\mathbf{1}$,
\[
\tilde Q(\boldsymbol{y})\ge\tilde Q(\boldsymbol{y}^*)+\nabla\tilde Q(\boldsymbol{y}^*)^{\top}(\boldsymbol{y}-\boldsymbol{y}^*)=s^2\,\mathbf{1}^{\top}\mathbf{A}\mathbf{1}+2s\,(\mathbf{A}\mathbf{1})^{\top}(\boldsymbol{y}-s\mathbf{1})\ge s^2\,\mathbf{1}^{\top}\mathbf{A}\mathbf{1},
\]
because $s>0$, $\mathbf{A}\mathbf{1}\ge\boldsymbol{0}$ by (a), and $\boldsymbol{y}-s\mathbf{1}\ge\boldsymbol{0}$. Since $\mathbf{1}^{\top}\mathbf{A}\mathbf{1}=\gamma_{\boldsymbol{\varsigma}}$, condition (b) gives $Q(\boldsymbol{e})>\chi^2_{d,1-\alpha}$ for every $\boldsymbol{e}\in\mathcal{O}_{\boldsymbol{\varsigma}}(s)$, that is, $\mathcal{O}_{\boldsymbol{\varsigma}}(s)\subseteq\mathcal{R}^c$ by Lemma~\ref{lem:hotelling}. Also $\mathcal{O}_{\boldsymbol{\varsigma}}(s)\subseteq\mathcal{M}^c$, so $\mathcal{O}_{\boldsymbol{\varsigma}}(s)\subseteq\mathcal{M}^c\cap\mathcal{R}^c=\mathcal{S}_G\subseteq\mathcal{S}$, and $\mathcal{O}_{\boldsymbol{\varsigma}}(s)\cap\mathcal{R}=\emptyset$.
\end{proof}

\begin{corollary}[Bivariate case]\label{cor:biv}
For $d=2$ and correlation $\rho\in(-1,1)$, condition (a) holds for all four sign patterns, and $\gamma_{\boldsymbol{\varsigma}}=2/(1+\varsigma_1\varsigma_2\rho)$. Hence, if
\[
s^2>(1+|\rho|)\ln(1/\alpha),
\]
every observation whose two standardized coordinates both lie outside $[-s,s]$ is signaled by the $C_n$-chart.
\end{corollary}

\begin{proof}
Here $\Sigma^{-1}=(1-\rho^2)^{-1}\begin{pmatrix}1&-\rho\\-\rho&1\end{pmatrix}$, and $\mathbf{D}_\varsigma\Sigma^{-1}\mathbf{D}_\varsigma$ has the same form with $\rho$ replaced by $\varsigma_1\varsigma_2\rho$. Hence
\[
\mathbf{D}_\varsigma\Sigma^{-1}\mathbf{D}_\varsigma\mathbf{1}=\frac{1-\varsigma_1\varsigma_2\rho}{1-\rho^2}\,\mathbf{1}=\frac{1}{1+\varsigma_1\varsigma_2\rho}\,\mathbf{1}>\boldsymbol{0}
\quad\text{and}\quad
\gamma_{\boldsymbol{\varsigma}}=\frac{2}{1+\varsigma_1\varsigma_2\rho}\ge\frac{2}{1+|\rho|}.
\]
With $\chi^2_{2,1-\alpha}=2\ln(1/\alpha)$ from Lemma~\ref{lem:hotelling}, condition (b) holds for every $\boldsymbol{\varsigma}$ whenever $2s^2/(1+|\rho|)>2\ln(1/\alpha)$. Every point with both standardized coordinates outside $[-s,s]$ belongs to some $\mathcal{O}_{\boldsymbol{\varsigma}}(s)$, up to the null set $\{|\xi_1|=s\}\cup\{|\xi_2|=s\}$, and Theorem~\ref{thm:orthant} applies.
\end{proof}

\begin{remark}
At the level $\alpha=0.05$, Corollary~\ref{cor:biv} holds for every $|\rho|<1$ with $3\sigma$ marginal limits ($s=3$), but only for $|\rho|<0.282$ with $95\%$ marginal limits ($s=1.96$). With $s=3$ it holds for $|\rho|<0.954$ when $\alpha=0.01$ and for $|\rho|<0.522$ when $\alpha=0.0027$. Outside these ranges, part of a joint-exceedance region can lie inside $\mathcal{R}$ and be absorbed.
\end{remark}

\begin{corollary}[Small-$\alpha$ asymptotics]\label{cor:asym}
In the setting of Theorem~\ref{thm:orthant}, let $\alpha\to0$ and let the per-margin false-alarm probability $a=a(\alpha)$ satisfy $a(\alpha)\le\kappa\,\alpha$ for some constant $\kappa>0$, with $s=s(\alpha)=\Phi^{-1}(1-a/2)$. If condition (a) holds and $\gamma_{\boldsymbol{\varsigma}}>1$, there is $\alpha_0>0$ such that $\mathcal{O}_{\boldsymbol{\varsigma}}(s(\alpha))\subseteq\mathcal{S}_G$ for all $\alpha\in(0,\alpha_0)$. For $d=2$ this holds for all sign patterns whenever $|\rho|<1$.
\end{corollary}

\begin{proof}
\emph{Lower bound for $s^2$.} For $x>0$, the Mills-ratio bounds $\frac{x}{1+x^2}\phi(x)\le1-\Phi(x)\le\frac{\phi(x)}{x}$ hold. As $\alpha\to0$, $a\to0$ and $s\to\infty$, so we may take $s\ge1$. Then $x/(1+x^2)\ge1/(2x)$ at $x=s$, so $a/2=1-\Phi(s)\ge\phi(s)/(2s)$ and $\ln a\ge-s^2/2-\tfrac12\ln(2\pi)-\ln s$, that is, $s^2\ge2\ln(1/a)-\ln(2\pi)-2\ln s$. The upper bound gives $a/2\le\phi(s)/s\le\phi(s)$, so $s^2\le2\ln(2/a)$ and $2\ln s\le\ln\{2\ln(2/a)\}$. Combining, $s^2\ge2\ln(1/a)-\ln(2\pi)-\ln\{2\ln(2/a)\}$, and since $\ln(1/a)\ge\ln(1/\alpha)-\ln\kappa$,
\[
\liminf_{\alpha\to0}\frac{s(\alpha)^2}{2\ln(1/\alpha)}\ge1.
\]

\emph{Upper bound for $\chi^2_{d,1-\alpha}$.} Let $x_\alpha=\chi^2_{d,1-\alpha}$ and $V\sim\chi^2_d$. For $t\in(0,\tfrac12)$, Markov's inequality gives $\mathbb{P}(V\ge x)\le(1-2t)^{-d/2}e^{-tx}$; for $x>d$, the choice $t=(1-d/x)/2$ gives $\mathbb{P}(V\ge x)\le(x/d)^{d/2}e^{-(x-d)/2}$. Since $x_\alpha\to\infty$, for small $\alpha$ we have $x_\alpha>d$ and
\[
\alpha=\mathbb{P}(V\ge x_\alpha)\le\Bigl(\frac{x_\alpha}{d}\Bigr)^{d/2}e^{-(x_\alpha-d)/2},
\]
and hence
\[
x_\alpha\Bigl\{1-\frac{d\ln(x_\alpha/d)}{x_\alpha}\Bigr\}\le2\ln(1/\alpha)+d.
\]
Because $\ln(x_\alpha/d)/x_\alpha\to0$, this yields $\limsup_{\alpha\to0}x_\alpha/\{2\ln(1/\alpha)\}\le1$.

\emph{Conclusion.} From the two bounds, $\liminf_{\alpha\to0}s(\alpha)^2\gamma_{\boldsymbol{\varsigma}}/x_\alpha\ge\gamma_{\boldsymbol{\varsigma}}>1$, so condition (b) of Theorem~\ref{thm:orthant} holds for all sufficiently small $\alpha$. For $d=2$, Corollary~\ref{cor:biv} gives condition (a) and $\gamma_{\boldsymbol{\varsigma}}\ge2/(1+|\rho|)>1$ for $|\rho|<1$.
\end{proof}

The tolerance region of the $C_n$-chart is defined through the joint density $h$ in \eqref{arl:h}, which involves the marginal densities $f_j$ as well as the copula density. Density level sets are not invariant under monotone transformations of the margins, so a region built from the copula density $c_{\boldsymbol{\psi}}$ alone, on the scale of $\boldsymbol{U}$, behaves differently.

\begin{proposition}[Copula scale versus data scale]\label{prop:scale}
Let $d=2$, let $C_{\boldsymbol{\psi}}$ be the Gaussian copula with $\rho\in(0,1)$, and consider the copula-scale region $\mathcal{R}^U=\{\boldsymbol{u}:c_{\boldsymbol{\psi}}(\boldsymbol{u})\ge c^U_\alpha\}$, with $c^U_\alpha>0$. For $s>0$, let
\begin{gather*}
\mathcal{T}_s^{+}=\{\boldsymbol{u}:\ \Phi^{-1}(u_j)\in[s,s+1],\ j=1,2\},\\
\mathcal{T}_s^{-}=\{\boldsymbol{u}:\ \Phi^{-1}(u_j)\in[-s-1,-s],\ j=1,2\}.
\end{gather*}
Then there is $s_0$ such that $\mathcal{T}_s^{+}\cup\mathcal{T}_s^{-}\subseteq\mathcal{R}^U$ for all $s\ge s_0$. Consequently, these joint exceedances never trigger a signal of a chart based on $c_{\boldsymbol{\psi}}(\boldsymbol{U}_{n+k})$ alone, under any law $\mathbb{P}_1$. With normal margins, $\Phi^{-1}(U_j)=\xi_j$, and by Corollary~\ref{cor:biv} the $C_n$-chart signals all of them whenever $s^2>(1+\rho)\ln(1/\alpha)$.
\end{proposition}

\begin{proof}
With $\xi_j=\Phi^{-1}(u_j)$, the bivariate Gaussian copula density is
\[
c_{\boldsymbol{\psi}}(\boldsymbol{u})=\frac{\phi_{2,\rho}(\xi_1,\xi_2)}{\phi(\xi_1)\phi(\xi_2)}=(1-\rho^2)^{-1/2}\exp\Bigl\{-\frac{\rho^2(\xi_1^2+\xi_2^2)-2\rho\xi_1\xi_2}{2(1-\rho^2)}\Bigr\}.
\]
Expanding shows $\rho^2(\xi_1^2+\xi_2^2)-2\rho\xi_1\xi_2=\rho^2(\xi_1-\xi_2)^2-2\rho(1-\rho)\xi_1\xi_2$. On $\mathcal{T}_s^{\pm}$, $(\xi_1-\xi_2)^2\le1$ and $\xi_1\xi_2\ge s^2$, so
\[
\ln c_{\boldsymbol{\psi}}(\boldsymbol{u})\ge-\tfrac12\ln(1-\rho^2)-\frac{\rho^2}{2(1-\rho^2)}+\frac{\rho\,s^2}{1+\rho},
\]
using $2\rho(1-\rho)/\{2(1-\rho^2)\}=\rho/(1+\rho)$. The right-hand side tends to $+\infty$ as $s\to\infty$ because $\rho>0$, so it exceeds $\ln c^U_\alpha$ for all $s\ge s_0$. Hence $\mathcal{T}_s^{\pm}\subseteq\mathcal{R}^U$, which is disjoint from the signal set $(\mathcal{R}^U)^c$. The last statement is Corollary~\ref{cor:biv}, since under normal margins $\mathcal{T}_s^{\pm}$ corresponds to joint exceedances in the standardized coordinates.
\end{proof}

%%%%%%%%%%%%%%%%%%%%%%%%%%%%%%%%%%%%%%%%%%%%%%%%%%%%%%%%%
\section{RESULTS}
\label{sec:verify}

This section illustrates the proposed methodology with two case studies: (i) water quality control, measuring the relationship between the potential of hydrogen (pH) levels and the phosphate (PO4) concentration (Subsection \ref{sec:ap_biv}); and (ii) brass-steel thermostats monitoring, considering three response variables: deflection, curvature, and resistivity (Subsection \ref{sec:ap_tri}). %Both data sets were originally available in the \texttt{MSQC} package (Santos-Fernández 2013), implemented in the R \textit{software} (R Core Team 2022), can be seen in Supplementary Material - \url{https://github.com/ProfNascimento/Cn-chart}. \nocite{santos2012,RCRAN} Both data sets are available at \url{https://github.com/ProfNascimento/Cn-chart}. 
A simulation study of run-length performance and computational cost follows (Subsection~\ref{sec:simul}). In the case studies, the $C_n$-chart is compared with the univariate (marginal) control charts (Figure~\ref{fig:MARGs}), whose joint acceptance region is the rectangular box $\mathcal{M}$ of Definition~\ref{def:charts}, and, in the trivariate case, with Hotelling's $T^2$ chart. Whatever the dimension of the copula, the $C_n$-chart always delivers a 2D chart that decision-makers can read directly.

\subsection{Water Quality Control Case}\label{sec:ap_biv}

The first data set, \texttt{water1.csv}, stored in the GitHub repository (\url{https://github.com/ProfNascimento/Cn-chart}) of the first author, comes from a water quality test and consists of 30 observations (Phase I) of five variables: pH levels, phosphates (mg/L), nitrates (mg/L), dissolved oxygen, and total solids (mg/L) measured in the water sample. For ease of exposition, this first example uses only two variables: $X=1/\text{pH}$, the reciprocal of the pH level, and $Y=1/\sqrt{\text{PO4}}$, the reciprocal square root of the PO4 concentration. These transformations make the relationship between the variables positive and the data more homogeneous; the transformed PO4 variable remains clearly non-normal. The data set \texttt{water2.csv}, from the same repository and process, contains 25 observations (Phase II).
%O conjunto de dados intitulado \textit{water1}, armazenado no pacote \textit{MSQC} do R, possui 30 observa\c{c}\~oes referentes aos n\'iveis de pH, fosfato (mg/L), nitrato (mg/L), oxig\^enio dissolvido e s\'olidos totais (mg/L) medidos em um teste de qualidade de \'agua. Para fins de ilustração, utilizamos apenas duas variáveis, a qual foram transformadas como $X=1/\text{pH}$ e a variável $Y=1/\sqrt{ fosfato}$. Exclu\'imos ainda a observa\c{c}\~ao 12, restando $n=29$ observações no conjunto de dados (tais tranforma\c{c}\~oes foram necess\'arias para manter a rela\c{c}\~ao positiva entre as vari\'aveis, apresentar uma proposta n\~ao-normal dos dados e mante-los mais homog\^eneos). 

Table \ref{TA:biv} shows the main summary statistics of the variables under study (Phase I data). Variable $X$ ranges from $0.1160$ to $0.1855$, with median and mean close to each other ($0.1456$ and $0.1472$, respectively, suggesting a symmetric distribution), and a standard deviation of $0.0179$. Variable $Y$ has a minimum value of $1.9610$ and a maximum value of $10.0000$. Its median and mean values are, respectively, $2.8870$ and $3.1560$; the maximum is an upper outlier, which produces positive (right) skewness.
%A Tabela \ref{t_a_biv} apresenta as principais medidas-resumo das vari\'aveis em estudo. Observe que $X$ variou de $0,1160$ a $0,1855$, com mediana e m\'edia muito pr\'oximas ($0,1456$ e $0,1462$, respectivamente) e desvio-padr\~ao igual a $0,0174$. A vari\'avel $Y$ apresentou valor m\'inimo de $1,9610$ e m\'aximo de $5,000$. Ademais, metade das observa\c{c}\~oes de $Y$ se encontram abaixo de $2,8870$ e seu valor m\'edio \'e de $2,9200$, indicando uma leve assimetria positiva.

\begin{table}[h!] %\footnotesize
\caption{Descriptive statistics of the \texttt{water1} data set ($n=30$). SD $=$ standard deviation.}
\label{TA:biv}
\begin{center}
\begin{tabular}{cccccccc}
\hline
\multicolumn{1}{c}{Variable} & {Minimum} & $1$st Quartile    & Median & Mean & $3$rd Quartile & Maximum  & SD     \\ \hline
$X$ &   0.1160    & 0.1346& 0.1456& 0.1472& 0.1575& 0.1855& 0.0179\\
$Y$  &  1.9610     & 2.6730  & 2.8870 & 3.1560 & 3.1620 & 10.0000& 1.4083\\
\hline
\end{tabular}
\end{center}
\end{table}

\begin{figure}[h!]
\begin{center}
\includegraphics[width=\textwidth]{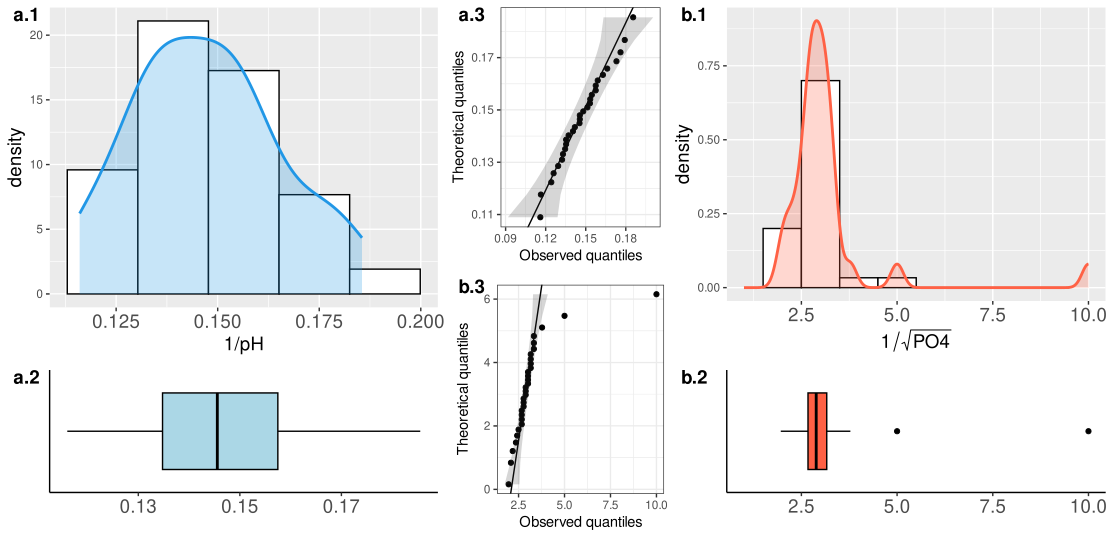}
\caption{Histograms, kernel densities, normal probability plots and boxplots. The panels labeled ``a'' are related to the pH rates, and the panels labeled ``b'' are related to the PO4 rates.}
\label{f_desc_a}
\end{center}
\end{figure}

%%%%%%%%%%%%%%%%%%%%%%%%%%%%%%%%%%%%%%%%%%%%%%%%%%%%
Visual representations, such as the histogram, kernel density plot, normal probability plot, and boxplot, for each adopted variable, are presented in Figure \ref{f_desc_a}. These plots suggest that the variable $X$ might have originated from a normal distribution, unlike the variable $Y$, whose distribution exhibits positive skewness (due to the presence of an upper outlier). Therefore, the Shapiro-Wilk (SW) normality test (Shapiro and Wilk 1965) was applied to each variable. In addition, the generalization of the SW test was also adopted, which is related to the hypothesis of multivariate normality (Villaseñor Alva and González Estrada 2009). Table \ref{t_a_norm} shows the SW test statistics, and the generalized SW test statistic, as well as their respective $p$-values, which confirm the visual patterns. \nocite{shapiro1965analysis,villasenor2009generalization}

%%%%%%%%%%%%%%%%%%%%%%%%%%%%%%%%%%%%%%%%%%%%%%%%%%%%
\begin{table}[h!] %\footnotesize
\caption{Normality tests for univariate and bivariate scenarios (\texttt{water1} data set).}
\label{t_a_norm}
\begin{center}
\begin{tabular}{ccc}
\hline
\multicolumn{1}{c}{Variable(s)} & {Test Statistic} & $p$-value      \\ \hline
$X$  &  $0.9790$   & 0.8045\\ 
$Y$  &  $0.5090$   & $6.69 \times 10^{-9}$ \\ 
$(X,Y)$  & $0.4443$ & $1.45 \times 10^{-9}$\\ 
\hline       
\end{tabular}
\end{center}
\end{table}
%%%%%%%%%%%%%%%%%%%%%%%%%%%%%%%%%%%%%%%%%%%%

After describing the distribution of each variable, the next question concerns their dependence. Kendall's $\tau$ ($\widehat{\tau}=0.5979$) indicates a moderate positive association between them.

Given this dependence, a copula-based model is appropriate. For variable $X$, due to the results presented in Figure \ref{f_desc_a} and Table \ref{t_a_norm}, the normal distribution was adopted. For variable $Y$, the selected univariate distribution was based on the Akaike information criterion (AIC; Akaike 1977), i.e., the best choice was the one with the lowest AIC value among all tested models. Among the candidate distributions were: normal ($\text{AIC}=52.5596$), gamma ($\text{AIC}=48.8620$), Weibull ($\text{AIC}=58.9891$), and logistic ($\text{AIC}={47.2012}$); the logistic distribution was selected. All distributions were fitted using the R  \texttt{fitdistrplus} package (Delignette-Muller and Dutang 2015). The Kolmogorov-Smirnov goodness-of-fit (GOF) test (Kolmogorov 1933; Smirnov 1939) also confirmed the suitability of the logistic distribution in representing the probabilistic behavior of $Y$ ($p$-value $=0.5537$). Once the appropriate marginal distributions are fitted, the next step is to fit the link function that fully describes the joint distribution of $X$ and $Y$, that is, the copula function. \nocite{akaike1977,fitdistrplus,kolmogorov1933sulla,smirnov1939estimate}
%Confirmada a relação de dependência positiva e moderada entre $X$ e $Y$. Inicialmente, selecionamos a distribui\c{c}\~ao univariada que melhor se adequa \`a variável $Y$ (adotamos a distribui\c{c}\~ao normal para a variável $X$ devido aos resultados apresentados na Figura \ref{f_desc_a} e Tabela \ref{t_a_norm}), com base no crit\'erio de Akaike (AIC, do inglês \textit{Akaike information criterion}) (Akaike, 1977). Entre as distribui\c{c}\~oes candidatas: normal ($\text{AIC}=52,5596$), gama ($\text{AIC}=48,8620$), Weibull ($\text{AIC}=58,9891$) e log\'istica ($\text{AIC}=47,2012$), esta última foi a que obteve o melhor ajuste segundo o AIC (isto é, possui o menor valor de AIC dentre todos os modelos testados). O teste de aderência de Kolmogorov-Smirnov confirma a boa adequabilidade da distribuição logística em representar os dados dessa variável ($D = 0,1501$, com $\text{p-valor} = 0,5537$). Once the appropriated marginal distributions were fitted, the next step is to adjust the link function that will fully describe the joint distribution, that is, the copula density.

\begin{figure}
    \centering
    \includegraphics[width=\textwidth]{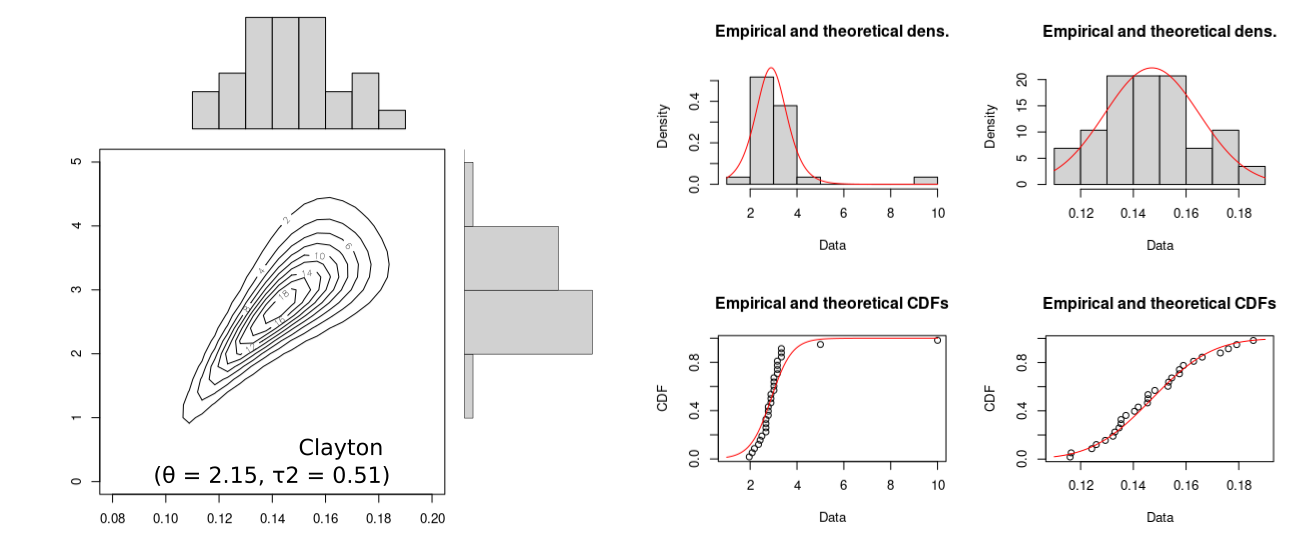}
    \caption{The left-hand panel shows the empirical marginal distributions, and the estimated copula density for the transformed sample, that is, $\text{Clayton}\left(\widehat{\theta}=2.1567\right)$. The right-hand panel, more specifically, the densities (top charts) and cumulative frequencies (bottom charts), show the fit for the two variables, $Y=1/\sqrt{\text{PO4}} \sim \text{Logistic}\left(\widehat{\mu}=2.896, \widehat{s}=0.445\right)$ and $X=1/\text{pH} \sim \text{Normal}\left(\widehat{\mu}=0.147,\widehat{\sigma}=0.018\right)$, as solid red lines.}
    \label{fig:MARG+COPULA}
\end{figure}

Thus, a set of parametric copula families was tested, and the best candidate model was selected via AIC. The fitted copulas were: Clayton ($\text{AIC}={-125.2316}$), Frank ($\text{AIC}=-122.3888$), Gumbel ($\text{AIC}=-109.5409$), Gaussian ($\text{AIC}=-122.3888$), and Student-t ($\text{AIC}=-120.0088$). Figure \ref{fig:MARG+COPULA} summarizes both steps: the contours of the best-fitting copula density, the estimated densities of $X$ and $Y$, and the empirical cumulative probabilities (empirical CDFs) with their model-based counterparts.

Among the tested copula families, the $\text{Clayton}\left(\widehat{\theta}=2.1567\right)$ copula was selected as the best one according to the AIC. This corresponds to an estimated Kendall correlation coefficient of $\widehat{\tau}=0.5188$.
A GOF test was carried out for copulas (for further details, see Genest et al. 2009), using the  \textit{gofCopula}($\cdot$) function of the R  \texttt{copula} package (Yan 2007), which did not reject the null hypothesis that the copula belongs to the Clayton family ($p$-value $= 0.3931$). \nocite{genest2009goodness,yan2007enjoy}
%O critério AIC para selecionar a família de cópulas mais adequada aos dados, assim a Clayton ($\hat{\theta}=2,1567$, $\hat{\tau}_2=0,5188$) foi a selecionada. Realizamos o teste de bondade do ajuste da c\'opula estimada por meio da fun\c{c}\~ao \textit{gofCopula(.)} (Genest \textit{et al}, 2009), assim n\~ao rejeitamos a hip\'otese de que os dados s\~ao oriundos da fam\'ilia de Clayton ($S_n=0,0321$, com $\text{p-valor} = 0,3931$).

%%%%%%%%%%%%%%%%%%%%%%%%%%%%%%%%%%%%%%%
\begin{figure}[h!]
\begin{center}
\includegraphics[width=\textwidth]{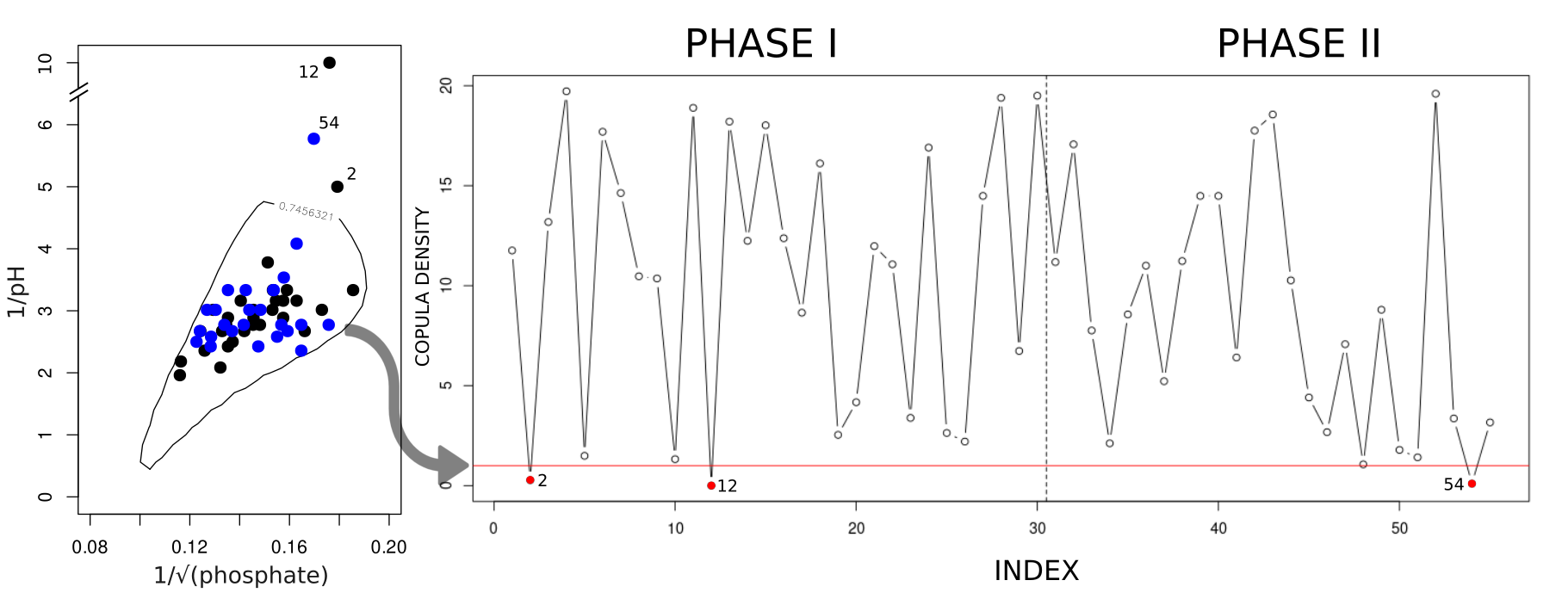}
\caption{The left-hand panel draws the  estimated tolerance region as a solid black line, at the 5\% level, from the joint density built with the Clayton copula and normal and logistic margins (lower bound $\hat{c}_n=0.7456$). Phases I and II data are plotted as black and blue dots (left-hand chart), respectively. The right-hand panel represents the obtained density level ($\hat{c}_n=0.7456$, as a solid red line), considering both Phases I and II data, and highlights the anomalous water samples \#2, \#12 and \#54 as red dots. Both representations lead to the same conclusion; however, the left-hand plot is limited to two dimensions, whereas the right-hand plot ($C_n$-chart) summarizes it regardless of the higher dimensions ($d \geq 2$).}
\label{f_rt_agua}
\end{center}
\end{figure}
%%%%%%%%%%%%%%%%%%%%%%%%%%%%%%%%%%%%%

Once the estimation process is completed, the proposed $C_n$-chart displays the level set of the estimated joint density (the copula combined with the fitted margins) as the tolerance region. Figure \ref{f_rt_agua} shows this tolerance region in two perspectives: a bivariate copula density with the observation points across these joint fields (left-hand panel), and plotted as a traditional 2D control chart (right-hand panel).
The density level of 0.7456 (tolerance region from the joint density based on the estimated Clayton copula) corresponds to a significance level of 5\%. Moreover, the pairs of observations $(x_1,y_1),\ldots,(x_{30},y_{30})$ were used in estimating this region, that is, only the information from Phase I data was used. Then, the following 25 observations from the same process were considered for monitoring purposes (Phase II). 

The estimated tolerance region based on the Clayton copula with normal and logistic marginal distributions fits the data well. Two out-of-control signals were generated in Phase I (samples \#2 and \#12), which is consistent with the 5\% level (1.5 false alarms expected among 30 observations). Additionally, only one Phase II observation was identified as out of control (sample \#54).
%Figure \ref{f_rt_agua} é exibida a regi\~ao de toler\^ancia construída sob a \'otica de conjunto de n\'iveis de densidade baseada em fun\c{c}\~oes c\'opula, com n\'ivel de signific\^ancia de $5\%$. Al\'em de plotar os pares de observações $(x_1,y_1),\ldots,(x_{29},y_{29})$ utilizados na estimação dessa regi\~ao (fase I), tamb\'em plotamos uma sequ\^encia de 25 novas observa\c{c}\~oes do mesmo processo (fase II), dispon\'iveis no conjunto de dados \textit{water2} do pacote \textit{MSQC}. Nota-se que a regi\~ao de tolerância estimada a partir da c\'opula de Clayton com distribuições marginais normal e logística, se adequou bem ao comportamento dos dados. Dois alarmes ``falsos'' foram identificados, as observa\c{c}\~oes 2 e 5 (quantidade aceit\'avel e, até mesmo esperada, mediante o n\'ivel de signific\^ancia adotado). Bem como, quatro novas observa\c{c}\~oes foram identificadas como fora de controle (observações 44, 47, 50 e 53).
 
\begin{figure}[!ht]
    \centering
    \includegraphics[scale=0.5]{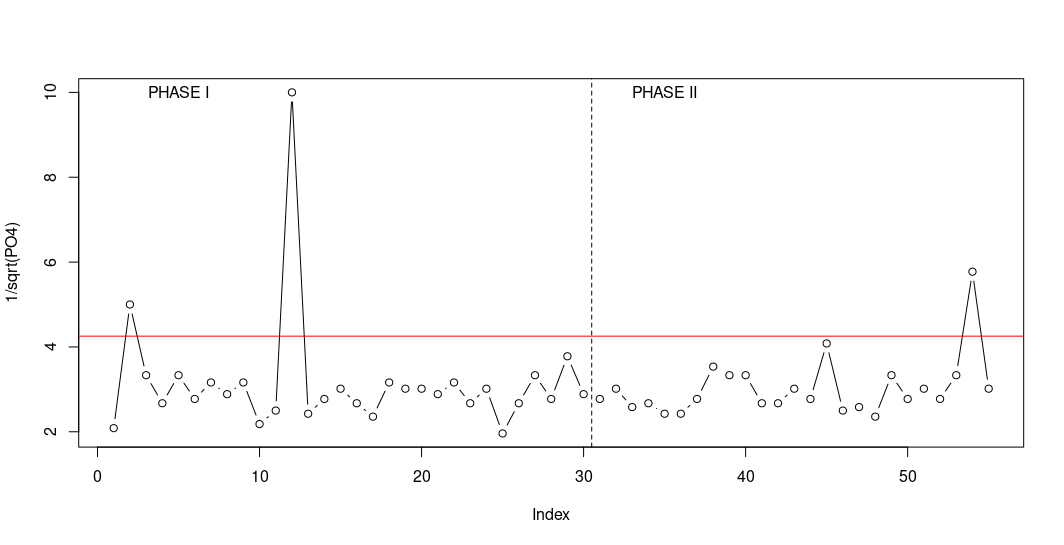} \quad
    \includegraphics[scale=0.5]{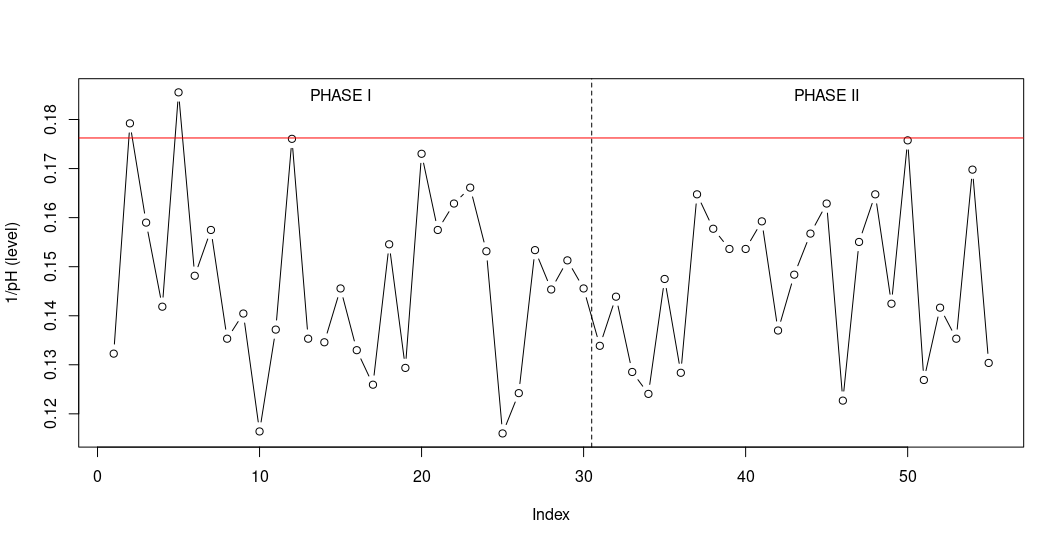}
    \caption{Marginal control charts based on the fitted models (top: logistic, variable $Y$; bottom: normal, variable $X$). The solid red lines are the 95\% probability limits.}
    \label{fig:MARGs}
\end{figure}

Examining the marginal charts at the same 5\% level (or equivalently, a tolerance level of 95\%), samples \#2, \#12 and \#54 are out of control for the fitted marginal distribution of $Y$ (logistic distribution), as shown in the top plot of Figure \ref{fig:MARGs}, whereas, for the fitted marginal distribution of $X$ (normal distribution), samples \#2, \#5 and \#12 are out of control, as can be seen in the bottom plot of Figure \ref{fig:MARGs}. Note that, with two marginal charts each at the 5\% level, the overall false-alarm probability of the marginal charts lies between 5\% and 10\%, whereas that of the $C_n$-chart is 5\% (Theorem~\ref{thm:decomp}).  

In summary, the proposed methodology links the marginal behaviors through their dependence, which would be ignored under an independence assumption. Notice that sample \#5, if judged only marginally, would be considered an out-of-control signal, although according to the proposed approach, this is an accepted pattern based on the joint behavior of the process: it lies in the absorbed region $\mathcal{M}^c\cap\mathcal{R}$ of Theorem~\ref{thm:decomp}(ii), whereas samples \#2, \#12 and \#54 lie in $\mathcal{M}^c\cap\mathcal{R}^c$ and are signaled by both approaches. Thus, copula-based tolerance regions offer flexible, non-rectangular shapes for the joint distribution. To illustrate further the flexibility of our proposal, the second part of this subsection considers a conditional mean (regression) model for one margin, combined with variable selection.
%%%%%%%%%%%%%%%%%%%%%%%%%%%%%%%%%%%%%%%%%%%%%%

In practice, a margin may depend on covariates, so that its conditional mean must be modeled before the joint analysis. For instance, consider again the water quality test, in which a researcher is interested in the association between the pH level and the nitrates (mg/L), dissolved oxygen and total solids (mg/L) measurements. Whatever remains unexplained (the residual) is then related to PO4 (mg/L) through a copula.

That is, as a prior step to the proposed methodology, variable selection is required. Any regression structure could be adopted (e.g., linear, polynomial, Least Absolute Shrinkage and Selection Operator (LASSO) or ridge, tree-based, support vector regression - SVR; for details, see, e.g., James et al. 2021), but for the sake of simplicity, the linear one was chosen. The full model is: \nocite{james2021introduction}
\begin{equation} \label{eq1}
    {\text{pH level}} = \beta_0 + \beta_1 \, {\text{nitrates}} + \beta_2 \, {\text{oxygen}} + \beta_3 \, {\text{solids}} + \epsilon,
\end{equation}
where $\beta_0$ is the intercept, $\beta_i$, for $i=1,2,3$, are parameters associated with the contribution of each covariate, and $\epsilon$ is the error (or residual) term. The residual standard error of model \eqref{eq1} is 0.6499 on 26 degrees of freedom (DF). Other results of such a multiple linear regression analysis include: multiple $R^{2} = 0.4611$, adjusted $R^{2} = 0.399$, and $F$-statistic $= 7.417$ on 3 and 26 DF, with $p$-value $=0.0009545$. Table \ref{reg1} shows the ordinary least-squares (OLS) estimates based on the model \eqref{eq1}.

\begin{table}[!ht]
\centering
\caption{Estimation results for the full regression model \eqref{eq1}. SE $=$ standard error.}
\begin{tabular}{l|cccc}
\hline
 Parameter & \multicolumn{1}{c}{ Estimate} & \multicolumn{1}{c}{ SE} & \multicolumn{1}{c}{ $t$-value} & \multicolumn{1}{c}{ $p$-value} \\ \hline
$\beta_0$ (intercept) & -10.937 & 5.669 & -1.929 & 0.0650 \\ 
$\beta_1$ (nitrates) & 1.194 & 1.014 & 1.177 & 0.2500 \\ 
$\beta_2$ (oxygen) & 0.155 & 0.065 & 2.387 & \textbf{0.0246} \\ 
$\beta_3$ (solids) & 0.011 & 0.027 & 0.386 & 0.7030 \\ \hline
\end{tabular}
\label{reg1}
\end{table}

At the 5\% significance level, only dissolved oxygen is significantly associated with the pH level. Then, the reduced model is:
\begin{equation} \label{red-mod}
    {\text{pH level}} = \beta^{*}_{0} + \beta^{*}_{2} \,  {\text{oxygen}} + \epsilon^{*},
\end{equation}

\begin{figure}[!ht]
    \centering
    \includegraphics[width=\textwidth]{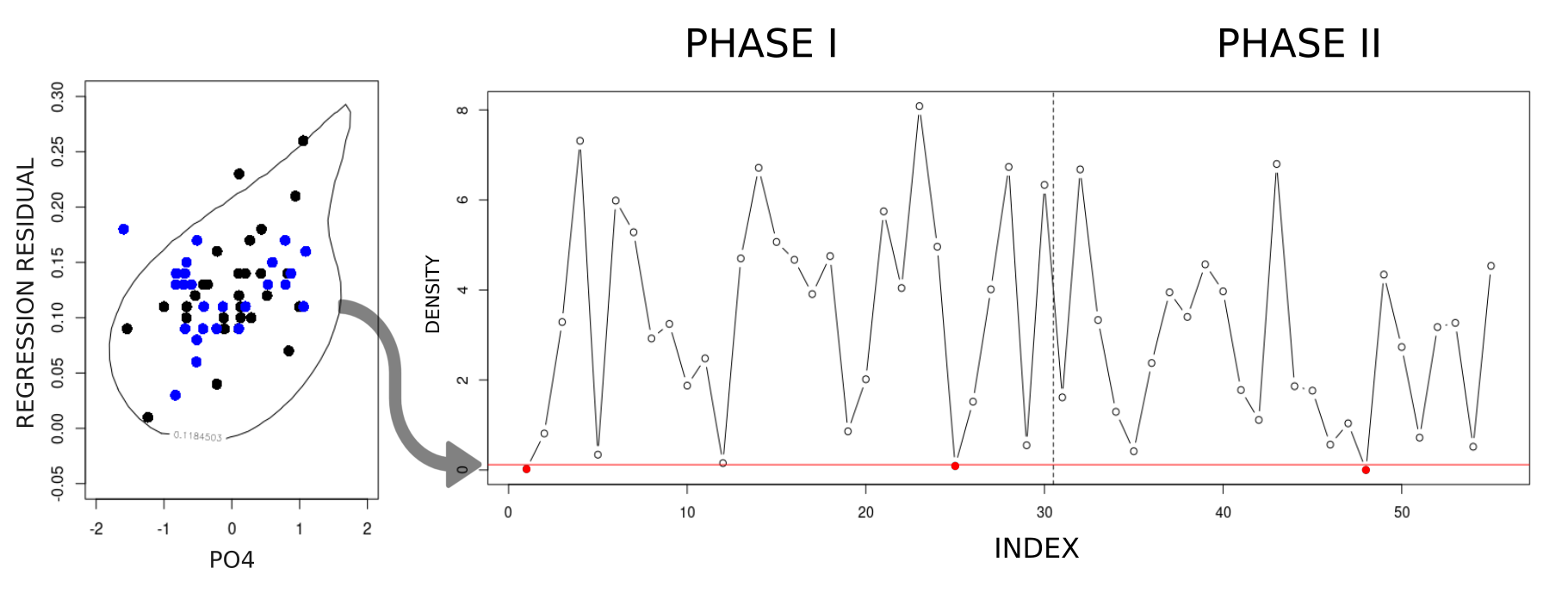}
    \caption{The left-hand panel draws the $\text{TT1}\left(\widehat{\boldsymbol{\theta}}=(2.37,0.51)\right)$ copula density, with the 5\% tolerance region shown as a solid black line. The right-hand panel represents the obtained density level ($\hat{c}_n=0.1185$, as a solid red line), considering both Phases I and II data, and highlights the anomalous water samples \#1, \#25 and \#48 as red dots.}
    \label{fig:COP2}
\end{figure}

\begin{table}[!ht]
\centering
\caption{Estimation results for the reduced regression model.}
\begin{tabular}{l|cccc}
\hline
Parameter & \multicolumn{1}{c}{ Estimate} & \multicolumn{1}{c}{ SE} & \multicolumn{1}{c}{ $t$-value} & \multicolumn{1}{c}{ $p$-value} \\ \hline
$\beta^{*}_{0}$ (intercept) & -14.283 & 4.753 & -3.005 & \textbf{0.0055} \\
$\beta^{*}_{2}$ (oxygen) & 0.214 & 0.048 & 4.457 & \textbf{0.0001} \\ \hline
\end{tabular}
\label{reg2}
\end{table}

\noindent which presents the following results: multiple $R^2 = 0.415$, adjusted $R^2 = 0.3941$, and $F$-statistic $= 19.86$ on 1 and 28 DF, with $p$-value $= 0.0001223$. Hence, the reduced model \eqref{red-mod} explains 39.41\% of the variability of the pH level (adjusted $R^2$). The OLS estimates based on this model are given in Table \ref{reg2}. For the $C_n$-chart, the residuals of model \eqref{red-mod} were modeled as $\text{Normal}\left(\widehat{\mu}=0, \widehat{\sigma}=0.63\right)$. The best-fitted copula family, according to the AIC, was Tawn type 1, that is,  $\text{TT1}\left(\widehat{\boldsymbol{\theta}}=(2.37,0.51)\right)$, which corresponds to a Kendall correlation coefficient of $\widehat{\tau} = 0.35$ between the residual and the PO4 level. Further details on the Tawn copula can be found in Kraus (2017). The remaining dependence between the residual and PO4 is captured by the copula (Figure \ref{fig:COP2}). \nocite{kraus2017d}
% ORIGINAL (removed: Kendall's tau is not a proportion of explained variance, and the procedure is not bagging): This bagging methodology explains, linearly, (39.41+60.59$\times \widehat{\tau}$) $\approx$ 60.62\% of the process' total variance (Figure \ref{fig:COP2}).

This second result shows that accounting for covariates changes which observations are flagged: once part of the process variation is explained by the regression, the chart targets changes in what remains unexplained. Moreover, once the regression is included, the previously flagged samples are explained by their dissolved oxygen levels and therefore now lie inside the tolerance region. For instance, by checking the tolerance bounds marginally, samples \#1, \#10 and \#25 are in the lower tail of the PO4 distribution, samples \#2, \#29 and \#42 are out of control in the dissolved oxygen covariate, and samples \#5, \#12, \#48 and \#51 are out of control when analyzing the regression model's residuals (based on the lower tail of their distribution). That is, looking only marginally (i.e., univariately), a larger number of out-of-control signals was obtained.

\subsection{Brass-Steel Thermostat Monitoring}\label{sec:ap_tri}

The next example explores a stronger association among three variables measured on brass-steel thermostats. Also stored in the GitHub repository (\url{https://github.com/ProfNascimento/Cn-chart}) of the first author of this work under the name \texttt{bimetal1.csv}, this data set has 28 records (Phase I) and contains the measurements of $X$ (deflection level, in $10^{-6}$ 1/K), $Y$ (curvature level, in $10^{-6}$ 1/K), and $Z$ (resistivity level, in $10^{-1}$ $\Omega \cdot \text{mm}^2/\text{m}$). The data set \texttt{bimetal2} is related to the same process and also contains 28 records (Phase II).  

Table \ref{t_m_tri} shows the main summary statistics of the three variables of the  data set \texttt{bimetal1}. The thermostats' deflection range of the observed process is $1.35$ ($10^{-6}$ 1/K), with a mean value of $21.02$ and a median value of $21.07$. The average curvature level is $40.02$ ($10^{-6}$ 1/K), and $50\%$ of the thermostats have a curvature lower than $40.03$ and a resistivity below $15.22$ ($10^{-1}$ $\Omega \cdot \text{mm}^2/\text{m}$). Both variables (deflection and curvature levels) seem to have a symmetric distribution, based on the comparison of their respective mean and median values.
%A amplitude da deflex\~ao dos termostatos foi de $1,35$, com m\'edia ($21,02$) e mediana ($21,07$) aproximadamente iguais. A curvatura m\'edia \'e de $40,02$ e, al\'em disso, $50\%$ dos termostatos apresentam curvatura inferior a $40,03$ e resistividade inferior a $15,22$. As tr\^es vari\'aveis aparentam simetria na distribuição dos dados, al\'em de terem baixa variabilidade.

\begin{table}[h!] %\footnotesize
\caption{Descriptive statistics of the \texttt{bimetal1} data set ($n=28$).}
\label{t_m_tri}
\begin{center}
\begin{tabular}{cccccccc}
\hline
\multicolumn{1}{c}{Variable} & {Minimum} & 1st Quartile & Median & Mean & 3rd Quartile & Maximum  & SD     \\ \hline
$X$  & 20.42 & 20.81 & 21.07 & 21.02 & 21.22 & 21.77 & 0.30\\
$Y$  &    39.73 &    39.94 &    40.03 &    40.02 &    40.11 &    40.31 &    0.14\\ 
$Z$ &      14.35 &      14.98 &      15.22 &      15.19 &      15.41 &      15.71 &      0.33\\
\hline    
\end{tabular}
\end{center}
\end{table}

\begin{figure}[!ht]
\begin{center}
  \includegraphics[scale=0.65]{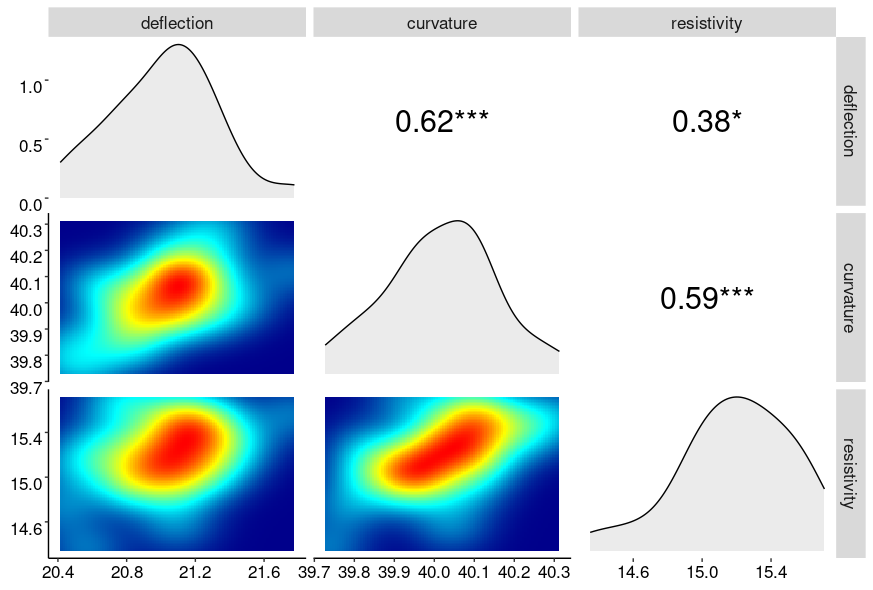}
  \caption{Kernel density plots (main diagonal), Pearson correlations (upper triangle), and 2D contour densities (lower triangle). The estimated contour densities in red represent the higher spots and the ones in dark blue have the lower values.}
  \label{t_desc_m}
  \end{center}
\end{figure}

Figure \ref{t_desc_m} shows, for this data set, the kernel density plots (on the main diagonal), the estimated Pearson correlations (on the upper triangle), and the 2D contour density estimates (on the lower triangle, where the highest spot is shown in red and the lowest in blue). Each correlation estimation conducted a hypothesis test considering the significance codes as:  ‘***’ 0.001, ‘**’ 0.01, and ‘*’ 0.05.
%Os histogramas, os gráficos de densidade Kernel, os gráficos de probabilidade normal e os \textit{boxplots} exibidos na Figura \ref{t_desc_m} mostram que as tr\^es vari\'aveis em estudo e suas relações dois-à-dois. 

Some asymmetry can be seen in the observed data, especially for $Z$. Therefore, the skewness and kurtosis of each variable were calculated. Table \ref{TAB:symm} shows that all variables are approximately mesokurtic (kurtosis close to 3), although the resistivity ($Z$) variable is moderately left-skewed (skewness below $-0.5$).

\begin{table}[!ht]
\caption{Skewness and kurtosis measures of the brass-steel thermostats' variables.}
\label{TAB:symm}
\begin{center}
\begin{tabular}{lccc}
\hline
 & \multicolumn{1}{l}{$X$ (Deflection)} & \multicolumn{1}{l}{$Y$ (Curvature)} & \multicolumn{1}{l}{$Z$ (Resistivity)} \\ \hline
Skewness & 0.083 & -0.073 & \textbf{-0.615} \\ 
Kurtosis & 3.042 & 2.749 & 3.143 \\ \hline
\end{tabular}
\end{center}
\end{table}

Nevertheless, the SW normality test and its generalized/multivariate version were applied. The univariate normality hypothesis (for each variable) was not rejected, as well as the multivariate normality hypothesis (among the three variables), at any conventional significance level. The test statistics and their corresponding $p$-values are shown in Table \ref{t_m_norm}. %The results associated with the variables deflection level ($SW = 0,9803$, equivalent $\text{p-value} = 0,8588$), curvature level ($SW = 0,9819$, that is $\text{p-value} = 0,8927$) and resistivity level ($SW = 0,9650$, with $\text{p-value} = 0,4562$). As well the multivariate normal hypothesis among the three of them ($MVW=0,9670$, with $\text{p-value} = 0,6062$) shows evidence in favor of the normality (Table \ref{t_m_norm}). 

\begin{table}[h!] %\footnotesize
\caption{Normality tests for univariate and trivariate scenarios (\texttt{bimetal1} data set).}
\label{t_m_norm}
\begin{center}
\begin{tabular}{ccc}
\hline
\multicolumn{1}{c}{Variable(s)} & {Test Statistic} & $p$-value   \\ \hline
$X$  &  $0.980$   &  0.858 \\ 
$Y$ &  $0.981$  &  0.892\\ 
$Z$  &  $0.965$   & 0.456\\ 
$(X,Y,Z)$  & $0.959$  & 0.332\\ \hline
\end{tabular}
\end{center}
\end{table}

There is clear evidence of dependence among the three variables. The pairwise correlations, based on the Pearson and Kendall coefficients, were positive. The estimated pairwise Kendall correlation coefficients were: $\widehat{\tau}\left(X,Y\right)=0.4400$, $\widehat{\tau}\left(X,Z\right)=0.2005$, and $\widehat{\tau}\left(Y,Z\right)=0.4249$.

\begin{figure}[!ht]
\begin{center}
    \includegraphics[width=12cm]{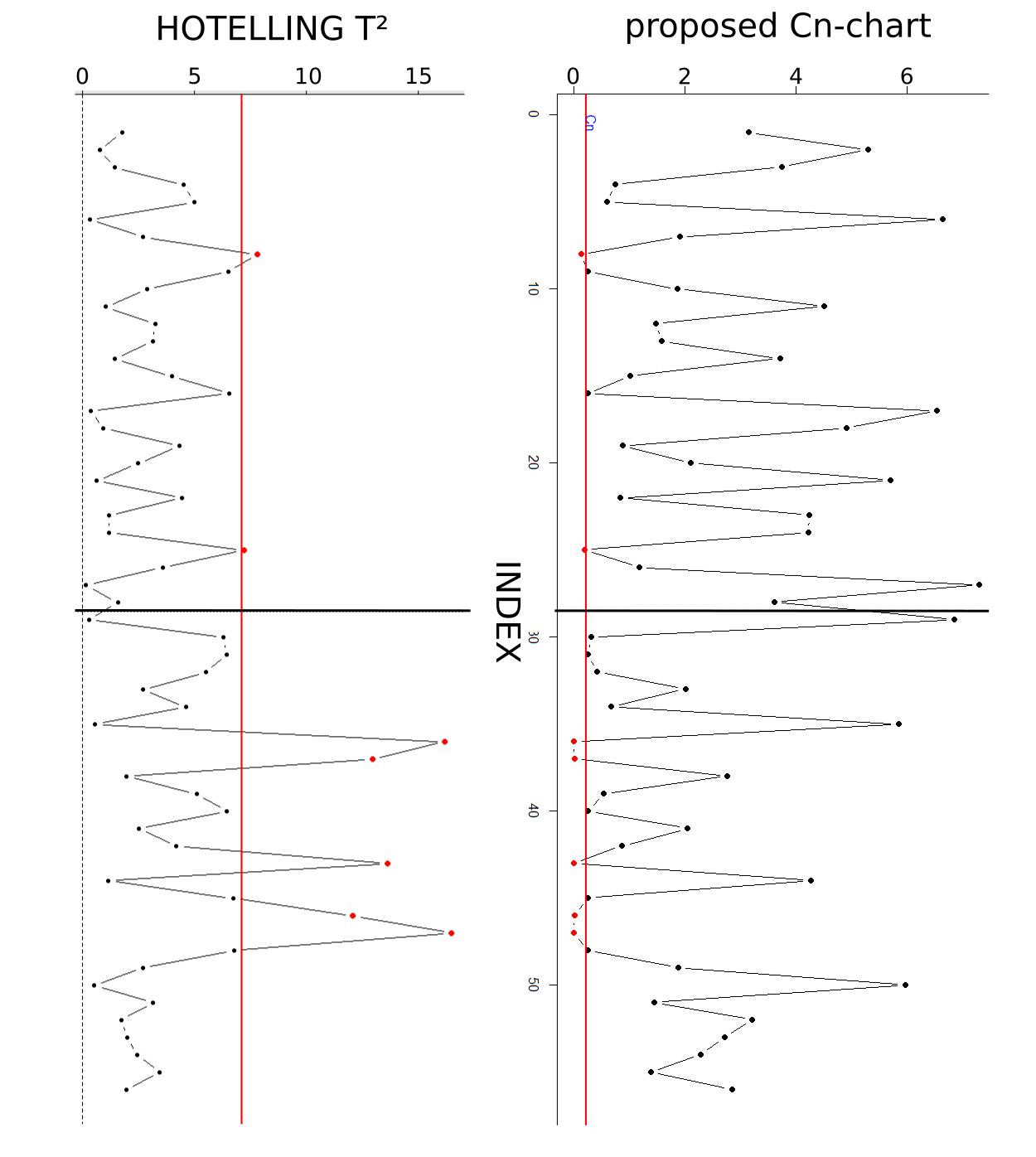}
   \caption{The left-hand panel shows the Hotelling's $T^2$ control chart, indicating seven out-of-control signals (two observations in Phase I and five observations in Phase II), and considering 5\% of tolerance upper limit. The right-hand panel presents the results from the proposed $C_n$-chart, whose lower limit highlights the same out-of-control observations. These findings are consistent with Lemma~\ref{lem:hotelling}, as the best-fitted copula was the autoregressive of order one (or AR(1)) Gaussian copula, assuming that the data have normal marginal distributions.}
  \label{cn_metal}
  \end{center}
\end{figure}

Under these conditions, the traditional Hotelling's $T^2$ control chart may be recommended as a tool for detecting changes in the process mean vector. Considering a significance level of 5\%, Figure \ref{cn_metal} (left-hand panel) plots the $T^2$ statistic on the vertical axis against the sample index on the horizontal axis, with the control limit as a solid red line. For Phase I, this control chart identified the sample points \#8 and \#25 as out-of-control (red dots). For Phase II monitoring, five alerts were issued (sample points \#36, \#37, \#43, \#46 and \#47).
%Sob tais condi\c{c}\~oes, o uso do gr\'afico de controle tradicional $T^2$ de Hotelling poderá ser recomendado como ferramenta de detec\c{c}\~ao de mudan\c{c}as no vetor de m\'edias do processo. Considerando um $\alpha = 0,05$, a Figura \ref{t2_cn_metal} representa na dimensão das ordenadas a densidade da normal multivariada e na dimensão das abscissas the sample index. Na fase I, esse gr\'afico identificou os pontos 8 e 25 como fora de controle. Durante o monitoramento na fase II, cinco alertas foram dados (observações 8, 9, 15, 18 e 19).

Considering normal marginal distributions, Table \ref{TAB:triCOPULA} shows the AIC results of eleven parametric copulas. For the elliptical copulas (Gaussian and Student-t), the dispersion structures tested were: exchangeable (``ex''), autoregressive of order one (AR(1), or ``ar1''), Toeplitz (``Toep''), and unstructured (``un''). The AR(1) Gaussian copula ($\widehat{\rho}_1=0.6049$) provided the best fit to the observed data ($\text{AIC}= -17.45$), and this was supported by the GOF test based on the Cramér-von Mises statistic (Genest et al. 2009) ($p$-value $=0.9036$). Therefore, the density level set approach was used to obtain $\hat{c}_n$ based on the selected copula model (right-hand panel of Figure \ref{cn_metal}). \nocite{genest2009goodness}

\begin{table}[!ht]
\caption{AIC values for the different candidate copula models.}
\resizebox{\columnwidth}{!}{%
\begin{tabular}{l|lllllllllll}
 & \rot{Gaussian (ar1)}& \rot{Gaussian (Toep)}& \rot{Student-t (ar1)}& \rot{Gaussian (ex)}& \rot{Gaussian (un)}& \rot{Student-t (Toep)}& \rot{Clayton}& \rot{Student-t (ex)}& \rot{Frank}& \rot{Student-t (un)}& \rot{Gumbel} \\ \hline
AIC & \multicolumn{1}{r}{-17.45} & \multicolumn{1}{r}{-15.47} & \multicolumn{1}{r}{-15.26} & \multicolumn{1}{r}{-14.09} & \multicolumn{1}{r}{-13.49} & \multicolumn{1}{r}{-13.33} & \multicolumn{1}{r}{-12.75} & \multicolumn{1}{r}{-12.01} & \multicolumn{1}{r}{-11.67} & \multicolumn{1}{r}{-11.38} & \multicolumn{1}{r}{-9.90} \\ \hline
\end{tabular}%
}
\label{TAB:triCOPULA}
\end{table}

Although the copula-based and Hotelling's $T^2$ tolerance regions are similar here, the conclusions can differ when the margins are not normal. The simulation studies in Subsection~\ref{sec:simul} and Supplementary Material B show that as the marginal distributions deviate from normality, the Hotelling's $T^2$ control chart becomes limited as a change detection tool. Furthermore, divergent results can be expected as the degree of association between the variables increases (see Supplementary Material B).
% Apesar dos resultados mostrarem similaridades entre regi\~oes de toler\^ancia com base na modelagem de c\'opulas em comparação com $T^2$ de Hotelling, estudos revelam que à medida em que as distribui\c{c}\~oes marginais se distanciavam da distribui\c{c}\~ao normal (further details Supplementary Material A), a regra $T^2$ revelou-se falha/limitada como ferramenta de detec\c{c}\~ao de mudança. Además é de se esperar resultado divergente conforme o grau de associa\c{c}\~ao entre as varia\'aveis aumenta.
 
%Regi\~oes de tolerância contru\'idas a partir de c\'opulas com marginais Kernel Gaussiano, bem como estimadas atrav\'es da c\'opula n\~ao-param\'etrica sob transforma\c{c}\~ao Kernel Cl\'assica, tamb\'em se mostraram melhor ajustadas à regi\~ao te\'orica do que a regi\~ao constru\'ida sob a abordagem $T^2$ de Hotelling para dados n\~ao-normais multivariados.

\subsection{Simulation Study: Shifts, Comparators and Computational Cost}\label{sec:simul}
%% ARL₀/ARL₁ and SDRL tables across mean shifts, scale shifts and dependence shifts + example with timings
%% Comparators: T², MEWMA, Non-Parametric Multivariate Control Chart Using Copula Entropy
%% Elliptical, Archimedean and Vine copulas generation

A simulation study was also conducted to assess the run-length performance and the computational tractability of the $C_n$-chart. For the computational study, the true model is a D-vine with mixed Clayton, Gumbel, Frank and Gaussian pair-copulas ($\tau=0.5$) and mixed skewed margins. The $C_n$-chart is fitted with AIC selection of the margins and AIC selection of the copula, both over the elliptical and Archimedean families and over vines, and the Monte Carlo level $\hat c_n$ is computed with $N=10^4$ and $N=10^5$ auxiliary draws (the total time uses $N=10^5$). Phase~II monitoring consists of $10^4$ observations. The comparators are the multivariate exponentially weighted moving average (MEWMA) chart (Lowry et al. 1992), with a one-off calibration of its limit $h$ ($\lambda=0.1$, ARL$_0=200$), and the nonparametric maximum-copula-entropy (MCE) chart (Fallah Mortezanejad et al. 2025), for which we report the effective sample size (ESS) of its quadrature; n.c.\ denotes not computable (ESS $<500$). Hotelling's $T^2$ needs less than $0.01$\,s for both fitting and monitoring and is omitted. Table~\ref{tab:timings} reports the median elapsed times (in seconds) of Phase~I fitting and Phase~II monitoring by dimension $d$ and Phase~I size $n$; the $C_n$-chart remains computationally tractable as the dimension grows.

% Requires: \usepackage{booktabs,multirow}
\begin{table}[!htbp]
\centering
\footnotesize
\setlength{\tabcolsep}{4pt}
\caption{Median elapsed times (in seconds) of Phase~I fitting and Phase~II monitoring for the $C_n$-chart, the MEWMA chart and the MCE chart.}
\label{tab:timings}
\begin{tabular}{rr rrr rr r r r rr}
\toprule
 & & \multicolumn{7}{c}{$C_n$-chart} & & \multicolumn{2}{c}{MCE} \\
\cmidrule(lr){3-9}\cmidrule(lr){11-12}
$n$ & $d$ & Margins & Ell.+Arch. & Vine & MC $10^4$ & MC $10^5$ & Total & Monitor
 & MEWMA $h$ & Fit & ESS \\
\midrule
\multirow{4}{*}{50}
 & 2  & 0.01 & 0.04 & 0.01 & $<$0.01 & 0.04  & 0.10  & 0.002 & 0.70 & 0.11 & 8\,077 \\
 & 5  & 0.02 & 0.05 & 0.07 & 0.09    & 0.87  & 1.00  & 0.014 & 1.43 & 0.44 & 871 \\
 & 10 & 0.04 & 0.11 & 0.72 & 0.31    & 3.14  & 4.00  & 0.066 & 2.36 & n.c. & 96 \\
 & 20 & 0.10 & 0.40 & 6.07 & 1.02    & 10.33 & 16.72 & 0.232 & 4.68 & n.c. & 1 \\
\midrule
\multirow{4}{*}{250}
 & 2  & 0.01 & 0.10 & 0.03 & $<$0.01 & 0.04  & 0.19  & 0.002 & 0.76 & 0.14 & 13\,256 \\
 & 5  & 0.02 & 0.13 & 0.31 & 0.10    & 1.03  & 1.54  & 0.015 & 1.43 & 0.42 & 1\,013 \\
 & 10 & 0.05 & 0.22 & 1.59 & 0.30    & 3.02  & 4.92  & 0.061 & 2.34 & n.c. & 150 \\
 & 20 & 0.11 & 0.58 & 7.38 & 1.02    & 10.24 & 18.29 & 0.206 & 4.52 & n.c. & 1 \\
\midrule
\multirow{4}{*}{1000}
 & 2  & 0.02 & 0.35 & 0.11  & $<$0.01 & 0.02 & 0.50  & 0.002 & 0.71 & 0.08 & 12\,206 \\
 & 5  & 0.05 & 0.64 & 1.29  & 0.11    & 1.10 & 3.19  & 0.016 & 1.41 & 0.33 & 893 \\
 & 10 & 0.09 & 1.07 & 6.92  & 0.30    & 2.99 & 10.93 & 0.068 & 2.33 & n.c. & 89 \\
 & 20 & 0.19 & 3.24 & 30.60 & 0.92    & 9.18 & 43.30 & 0.208 & 4.55 & n.c. & 1 \\
\bottomrule
\end{tabular}
\end{table}

The complete Phase~I fit takes 0.1--0.5\,s for $d=2$ and 1.0--3.2\,s for $d=5$; for $d=10$ it takes 4.0--10.9\,s, and for $d=20$ it takes 16.7--43.3\,s, across Phase~I sizes $n\in\{50,250,1000\}$. Two terms dominate. The Monte Carlo level $\hat c_n$ ($N=10^5$) does not depend on $n$: it costs about 3\,s at $d=10$ and 10\,s at $d=20$, and $N=10^4$ cuts this roughly tenfold (about 1\,s at $d=20$). Vine selection is the only step whose cost grows noticeably with $n$; at $d=20$ it rises from 6.1\,s ($n=50$) to 30.6\,s ($n=1000$). Phase~II monitoring is negligible: $10^4$ observations take at most 0.23\,s at $d=20$ (about $20\,\mu$s per observation). Because the fit is done once in Phase~I, the chart is suitable for online monitoring of multi-sensor streams. The MCE chart, by contrast, is cheap for $d\le5$ (below 0.5\,s), but its fitted density degenerates as $d$ grows. The effective sample size of its quadrature falls from about $10^4$ ($d=2$) to about $10^3$ ($d=5$), then to 89--150 ($d=10$) and to 1 ($d=20$), so its UCL cannot be computed for $d\ge10$ at any of the Phase~I sizes considered. Hotelling's $T^2$ and the MEWMA chart are the cheapest; the MEWMA limit $h$ takes 0.7--4.7\,s to calibrate. Neither chart, however, adapts to the shape of the dependence. The simulations support three results:

\textit{(i) In-control calibration.} With $n=100$ and $d=2$, the $C_n$-chart attains in-control ARLs of 47 (target 50), 146--164 (target 200) and 764 (target 1000). For $d=3$ the ARL$_0$ falls to about half the target (98--108 for target 200). This reflects Phase~I estimation: by Theorem~\ref{thm:phase1}, the conditional ARL varies around $1/\alpha$ with a relative spread of order $\sigma_\varrho/(\alpha\sqrt{n})$, and copula and vine selection, which that theorem does not cover, add model uncertainty that shrinks the estimated acceptance region. The effect diminishes as $n$ grows (ARL$_0=236$ with $n=250$). Under non-normal margins the normal-theory charts are badly calibrated. Hotelling's $T^2$ attains ARL$_0\approx$ 42--58 for a target of 200, and only 57--249 for a target of 1000. The MEWMA chart signals too early, and the MCE chart signals far too late (ARL$_0$ of 430--663 for a target of 200); in up to 14\% of Phase~I samples with $n=50$ the MCE chart could not be fitted.

\textit{(ii) Dependence shifts.} The $C_n$-chart is by far the fastest chart to detect changes in the dependence structure while the margins stay fixed (Corollary~\ref{cor:dep} and Example~\ref{ex:counter}). For ARL$_0=200$ ($d=2$, Clayton), the ARL$_1$ under a reversal of dependence is 8.1, against 48.0 for the calibrated (oracle) $T^2$, 28.5 for MEWMA, 57.7 for MCE and 227 for the marginal chart, which is essentially blind to this shift. Similar gains occur when the dependence weakens ($\tau$: $0.5\to0.2$; 36.3 versus 141.7 for oracle $T^2$ and 61.2 for MEWMA) and for a switch from lower- to upper-tail dependence (57.2 versus 281.9 and 161.4). The pattern persists for $d=3$ and for the larger ARL$_0$.

\textit{(iii) Mean and scale shifts.} For large shifts the $C_n$-chart is competitive. For small, sustained mean shifts ($\delta\le1$) MEWMA is superior, as expected for a chart with memory; the $C_n$-chart is a Shewhart-type chart. With skewed margins, the $C_n$-chart nevertheless clearly outperforms the calibrated $T^2$ (e.g.\ 7.0 versus 98.9 for $\delta=2$), because its acceptance region follows the shape of the joint density rather than an ellipsoid. It is also the best chart for scale increases (18.1 versus 35.2 for MEWMA at $\gamma=1.5$).

%\textit{(iv) Limitation.} When the dependence becomes stronger ($\tau$: $0.5\to0.8$), probability mass concentrates inside the highest-density region. The $C_n$-chart is then ARL-biased, with ARL$_1\ge$ ARL$_0$. Combining the density statistic with an EWMA-type memory, or adding an upper threshold on $\hat h$, is a natural extension.

Overall, the $C_n$-chart is reasonably well calibrated when the Phase~I sample is moderately large (e.g., $n\geq250$ for $d=3$), which is common in the Industry 4.0 era, where large data sets predominate. It is the most effective chart for shifts in the dependence structure and for non-elliptical joint distributions, and it remains computationally tractable, tested up to $d=20$. MEWMA remains competitive for small mean shifts.

%%%%%%%%%%%%%%%%%%%%%%%%%%%%%%%%%%%%%%%%%%%%%%%%%%%%%
\section{CONCLUSION}
\label{sec:conc}

Most multivariate SPC procedures are based on the assumption of multivariate normality of the process variables, as is the case, for example, with the traditional Hotelling's $T^2$ control chart (Hotelling 1947). However, when the distribution of the random vector is not multivariate normal, such a method is not recommended (Chou et al. 2001). This work proposes the use of tolerance regions, estimated under the density level set approach, as a tool for detecting changes in the characteristic vector of the process. The $C_n$-chart is shown to be useful even if the marginal data originate from non-normal distributions, and, consequently, the joint (or multivariate) distribution is not normal. To this end, we suggested the use of copula functions to model the dependence among the marginal distributions, also measuring the degree of association between the variables under study, making it possible to more flexibly observe the magnitude of changes, adopting the parametric copula families as an example. \nocite{hotelling1947multivariate,chou2001control} %\textcolor{blue}{(TALVEZ ACRESCENTAR AQUI UMA FRASE REFERENTE AO CN-CHART)}

We also derived exact and asymptotic run-length properties of the $C_n$-chart. Its in-control run length is geometric with $\mathrm{ARL}_0=1/\alpha$, whatever the copula family, the margins and the dimension; unlike univariate charts, it detects shifts in the dependence structure alone; it coincides with Hotelling's $T^2$ chart under a Gaussian copula with normal margins; and Phase~I estimation makes the conditional in-control ARL deviate from $1/\alpha$ at rate $n^{-1/2}$. The simulation study confirmed these results and showed that the chart remains computationally tractable up to $d=20$.

%Grande parte dos procedimentos de CEP Multivariado baseia-se na suposi\c{c}\~ao de normalidade multivariada das variáveis do processo, como é o caso, por exemplo, do tradicional gráfico de controle $T^2$ de Hotelling [CITE]. Entretanto, quando a distribui\c{c}\~ao do vetor aleat\'orio não está basada na normal multivariada, tal procedimento n\~ao \'e recomendado segundo à literatura [CITE]. O presente trabalho prop\~oe o uso de regi\~oes de toler\^ancia estimadas sob a abordagem de conjunto de n\'iveis de densidades comunando na função de cópula como ferramenta de detec\c{c}\~ao de mudan\c{c}as no vetor característico do processo, mesmo que os dados sejam originados de distribui\c{c}\~oes n\~ao-normais, e, por consequência, a distribuição conjunta (ou multivariada) não será normal. Para tanto, sugerimos o uso de fun\c{c}\~oes c\'opula para modelar a depend\^encia entre as distribuições marginais, medindo também o grau de associa\c{c}\~ao entre as vari\'aveis em estudo, possibilitando observar mais flexivelmente a magnitude das mudan\c{c}as, adotando a fam\'ilia de c\'opulas param\'etricas como exemplo.

%Mais que apenas uma ferramenta para detecção de out-samples, esta metodologia detecta a mudança conjunta anomalo considerando limites menos tolerantes conjuntos que não necessariamente tenham formas retangulares.

In the bivariate application to the pH and PO4 levels measured in a water quality test, one of the marginal distributions was asymmetric, and the tolerance region was built from the Clayton copula, which fitted the data well. The $C_n$-chart flagged the same observations as the tolerance region. Continuing with the bivariate exemplification, it was also shown that the adoption of regression structures for the multivariate SPC context is feasible.

%O resultado da aplicação da metodologia proposta no conjunto de dados bivariados referentes ao nível de pH e de fosfato medidos em um teste de qualidade de \'agua, uma das distribui\c{c}\~oes marginais era assim\'etrica, o que permitiu a constru\c{c}\~ao da regi\~ao de toler\^ancia baseada na cópula de Clayton, a qual se ajustou bem aos dados. Apresentamos tamb\'em uma proposta de Gr\'afico de Controle $c_n$, que identificou mudan\c{c}as no processo de maneira an\'aloga \`a regi\~ao de toler\^ancia. Seguindo com a exemplificação bivariado, também se mostrou que a adoção de estruturas de regressão para o contexto de CEP multivariado é factivel.

The trivariate case was explored in a data set related to the deflection, curvature, and resistivity measurements of brass-steel thermostats approximately following a multivariate normal distribution. We applied the traditional Hotelling's $T^2$ rule and compared it with the proposed $C_n$-chart, and both identified the same changes occurring in the process characteristic vector (i.e., mean vector). We emphasize, however, that the new control chart presented in this work produced satisfactory results in situations with or without multivariate normality of the process variables.

%O caso trivariado foi explorado num conjunto de dados relacionados \`a medidas de deflex\~ao, curvatura e resistividade de termostatos bimet\'alicos em lat\~ao e a\c{c}o com distribui\c{c}\~ao normal multivariada. Aplicamos a tradicional regra $T^2$ de Hotelling e comparamos com o gr\'afico $c_n$, sendo que ambos indentificaram as mesmas mudan\c{c}as ocorridas no vetor de m\'edias do processo. Ressaltamos, no entanto, que a poposta de gráfico de controle apresentada nesta disserta\c{c}\~ao produziu resultados satisfat\'orios em situa\c{c}\~oes com ou sem normalidade multivariada dos dados.

This study has limitations regarding the shapes of the parametric copula families used in the case studies, whose density contours are relatively simple (bulging). Hence, future work could extend the proposed control chart (via copula density) considering other more flexible families, such as semiparametric or nonparametric copulas (Nagler 2018). Furthermore, it is worth noting that the $C_n$-chart is essentially a Shewhart chart, which is primarily effective in detecting large shifts in the process. Hence, it may not be sensitive enough to detect small and persistent shifts. For this reason, further investigation can involve generalizing the method to more sensitive control charts, such as the Cumulative Sum (CUSUM) and Exponentially Weighted Moving Average (EWMA) charts (Montgomery 2013). Finally, the run-length results assume independent observations; extending the chart to autocorrelated sensor streams, for instance with time-series models for the margins, is a natural next step. \nocite{nagler2018kdecopula}

%Este estudo apresenta limitações no tocante do formatos da função de cópulas restrita a formas abauladas. Assim, como trabalhos futuros poderá ser estendido a carta de controle proposta (via densidade de cópula) considerando outras famílias mais flexíveis como a de cópulas semi-paramétrica ou não-paramétrica.

\section*{CRediT authorship contribution statement}

\textbf{Diego C. Nascimento:} Writing -- original draft, Methodology, Software, Visualization, Formal analysis, Data curation, Project administration, Funding acquisition.
\textbf{Ana C. S. Batista:} Software, Methodology, Writing -- review \& editing.
\textbf{Francisco Louzada}: Conceptualization, Validation, Writing – review \& editing.
\textbf{Paulo H. Ferreira:} Conceptualization, Methodology, Validation, Supervision, Writing – review \& editing.

\section*{Data Availability Statement}

All data and R code used in this study are available in the GitHub repository: \url{https://github.com/ProfNascimento/Cn-chart}.

\section*{Conflict of interest}

The authors declare no potential conflict of interest.

\section*{Declaration of Generative AI and AI-assisted technologies in the writing process}

During the preparation of this work, the authors used Claude to improve their grammar and readability. After using this tool/service, the authors reviewed and edited the content as needed and assume full responsibility for the content of the published article.

\section*{Funding}
Paulo H. Ferreira acknowledges support from CNPq Grant number 302620/2022-2. Francisco Louzada is supported by CNPq Grant number 308849/2021-3 and Innovation and Diffusion of Mathematical Sciences Center Applied to Industry (CEPID-CeMEAI) of FAPESP Grant number 2013/07375-0.

%%%%%%%%%%%%%%%%%%%%%%%%%%%%%%%%%%%%%%%%%%%
%\newpage
\section*{SUPPLEMENTARY MATERIAL A} \label{sA}

This supplementary material reviews the theory of copulas and its applicability. Typically, it is of interest to know the joint distribution of two or more random variables. Naturally, it is easier to obtain the marginal distributions of the variables in question. Alternatively, we will discuss the use of copula functions to model the joint distribution of random variables.
%This section reviews the Copulas' theory functions and its applicability. Usualmente, \'e de interesse conhecer a fun\c{c}\~ao de distribui\c{c}\~ao conjunta de duas ou mais vari\'aveis aleat\'orias. Naturalmente, \'e mais f\'acil de se obter as fun\c{c}\~oes de distribui\c{c}\~ao marginais das vari\'aveis em quest\~ao. Como alternativa, abordaremos o uso de fun\c{c}\~oes c\'opula para modelar a fun\c{c}\~ao de distribui\c{c}\~ao conjunta de vari\'aves aleat\'orias.

According to Nelsen (2006), copulas are mathematical functions that link multivariate distribution functions to their one-dimensional marginal distribution functions. Alternatively, they are multivariate distribution functions whose one-dimensional marginals are standard uniform, i.e., Uniform(0,1). Furthermore, Fisher (1997) stated that copulas are of interest to statisticians as a way to study scale-free measures of dependence, and also as a starting point for constructing families of multivariate distributions. \nocite{nelsen2006introduction,fisher1997copulas}
%Segundo Nelsen (2006), as c\'opulas s\~ao fun\c{c}\~oes que unem fun\c{c}\~oes de distribui\c{c}\~ao multivariadas a suas fun\c{c}\~oes de distribui\c{c}\~ao marginais unidimensionais. Ou alternativamente, s\~ao fun\c{c}\~oes de distribui\c{c}\~ao multivariadas cujas marginais unidimensionais s\~ao uniformes padrão, isto é, Uniforme(0,1). Além disso, Fisher (1997) diz que as c\'opulas interessam aos estat\'isticos por serem uma forma de estudo de medidas de depend\^encia livres de escala, e também por servirem como ponto de partida para a constru\c{c}\~ao de fam\'ilias de distribui\c{c}\~oes multivariadas.

A copula $C$ is a $d$-dimensional distribution function $C$: [0,1]$^d \rightarrow [0,1]$, whose univariate marginals are uniform on the interval $[0,1]$, i.e., $U_{j} \sim \text{Uniform}(0,1)$, for $j=1,2,\ldots,d$. That is,
\begin{equation}
C\left(u_1,u_2,\ldots,u_d\right) = \mathbb{P}\left(U_1 \le u_1,\mbox{ }U_2 \le u_2,\ldots,\mbox{ }U_d \le u_d\right).
\end{equation}

In this case, the variables $U_1,U_2,\ldots,U_d$ are obtained from a special transformation known as the probability integral transform, where $U_j=F_j(X_j)$, and $F_j$ represents the cumulative distribution function (CDF) of the random variable $X_j$, for $j=1,2,\ldots,d$.
%Neste caso, as vari\'aveis $U_1, U_2, \cdots, U_d \sim \text{Uniforme}(0,1)$ s\~ao originadas por uma transforma\c{c}\~ao especial, chamada de transforma\c{c}\~ao integral de probabilidade, segundo a qual $U_j=F_j(Y_j)$, em que $F_j$ representa a fda da vari\'avel aleat\'oria $Y_j$.

Sklar's theorem (Sklar 1959) states that, if $H(x_1, x_2, \ldots, x_d)$ is a multivariate distribution function with marginal distributions $F_1(x_1), F_2(x_2), \ldots, F_d(x_d)$, then there exists a copula $C: \left[0,1\right]^d \rightarrow \left[0,1\right]$, unique when all the marginals are continuous, such that \nocite{sklar1959fonctions}
%O Teorema de Sklar (1959) complementa: seja $H\left(x_1,x_2,\ldots,x_d\right)$ uma fun\c{c}\~ao de distribui\c{c}\~ao multivariada com distribui\c{c}\~oes marginais $F_1\left(x_1\right)$, $F_2\left(x_2\right)$,$\ldots$, $F_d\left(x_d\right)$. Se estas forem todas cont\'inuas, ent\~ao existe uma \'unica c\'opula $C : \mbox{ } \left[0,1\right]^{d} \rightarrow \left[0,1\right]$, tal que:\\
\begin{equation} \label{func_H}
H\left(x_1,x_2,\ldots,x_d\right)=C\left(F_1\left(x_1\right), F_2\left(x_2\right),\ldots, F_d\left(x_d\right)\right),
\end{equation}

\noindent for all $\left(x_1,x_2,\ldots,x_d\right)' \in \left(-\infty,\infty \right)^{d}$. Taking $F_j\left(x_j\right)=u_j$, with $u_j \in \left[0,1\right]$, we obtain $x_j=F_j^{-1}\left(u_j\right)$, for $j=1,2,\ldots,d$. Substituting these into \eqref{func_H}, we obtain the copula:
\begin{equation}
C\left(u_1,u_2,\ldots,u_d\right)=H\left(F_1^{-1}\left(u_1\right),F_2^{-1}\left(u_2\right),\ldots,F_d^{-1}\left(u_d\right)\right),
\end{equation}
for all $\left(u_1,u_2,\ldots,u_d\right)' \in \left[0,1\right]^d$.

%%%%%%%%%%%%%%%%%%%%%%%%%%%%%%%%%%%%%%%%%%
\subsection*{Density Function}
 
The copula density function $c$  is calculated as:
%\begin{equation}
\begin{align} \label{eq_dcopula2}
c(u_1,u_2\ldots,u_d)&=\frac{\partial^{d}}{\partial u_1 \partial u_2 \cdots \partial u_d }C(u_1,u_2,\ldots,u_d) \nonumber \\ &=\frac{\partial^{d}}{\partial u_1 \partial u_2  \cdots \partial u_d }H\left(F_1^{-1}(u_1),F_2^{-1}(u_2),\ldots,F_d^{-1}(u_d)\right) \nonumber \\
&=\frac{h\left(F_1^{-1}(u_1),F_2^{-1}(u_2),\ldots,F_d^{-1}(u_d)\right)}{\prod_{j=1}^{d}f_j(F_j^{-1}(u_j))}.
\end{align}
%\end{equation}

If the copula density $c$ and the marginal densities $f_j$ are known, then the joint density follows directly, in closed form: 
\begin{equation}
h(x_1,x_2\ldots,x_d)=c\left(F_1(x_1),F_2(x_2),\ldots,F_d(x_d)\right)\prod_{j=1}^{d}f_j(x_j).
\end{equation}

%%%%%%%%%%%%%%%%%%%%%%%%%%%%%%%%%%%%%%%%%%
\subsection*{Association Measures}

When one wishes to study the degree of dependence or association between random variables, it is common to use the Pearson correlation coefficient (Pearson 1895; Spearman 1907). It is also the parameter that measures the degree of dependence between the marginal distributions when using the Gaussian (or normal) and Student-t elliptical copulas. On the other hand, Trivedi and Zimmer (2007) pointed out that the Pearson correlation coefficient cannot capture nonlinear dependence. \nocite{pearson1895vii,spearman1907demonstration,trivedi2007copula}
%Quando se deseja estudar o grau de associa\c{c}\~ao entre vari\'aveis aleat\'orias, \'e comum utilizar o coeficiente de correla\c{c}\~ao de Pearson. Este, tamb\'em \'e o par\^ametro que mede o grau de depend\^encia entre as distribuições marginais quando usamos as c\'opulas el\'ipticas Gaussiana (ou normal) e t-Student. Em contrapartida, Trivedi e Zimmer (2005) lembram que o coeficiente de correla\c{c}\~ao de Pearson n\~ao pode ser utilizado para mensurar depend\^encia de fun\c{c}\~oes n\~ao-lineares. 

Measures of association between random variables $X_1,X_2,\ldots,X_d$, such as Kendall's $\tau$ and Spearman's $\rho$ rank correlation coefficients (Nelsen 2006; Joe 2014), can be represented using copulas, and they depend only on the copula of $X_1,X_2,\ldots,X_d$. In other words, they are scale-invariant measures, meaning that their values do not change under strictly increasing transformations of $X_1,X_2,\ldots,X_d$. Therefore, they are preferable to the Pearson correlation coefficient. In this study, we adopt Kendall's $\tau$ as the measure of association between the process variables. \nocite{nelsen2006introduction,joe2014dependence}
%As medidas de  associa\c{c}\~ao entre  vari\'aveis aleat\'orias $X_1,X_2,\ldots,X_d$, denominadas $\tau$ de Kendall e $\rho$ de Spearman, podem ser representadas por meio de c\'opulas e, al\'em disso, dependem apenas da c\'opula de $X_1,X_2,\ldots,X_d$.  Ou seja, s\~ao medidas invariantes em escala, logo, seus valores n\~ao se alteram sob transforma\c{c}\~oes estritamente crescentes de $X_1,X_2,\ldots,X_d$. Portanto, elas s\~ao prefer\'iveis em rela\c{c}\~ao ao coeficiente de correla\c{c}\~ao de Pearson. Neste trabalho, adotaremos o $\tau$ de Kendall como medida de depend\^encia entre as vari\'aveis do processo. 
 
Kendall's $\tau$ can be calculated from the copula distribution function as follows:
%O $\tau$ de Kendall pode ser calculado a partir da fun\c{c}\~ao de distribuição da c\'opula como segue:
\begin{equation}
\tau_d=\frac{1}{2^{d-1}-1}\left(2^{d}\int \cdots \int_{[0,1]^d}C\left(u_1,u_2,\cdots,u_d\right)dC\left(u_1,u_2,\cdots,u_d\right)-1\right),
\end{equation}
where $u_{1},u_{2},\ldots,u_{d} \in [0,1]$.
 
 %Na vers\~ao populacional, o $\tau$ de Kendall tamb\'em pode ser definido como sendo a diferen\c{c}a entre as probabilidades de concord\^ancia e discord\^ancia entre as vari\'aveis aleat\'orias, isto \' e:
%\begin{equation*}

%\tau(X_1,X_2,X_3,..., X_d)=P[(X_{1}-X_{2})(Y_{1}-Y_{2})>0]-P[((X_{1}-X_{2})(Y_{1}-Y_{2})<0],
%\end{equation*}

%sendo $(X_{1},Y_{1})$ e $(X_{2},Y_{2})$, pares de vari\'aveis aleat\'orias independentes e identicamente distribu\'idos.

For instance, for the bivariate case, Kendall's $\tau$ is expressed as:
\begin{equation}
\tau_2=4\int_{0}^{1} \int_{0}^{1} C(u_{1},u_{2})c(u_{1},u_{2}) du_{1} du_{2} -1, 
\end{equation}
whereas in the trivariate case it is given by
\begin{equation} \label{eq_tau3}
\tau_3=\frac{8}{3}\int_{0}^{1}\int_{0}^{1} \int_{0}^{1} C(u_{1},u_{2},u_3)c(u_{1},u_{2},u_3) du_{1} du_{2} du_{3} -\frac{1}{3}. 
\end{equation}

%\textbf{DEPENDENCIA CAUDAL}

%%%%%%%%%%%%%%%%%%%%%%%%%%%%%%%%%%%%%%%%%%
\subsection*{Parametric Families}

It is of paramount importance to choose a suitable copula to model the dependence structure between the process variables. The most commonly used parametric families of copulas in the literature are the Archimedean and elliptical families (especially their respective parameterizations for bivariate and trivariate cases).
%\'E de suma import\^ancia selecionar uma c\'opula adequada para modelar a estrutura de depend\^encia entre as variáveis do processo, sendo que as famílias paramétricas de c\'opulas mais comumente utilizadas na literatura s\~ao as Arquimedianas e as el\'ipticas (especialmente as suas respectivas parametrizações para os casos bivariado e trivariado).

\subsubsection*{Archimedean Copulas}

Consider $\varphi$ a function called the copula generating function. We say that a copula $C$ belongs to the Archimedean family if:
\begin{equation}
C(u_1,u_2,\ldots,u_d) = \varphi^{-1}\left(\varphi(u_1)+\varphi(u_2)+\ldots+\varphi(u_d)\right),
\end{equation}
for $(u_1,u_2,\ldots,u_d) \in [0,1]^{d}$, where $\varphi:[0,1]\rightarrow[0,\infty]$ is continuous, strictly decreasing and convex, with $\varphi(1)=0$ (for $d\geq3$, $\varphi^{-1}$ must also be $d$-monotone; see Nelsen 2006).

The most used Archimedean copulas are (for further details on them, see, e.g., Joe 2014): \nocite{joe2014dependence}
\begin{itemize}
\item \textbf{Clayton copula}:
\begin{equation} \label{eq_clayton}
C(u_1,u_2,\ldots,u_d; \theta) = \left[u_1^{-\theta}+u_2^{-\theta}+\ldots+u_d^{-\theta}-(d-1)\right]^{-\frac{1}{\theta}},
\end{equation}
where $\varphi(t)=\frac{1}{\theta}(t^{-\theta}-1)$, and $\theta \in [-1,\infty) \backslash \{0\}$ for $d=2$ ($\theta\geq-1/(d-1)$, $\theta\neq0$, for $d\geq3$).
 
Kendall's $\tau$ derived from Clayton copula \eqref{eq_clayton} is given by (Genest et al. 2011): \nocite{genest2011}
 \begin{equation}
\tau_d=\frac{1}{2^{d-1}-1}\left[-1+ 2^d\prod_{w=0}^{d-1}\left(\frac{1+w\theta}{2+w\theta}\right)\right].
\end{equation}
  
\begin{itemize}
\item \textbf{Bivariate case:}
\begin{equation} \label{clayton_biv}
C(u_1,u_2; \theta) = \left[u_1^{-\theta}+u_2^{-\theta}-1\right]^{-\frac{1}{\theta}},
\end{equation}
with the relationship between Kendall's $\tau$ and its parameter $\theta$ given by $\theta = 2\tau_2 / (1-\tau_2)$.

\item \textbf{Trivariate case:}
\begin{equation} \label{clayton_tri}
C(u_1,u_2,u_3; \theta) = \left[u_1^{-\theta}+u_2^{-\theta}+u_3^{-\theta}-2\right]^{-\frac{1}{\theta}}.
\end{equation}
After some calculations, Ferreira and Louzada (2020) found that the relationship between Kendall's $\tau$ and the parameter $\theta$, for the trivariate case, is similar to the bivariate one, that is, $\theta = 2\tau_3 / (1-\tau_3)$. \nocite{ferreira2020extending} 
\end{itemize}

\item \textbf{Frank copula}:
\begin{equation}
C(u_1,u_2,\ldots,u_d; \theta) = -\frac{1}{\theta} \log\left(1-\frac{\prod_{j=1}^{d}(1-\exp\{-\theta u_j\})}{(1-\exp\{-\theta\})^{d-1}}\right),
\end{equation}
where the copula generating function is $\varphi(t)=-\log \left(\frac{\exp\{-\theta t\}-1}{\exp\{-\theta\}-1}\right)$, and $\theta \in (-\infty,\infty) \backslash \{0\}$ for $d=2$ ($\theta>0$ for $d\geq3$). 
 
\begin{itemize}
    \item \textbf{Bivariate case:}
    \begin{equation}
C(u_1,u_2; \theta) = -\frac{1}{\theta} \log\left(1+\frac{(\exp\{-\theta u_1\}-1)(\exp\{-\theta u_2\}-1)}{\exp\{-\theta\}-1}\right).
\end{equation}
The relationship between Kendall's $\tau$ and the parameter $\theta$ is given by $\frac{1-D_{1}(\theta)}{\theta}$ $= \frac{1-\tau_2}{4}$, where $D_{1}(\theta)=\frac{1}{\theta}\int_{0}^{\theta} \frac{t}{\exp\{t\}-1} dt$ is the first-order Debye function.

    \item \textbf{Trivariate case:}
\begin{equation}
C(u_1,u_2,u_3; \theta) = -\frac{1}{\theta} \log\left(1+\frac{(\exp\{-\theta u_1\}-1)(\exp\{-\theta u_2\}-1)(\exp\{-\theta u_3\}-1)}{(\exp\{-\theta\}-1)^2}\right).
\end{equation}

Kendall's $\tau$ does not have a closed-form expression, and its numerical (approximate) calculation can be done using the general expression for $\tau_{3}$ in equation \eqref{eq_tau3} and the application of the \textit{adaptIntegrate}($\cdot$) function from the R \texttt{cubature} package (Narasimhan et al. 2022). \nocite{cubature}
%O $\tau$ de Kendall n\~ao possui forma fechada e seu c\'alculo num\'erico (aproximado) pode ser feito mediante o uso da express\~ao geral de $\tau_{3}$ \eqref{eq_tau3} e a aplica\c{c}\~ao da fun\c{c}\~ao \textit{adaptIntegrate(.)} do pacote \textit{cubature} do R.
\end{itemize}

\item \textbf{Gumbel copula}:
\begin{equation}
C(u_1,u_2,\ldots,u_d; \theta) = \exp\left\{-\left[\sum_{j=1}^{d}\left(-\log (u_j) \right) ^{\theta}\right]^{\frac{1}{\theta}}\right\},
\end{equation}
where the copula generating function is $\varphi(t)=(-\log t)^{\theta}$, and $\theta \in [1,\infty)$.

\begin{itemize}
    \item \textbf{Bivariate case:}
\begin{equation}
C(u_1,u_2; \theta) = \exp\left\{-\left[(-\log (u_1))^{\theta}+(-\log (u_2))^{\theta}\right]^{\frac{1}{\theta}}\right\},
\end{equation}
where the relationship between Kendall's $\tau$ and the parameter $\theta$ of this copula is given by $\theta = 1/(1-\tau_2)$.
    
    \item \textbf{Trivariate case:}
\begin{equation}
C(u_1,u_2,u_3; \theta) = \exp\left\{-\left[(-\log (u_1))^{\theta}+(-\log (u_2))^{\theta}+(-\log (u_3))^{\theta}\right]^{\frac{1}{\theta}}\right\}.
\end{equation}
\end{itemize}
For the trivariate case, Kendall's $\tau$ does not have a closed-form expression, and its numerical calculation can, again, be done using the general expression for $\tau_{3}$ and the \textit{adaptIntegrate}($\cdot$) function from the R \texttt{cubature} package.

\end{itemize}

\subsubsection*{Elliptical Copulas}
Derived from multivariate elliptical distributions, the most popular elliptical copulas are the Gaussian (or normal) and the Student-t (or simply, t-copula). An elliptical copula is obtained through the expression:
\begin{equation} 
C\left(u_1,u_2,\ldots,u_d;\Sigma\right)=H_{d}\left(H^{-1}\left(u_1\right),H^{-1}\left(u_2\right),\ldots, H^{-1}\left(u_d\right); \Sigma\right),
\end{equation}
where $H^{-1}(\cdot)$ represents the quantile function of the underlying univariate distribution, and $H_{d}\left(.; \Sigma\right)$ is the joint distribution function of the underlying multivariate distribution, which typically has a zero location vector and a $d\times d$ correlation matrix $\Sigma$.

For the trivariate cases, one can use the most well-known dispersion structures (i.e., correlation matrices $\Sigma$), namely: exchangeable (ex), first-order autoregressive (ar1), Toeplitz (Toep), and unstructured (un), which have the following structures, respectively:
\begin{equation} \label{estrut}
\left[ {\begin{array}{ccc} 1 & \rho_{1} & \rho_{1} \\ \rho_{1} & 1 & \rho_{1} \\ \rho_{1} & \rho_{1} & 1 \\ \end{array} } \right], \quad 
\left[ {\begin{array}{ccc} 1 & \rho_{1} & \rho_{1}^{2} \\ \rho_{1} & 1 & \rho_{1} \\ \rho_{1}^{2} & \rho_{1} & 1 \\ \end{array} } \right],
\quad
\left[ {\begin{array}{ccc} 1 & \rho_{1} & \rho_{2} \\ \rho_{1} & 1 & \rho_{1} \\ \rho_{2} & \rho_{1} & 1 \\ \end{array} } \right]
\quad \mbox{and} \quad
\left[ {\begin{array}{ccc} 1 & \rho_{1} & \rho_{2} \\ \rho_{1} & 1 & \rho_{3} \\ \rho_{2} & \rho_{3} & 1 \\ \end{array} } \right],
\end{equation}
with $\rho_j \in [-1,1]$, for $j=1,2,3$, subject to $\Sigma$ being positive definite.

 \begin{itemize}
     \item \textbf{Gaussian copula:}
  \begin{equation} \label{cop3}
C\left(u_1,u_2,\ldots,u_d;\Sigma\right)=\Phi_{d}\left(\Phi^{-1}\left(u_1\right),\Phi^{-1}\left(u_2\right),\ldots, \Phi^{-1}\left(u_d\right);\Sigma\right),
\end{equation}
where $\Phi^{-1}(\cdot)$ is the quantile function of the $\text{Normal}(0,1)$ distribution, and $\Phi_{d}(\cdot;\Sigma)$ represents the joint distribution function of the multivariate normal distribution with mean vector $\boldsymbol{0}$ and correlation (or covariance) matrix $\Sigma$. According to Genest et al. (2011), the formula for calculating Kendall's $\tau$ is: \nocite{genest2011}
\begin{equation}
    \tau_d=\frac{1}{2^{d-1}-1}\left[-1+2^d \mathbb{P}\left(\Sigma^{\frac{1}{2}}\boldsymbol{U}\geq 0\right)\right],
\end{equation}
where $\boldsymbol{U}$ is a $d$-dimensional random vector uniformly distributed.
     
     \begin{itemize}
         \item \textbf{Bivariate case:}
\begin{equation}
C(u_1,u_2; \rho)=\int_{-\infty}^{\Phi^{-1}(u_1)} \int_{-\infty}^{\Phi^{-1}(u_2)} \phi_{2,\rho}(t_1,t_2) dt_1 dt_2,
\end{equation}
where $\phi_{2,\rho}(\cdot)$ is the density of the bivariate normal distribution with the mean vector $(0,0)^{'}$ and correlation matrix 
$ \begin{bmatrix}
    1 & \rho \\
    \rho & 1
  \end{bmatrix}$, with $\rho \in [-1,1]$.

The relationship between Kendall's $\tau$ and the Pearson correlation coefficient $\rho$ is given by $\rho = \sin \left(\frac{\pi}{2}\tau_2\right)$.

 \item \textbf{Trivariate case:}
 \begin{equation}
C(u_1,u_2,u_3; \Sigma)=\int_{-\infty}^{\Phi^{-1}(u_1)} \int_{-\infty}^{\Phi^{-1}(u_2)}\int_{-\infty}^{\Phi^{-1}(u_3)}  \phi_{3,\Sigma}(t_1,t_2,t_3) dt_1 dt_2 dt_3,
\end{equation}
where $\phi_{3,\Sigma}(\cdot)$ denotes the density of the trivariate normal distribution with the mean vector $(0,0,0)^{'}$ and correlation matrix $\Sigma$ as shown in \eqref{estrut}.

The relationship between Kendall's $\tau$ and the correlation coefficients vector $\boldsymbol{\rho}=\left(\rho_{12},\rho_{13},\rho_{23}\right)'$, with $\rho_{ij}=\text{Corr}\left(X_i,X_j\right)$, according to Genest et al. (2011), is given by \nocite{genest2011}
\begin{equation}
 \tau_3 = \frac{2}{3\pi}\left[\arcsin(\rho_{12})+\arcsin(\rho_{13})+\arcsin(\rho_{23})\right].
\end{equation}

  \end{itemize}
  
\item \textbf{t-copula:}
\begin{equation} \label{cop2}
C\left(u_1,u_2,\ldots,u_d;\Sigma,\nu \right)=T_{d,\nu}\left(T^{-1}_{1,\nu}\left(u_1\right),T^{-1}_{1,\nu}\left(u_2\right),\ldots, T^{-1}_{1,\nu}\left(u_d\right);\Sigma\right),
\end{equation}
where $T_{1,\nu}^{-1}(\cdot)$ and $T_{d,\nu}(\cdot)$ are, respectively, the quantile function of the Student-t distribution with degrees of freedom $\nu > 0$, and the $d$-variate CDF of the Student-t distribution with degrees of freedom $\nu>0$ and correlation matrix $\Sigma$ as in \eqref{estrut}.

According to Genest et al. (2011), the calculation of Kendall's $\tau$ for the t-copula is done in an analogous manner to that of the Gaussian copula. \nocite{genest2011}

\begin{itemize}
  \item  \textbf{Bivariate case:}
\begin{equation}
C(u_1,u_2;\rho,\nu)=T_{2,\nu}\left(T^{-1}_{1,\nu}(u_1),T^{-1}_{1,\nu}(u_2);\rho\right).
\end{equation}
The parameter $\nu$ controls the weight in the tails, meaning that smaller values of $\nu$ result in heavier tails. A bivariate Gaussian copula with correlation coefficient $\rho$ can then be considered as the limiting case of a bivariate t-copula with the same parameter $\rho$ when $\nu \rightarrow \infty$. Furthermore, $\tau_2 = \frac{2}{\pi}\arcsin\left(\rho\right)$, which is the same result observed for the bivariate Gaussian copula.

  \item \textbf{Trivariate case:}
\begin{equation}
C(u_1,u_2,u_3;\Sigma,\nu)=T_{3,\nu}\left(T^{-1}_{1,\nu}(u_1),T^{-1}_{1,\nu}(u_2),T^{-1}_{1,\nu}(u_3);\Sigma\right),
\end{equation}
whereas Kendall's $\tau$ is identical to that of the trivariate Gaussian copula.
\end{itemize}
\end{itemize}

%%%%%%%%%%%%%%%%%
\subsection*{Parameter Estimation}
There are situations where parameter estimation is straightforward because the parameters' estimators are functions of the measure of association, in our case, Kendall's $\tau$. However, for some more complex multivariate copulas, other parametric estimation methods are required.
%Existem algumas situa\c{c}\~oes em que a estima\c{c}\~ao dos par\^ametros \'e feita de forma simples e direta, visto que seus estimadores s\~ao fun\c{c}\~oes da medida de depend\^encia. No nosso caso, o $\tau$ de Kendall. Entretanto, para algumas c\'opulas mais complexas, multivariadas, s\~ao necess\'arios outros m\'etodos de estima\c{c}\~ao param\'etrica.

The maximum likelihood (ML) method is commonly used to estimate the parameter vector of marginal distributions and the copula under study. Let $\boldsymbol{X}_i=\left(X_{i1},\ldots,X_{id}\right)'$, for $i=1,2,\ldots,n$, be independent observations from a $d$-variate distribution. Suppose that the marginals have CDFs $F_j$ and probability density functions (PDFs) $f_j$, for $j= 1,\ldots,d$, and the copula with density $c$ is used to model their dependence structure. Let $\boldsymbol{\theta}_j$ be the parameter vector of margin $j$, and $\boldsymbol{\psi}$ be the copula parameter vector, as in the main text. Then, the vector to be estimated is $\boldsymbol{\vartheta}=\left(\boldsymbol{\theta}_{1}^{'},\ldots,\boldsymbol{\theta}_{d}^{'},\boldsymbol{\psi}^{'}\right)^{'}$. The likelihood function, according to Yan (2007), is given by \nocite{yan2007enjoy}
%\'E comum utilizar o m\'etodo de m\'axima verossimilhan\c{c}a (MV) para estimar o vetor de par\^ametros das distribuições marginais e da c\'opula em estudo. Seja $\boldsymbol{X}_i=\left(X_{i1},\ldots,X_{id}\right)'$, $i=1,\ldots,n$, observa\c{c}\~oes independentes de uma distribui\c{c}\~ao $d$-variada. Suponha que as marginais tenham fda $F_j$ e fdp $f_j$, $j= 1,\ldots,d$, e a c\'opula de densidade $c$ seja empregada para modelar a estrutura de dependência delas. Seja $\boldsymbol{\beta}_j$ o vetor de par\^ametros da marginal $j$, $j= 1,\ldots,d$, e $\boldsymbol{\alpha}$ o vetor de par\^ametros da c\'opula. Ent\~ao, o vetor a ser estimado ser\'a $\boldsymbol{\theta}=(\boldsymbol{\beta}_{1}',\ldots,\boldsymbol{\beta}_{d}',\boldsymbol{\alpha}')'$. A fun\c{c}\~ao de verossimilhan\c{c}a, segundo Yan (2007), \'e dada por:
    \begin{equation}
      L(\boldsymbol{\vartheta})=\prod_{i=1}^n c\left(F_1(X_{i1};\boldsymbol{\theta}_1), \ldots,F_d(X_{id};\boldsymbol{\theta}_d);\boldsymbol{\psi} \right)\prod_{i=1}^n\prod_{j=1}^d f_j(X_{ij};\boldsymbol{\theta}_j). 
      \label{eq:mv}
  \end{equation}
By applying the logarithm, we obtain:
    \begin{align}
      \ell(\boldsymbol{\vartheta})&=\log\left(L(\boldsymbol{\vartheta})\right) \nonumber \\ &=\sum_{i=1}^n\log \left(c\left(F_1(X_{i1};\boldsymbol{\theta}_1),\ldots,F_d(X_{id};\boldsymbol{\theta}_d);\boldsymbol{\psi} \right)\right) +\sum_{i=1}^n\sum_{j=1}^d\log \left(f_j(X_{ij};\boldsymbol{\theta}_j)\right). 
      \label{eq:logmv}
  \end{align}
  
Then, the ML estimator for $\boldsymbol{\vartheta}$ is:
\begin{equation}
\widehat{\boldsymbol{\vartheta}}_{\text{ML}} = \arg\max_{\boldsymbol{\vartheta} \in \Theta} \ell(\boldsymbol{\vartheta}),
\end{equation}
where $\Theta\subseteq\mathbb{R}^p$ is the parameter space. Estimation using the ML method involves finding the values of the vector $\boldsymbol{\vartheta}$ that maximize the likelihood function \eqref{eq:mv}. To do this, we simply take the logarithm of the likelihood function, or log-likelihood function, as given in equation \eqref{eq:logmv}, differentiate it with respect to each parameter, and set the result equal to zero. In general, these ML estimators do not have closed-form expressions, thus necessitating the use of numerical methods to obtain their estimates. In the R software, parameters can be estimated using the ML method with the \textit{fitCopula}($\cdot$) and \textit{fitMvdc}($\cdot$) functions from the \texttt{copula} package created by Yan (2007). Both functions are linked to the \textit{optim}($\cdot$) function. Nevertheless, this approach results in the simultaneous estimation of all parameters, requiring a high computational effort. \nocite{yan2007enjoy}

An alternative method for estimating copula-based models, known as the Inference Function for Margins (IFM) method, was proposed by Joe and Xu (1996). In this procedure, parameter estimation is performed in two steps: in the first one, the parameters of the marginal distributions are estimated (e.g., via the ML method for fully parametric cases, or the kernel approach for semiparametric cases). In the second step, the parameter vector of the copula is estimated through a pseudo-likelihood approach. The IFM method is also known as the two-stage estimation method in the context of survival and reliability data, that is, for censored data (Shih and Louis 1995). Yan (2007) noted that as the dimension $d$ becomes larger, the number of parameters increases, and consequently, the optimization problem becomes more difficult. This highlights the computational advantage of using the IFM method, because each maximization task involves a small number of parameters at a time. \nocite{joe1996estimation,shih1995inferences,yan2007enjoy} %Other authors, such as Anjos et al. (2004), also noted that the IFM method is more efficient when compared to exact ML.

When each marginal distribution has its own parameters, the first step involves estimating them via ML for each one. That is,
   \begin{equation}
\widehat{\boldsymbol{\theta}}_{j, \,\text{IFM}} = \arg\max_{\boldsymbol{\theta}_j}\sum_{i=1}^n\log \left(f_j(X_{ij};\boldsymbol{\theta}_j)\right), \quad \text{for } j=1,\ldots,d.
\end{equation}
Then, given $\widehat{\boldsymbol{\theta}}_{1,\, \text{IFM}},\ldots,\widehat{\boldsymbol{\theta}}_{d,\,\text{IFM}}$, the association parameter $\boldsymbol{\psi}$ of the copula is estimated using the pseudo-log-likelihood function. That is,
\begin{equation}
\widehat{\boldsymbol{\psi}}_{\text{IFM}}=\arg \max_{\boldsymbol{\psi}} \sum_{i=1}^{n} \log\left( c\big(F_1(X_{i1};\widehat{\boldsymbol{\theta}}_{1,\,\text{IFM}}),\ldots,F_d(X_{id};\widehat{\boldsymbol{\theta}}_{d,\,\text{IFM}});\boldsymbol{\psi}\big)\right).
\label{eq:theta}
\end{equation}

The use of the term ``pseudo'' in this case arises from the use of $\widehat{\boldsymbol{\theta}}_{j,\,\text{IFM}}$ instead of the true ${\boldsymbol{\theta}}_{j}$, for $j=1,\ldots,d$. %A possible alternative for parameter estimation consists of applying the IFM method to parametric models. 
Thus, in the R software, the \textit{optim}($\cdot$) function obtains parameter estimates of the marginal distributions  via the ML method (using the quasi-Newton numerical optimization method ``L-BFGS-B''), and the \textit{fitCopula}($\cdot$) function estimates the copula parameter(s).
%O emprego do termo \textit{pseudo}, neste caso, decorre do uso de $\hat{\boldsymbol{\beta}}_{j}$, ao inv\'es do verdadeiro ${\boldsymbol{\beta}}_{j}$, $j=1,\ldots,d$. Uma possivel alternativa para a estima\c{c}\~ao dos par\^ametros, aplicado o m\'etodo IFM para os modelos param\'etricos. Assim, a fun\c{c}\~ao \textit{optim(.)} obtem as estimativas via m\'etodo de MV das distribuições marginais param\'etricas (com uso do m\'etodo de otimiza\c{c}\~ao num\'erica de quasi-Newton ``L-BFGS-B''), e a fun\c{c}\~ao \textit{fitCopula(.)} estima o(s) par\^ametro(s) da c\'opula.

%\newpage
\section*{SUPPLEMENTARY MATERIAL B} 
\label{sB}

Simulation studies, analogous to that of Verdier (2013), were carried out to compare the tolerance regions estimated with copula functions with that of the traditional Hotelling's $T^2$ rule (or Hotelling's $T^2$ control chart), considering four scenarios. \nocite{verdier2013application}

%Estudo de simula\c{c}\~ao foram realizado com intuito de mensurar e comparar a qualidade dos ajustes feitos atrav\'es das fun\c{c}\~oes c\'opula com o tradicional gr\'afico $T^2$ de Hotelling considerando 4 cenários.

The following parametric cases were considered: bivariate Clayton copula (equation \eqref{clayton_biv}), with parameter $\theta=2/3$, $2$ and $6$ (which correspond to Kendall's $\tau$ measures equal to $0.25$, $0.50$ and $0.75$, respectively, that is, to weak, moderate and strong dependence degrees between the margins), and varied marginal distributions.

%Foi considerado os casos param\'etricos: c\'opula de Clayton bivariada, equation \eqref{clayton_biv}, com parâmetro $\theta=2/3$, $2$ e $6$ (que correspondem a medidas $\tau$ de Kendall iguais a $0,25$; $0,50$ e $0,75$, respectivamente, isto é, a graus de depend\^encia fraca, moderada e forte entre as marginais), e distribui\c{c}ões marginais variadas.

The Phase~I (learning) sample, generated from the Clayton copula, had size $n = 250$, with marginal distributions varying according to {Cases \#1}, {\#2}, {\#3} and {\#4}. Tolerance regions were estimated based on the density level set approach, considering the Clayton, Frank, Gumbel, Gaussian, and Student-t copulas (with the degrees of freedom fixed at four), in addition to the Hotelling's $T^2$ rule (bivariate normal distribution). Every simulation scenario considered a significance level of 5\% (or a tolerance level of 95\%).

%O tamanho da amostra artificial de aprendizagem da cópula de Clayton foi de $n = 250$, com distribuições marginais variando conforme os \textbf{Casos 1}, \textbf{2}, \textbf{3} e \textbf{4}. Foram estimadas as regi\~oes de toler\^ancia baseadas na abordagem de conjunto de n\'iveis de densidade, considerando as cópulas de Clayton, Frank, Gumbel, Gaussiana e t (adotamos o par\^ametro graus de liberdade fixo igual a 4) , al\'em da regra $T^2$ de Hotelling (distribuição normal bivariada).

\bigskip

\noindent \textbf{Case \#1:} Artificial sample generated from the bivariate Clayton copula with symmetric marginal distributions (normal and logistic) specified below.
 \begin{itemize}
     \item \textit{Marginal distribution 1}: $X \sim \text{Normal}(\mu,\sigma^2)$, with mean $\mu=0$ and variance $\sigma^2=1$, and PDF given by
     \begin{equation}\label{eq:dnorm}
         f(x)=\frac{1}{\sqrt{2\pi\sigma^2}}\exp\left\{-\frac{(x-\mu)^2}{2\sigma^2}\right\}, \quad -\infty < x < \infty.
     \end{equation}
     %sendo $-\infty< \mu < \infty$ a média e $\sigma^2 > 0$ a variância de $X_{1i}$. Aqui, consideramos  %do pacote \textit{stats} do R.
     
     \item \textit{Marginal distribution 2}: $Y \sim \text{Logistic}(\mu,s)$, with location $\mu = 1$ and scale $s = 2$, and PDF defined as:
     \begin{equation}\label{eq:dlog}
         f(y)=\frac{\exp\left\{-(y-\mu)/s\right\}}{s\left[1+\exp\left\{-(y-\mu)/s\right\}\right]^{2}}, \quad -\infty < y < \infty.
         \end{equation}
     In this case, $\mathbb{E}[Y]=\mu = 1$ and $\mathbb{V}\text{ar}[Y]=\pi^2s^2/3\approx 13.16$.
    
 \end{itemize}
%%%%%%%%%%%% 

\begin{figure}[h!]
\begin{center}
\includegraphics[width=\textwidth]{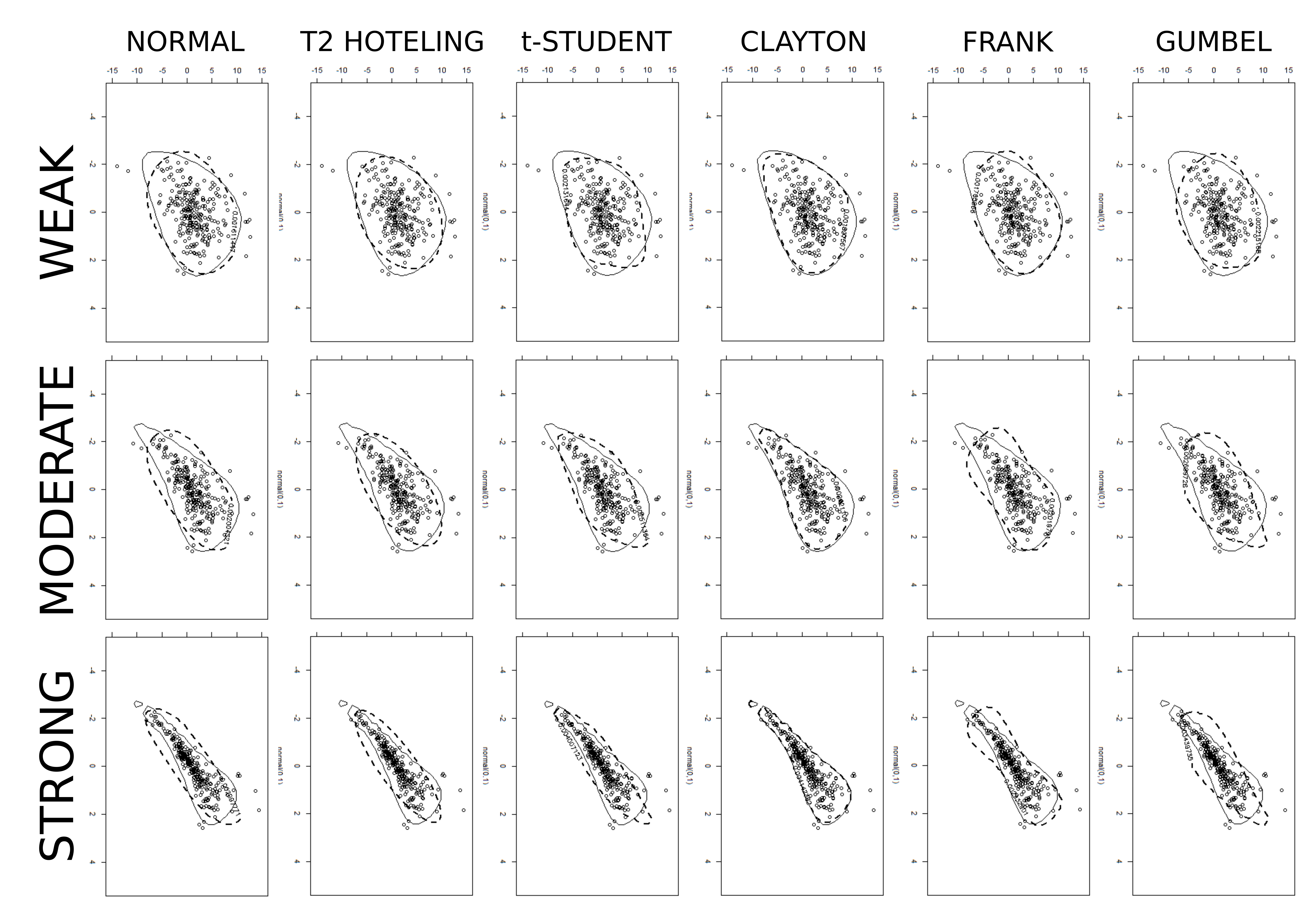}
\caption{Illustration of the $\text{Normal}(0,1)$ and $\text{Logistic}(1,2)$ marginal distributions, comparing the real tolerance region (continuous line) with the estimated one based on the parametric copula family (dashed line).}
\label{f_rt1_fr_mo_fo}
\end{center}
\end{figure}

The tolerance regions for Case \#1, when dependence is weak, are shown in Figure \ref{f_rt1_fr_mo_fo} (upper panels). It can be observed that the tolerance region estimated by the Clayton copula is very similar to the theoretical one. The traditional Hotelling's $T^2$ rule also showed a tolerance region similar to the theoretical one, as well as the other copula families tested. When the degree of dependence between the marginals increases (Figure \ref{f_rt1_fr_mo_fo}, middle panels), the Frank, Gumbel and Student-t copulas begin to move further apart from the expected region. Finally, for the strong degree of dependence (Figure \ref{f_rt1_fr_mo_fo}, lower panels), it is shown that the tolerance region estimated by the Clayton copula is even closer to that expected.

%As regi\~oes de toler\^ancia para o \textbf{Caso 1}, em que temos duas marginais sim\'etricas, quando a depend\^encia \'e fraca, s\~ao apresentadas na Figura \ref{f_rt1_fr}. Nela, \'e poss\'ivel perceber que a regi\~ao estimada pela c\'opula de Clayton \'e muito similar à regi\~ao te\'orica. A tradicional regra $T^2$ de Hotelling tamb\'em apresentou regi\~ao similar \`a te\'orica, bem como as demais fam\'ilias de c\'opulas testadas. Quando o n\'ivel de depend\^encia entre as marginais aumenta ($\tau_2=0,50$), as c\'opulas de Frank, Gumbel e \textit{t} come\c{c}am a se distanciar da regi\~ao esperada (Figura \ref{f_rt1_mo}). Para o  grau de depend\^encia forte, a Figura \ref{f_rt1_fo} mostra que a regi\~ao de toler\^ancia estimada pela cópula de Clayton \'e ainda mais pr\'oxima da esperada.

%%%%%%%%%%%%%%%%%%%%%%
% RT -CASO 1

\bigskip

\noindent \textbf{Case \#2}: Artificial sample generated from the bivariate Clayton copula with the marginal distributions specified below. 
 
  \begin{itemize}
     \item \textit{Marginal distribution 1}: $X \sim \text{Normal}(\mu,\sigma^2)$, with $\mu = 0$ and $\sigma^2=1$, and PDF given by \eqref{eq:dnorm}.
     
     \item \textit{Marginal distribution 2}: $Y \sim \text{Weibull}(a,b)$, with shape $a = 1$ and scale $b = 1$, and PDF defined as:
        \begin{equation}\label{eq:dwei}
         f(y)=\frac{a}{b}\left(\frac{y}{b}\right)^{a-1}\exp\left\{-\left(\frac{y}{b}\right)^a\right\}, \quad y > 0.
         \end{equation}
     In this case, $\mathbb{E}[Y]= b\Gamma(1+1/a)=1$ and $\mathbb{V}\text{ar}[Y]=b^2[\Gamma(1+2/a)-(\Gamma(1+1/a))^2] = 1$, where $\Gamma(t)=\int_{0}^{\infty}w^{t-1}\exp\{-w\}dw$ represents the gamma function.
  \end{itemize}

% RT -CASO 2
\begin{figure}[h!]
\begin{center}
\includegraphics[width=\textwidth]{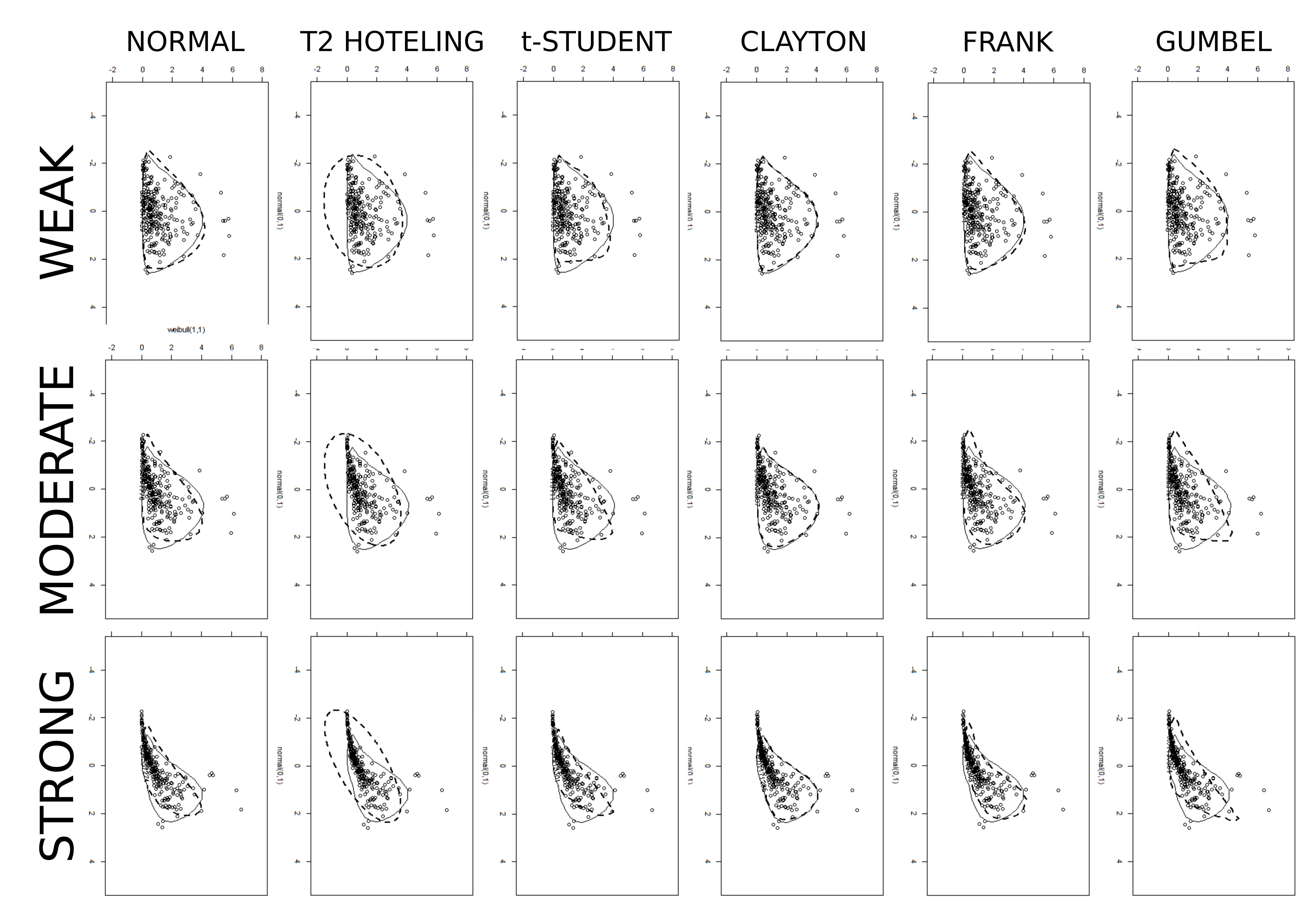}
\caption{Illustration of the $\text{Normal}(0,1)$ and $\text{Weibull}(1,1)$ marginal distributions, comparing the real tolerance region (continuous line) with the estimated one based on the parametric copula family (dashed line).}
\label{f_rt2_fr-mo-fo}
\end{center}
\end{figure}

When we analyze the theoretical tolerance region generated by Case \#2, in which we have a symmetric marginal and an asymmetric one, we see that the tolerance regions estimated using all the copula families fit well to the theoretical one, while the region generated by the Hotelling's $T^2$ rule is distant from the theoretical one even when the degree of dependence between the marginals is weak (Figure \ref{f_rt2_fr-mo-fo}, upper panels). For the moderate degree of dependence (Figure \ref{f_rt2_fr-mo-fo}, middle panels), in addition to the Clayton copula, the Frank and Gaussian copulas continue to fit well. Finally, for the strong degree of dependence (Figure \ref{f_rt2_fr-mo-fo}, lower panels), we see that the Clayton and Frank copulas continue to show a well-estimated tolerance region close to the theoretical one.

%Quando analisamos a regi\~ao te\'orica gerada pelo \textbf{Caso 2}, em que temos uma marginal sim\'etrica e outra assim\'etrica, vemos que as regi\~oes estimadas atrav\'es de todas as fam\'ilias de c\'opulas se adequam bem \`a mesma, enquanto a regi\~ao gerada pela regra $T^2$ de Hotelling j\'a se distancia da teórica mesmo quando o grau de associa\c{c}\~ao entre as distribuições marginais é fraca (Figura \ref{f_rt2_fr}). Para o grau de associa\c{c}\~ao moderada (Figura \ref{f_rt2_mo}), al\'em da c\'opula de Clayton, as c\'opulas de Frank e Gaussiana permanecem se ajustando bem. Por fim, para o grau de associa\c{c}\~ao forte (Figura \ref{f_rt2_fo}), vemos que as c\'opulas de Clayton e Frank continuam apresentando regi\~ao de toler\^ancia estimada bem pr\'oxima da te\'orica.

%%%%%%%%%%%%%%%%%%%%%%%%%%%%%%%%%%%%%%%%%%
% RT -CASO 3

\bigskip

\noindent \textbf{Case \#3}: Artificial sample generated from the bivariate Clayton copula with both asymmetric marginal distributions $\text{Weibull}(a,b)$, with parameters $a$ = 1 and $b$ = 1, and PDF given in \eqref{eq:dwei}.

\begin{figure}[h!]
\begin{center}
\includegraphics[width=\textwidth]{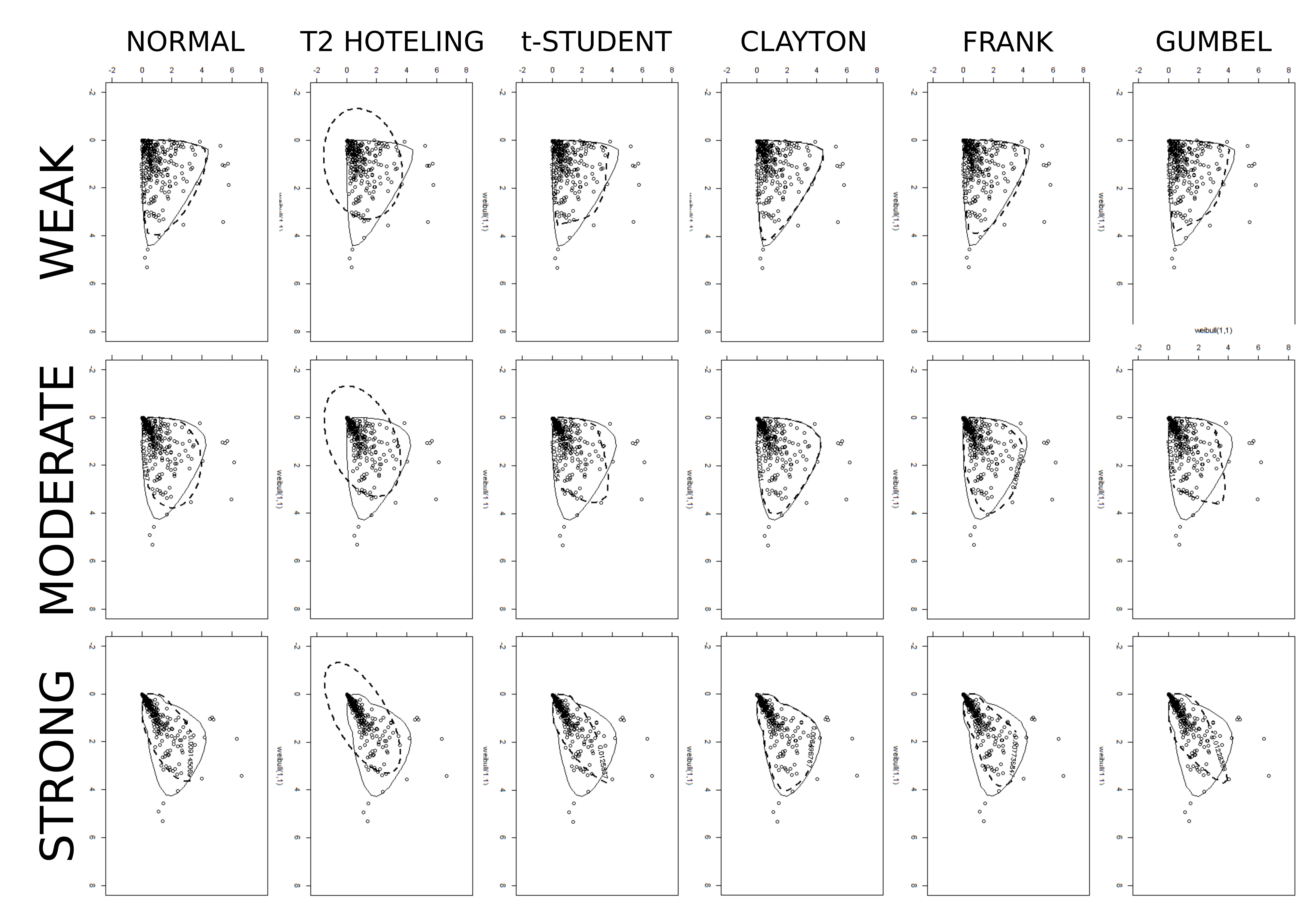}
\caption{Illustration of the $\text{Weibull}(1,1)$ marginal distributions, comparing the real tolerance region (continuous line) with the estimated one based on the parametric copula family (dashed line).}
\label{f_rt3_fr-mo-fo}
\end{center}
\end{figure}

The theoretical and estimated tolerance regions for Case \#3, in which both marginal distributions are asymmetric, are presented in the upper, middle, and lower panels of Figure \ref{f_rt3_fr-mo-fo}, for the degrees of dependence $\tau_2$ = 0.25, 0.50 and 0.75, respectively. As the dependence increases, at first ($\tau_2=0.25$) only the Student-t copula and the Hotelling's $T^2$ rule did not show tolerance regions close to the theoretical one. At $\tau_2=0.50$, the tolerance region estimated by the Gumbel copula also moved away. Finally, at $\tau_2=0.75$, only the Clayton copula presented an estimated tolerance region similar to that expected.

%As regi\~oes de toler\^ancia te\'oricas e estimadas para o \textbf{Caso 3}, em que as duas distribuições marginais s\~ao assim\'etricas, s\~ao apresentadas nas Figuras \ref{f_rt3_fr}, \ref{f_rt3_mo} e \ref{f_rt3_fo}, para os graus de associação $\tau_2$ = 0,25; 0,50 e 0,75, respectivamente. Levando em considera\c{c}\~ao o n\'ivel de depend\^encia, inicialmente apenas a c\'opula t e a regra $T^2$ de Hotelling n\~ao apresentaram regi\~oes pr\'oximas \`a te\'orica. Posteriormente, a regi\~ao estimada pela c\'opula de Gumbel se distanciou. Por fim, apenas a cópula de Clayton apresentou regi\~ao estimada similar à esperada.

%%%%%%%%%%%%%%%%%%%%%%%%%%%%%%%%%%%%%%%%%%

\bigskip
 
\noindent \textbf{Case \#4}: Artificial sample originated from the bivariate normal distribution with mean vector $\boldsymbol{\mu}=(0,0)^{'}$ and covariance (or correlation) matrix $\Sigma$ = $
  \begin{bmatrix}
    1 & \rho \\
    \rho & 1
  \end{bmatrix}$, with $\rho = 0.25$, $0.50$ and $0.75$ (weak, moderate and strong associations, respectively), and joint PDF given by
  \begin{equation} \label{eq:dmnorm}
     f(\textbf{x})=\frac{1}{(2\pi)^{d/2}|\Sigma|^{1/2}}\exp\left\{-\frac{1}{2}(\textbf{x}-\boldsymbol{\mu})^{'}\Sigma^{-1}(\textbf{x}-\boldsymbol{\mu})\right\},
 \end{equation}
 where $d=2$.

\begin{figure}[h!]
\begin{center}
\includegraphics[width=\textwidth]{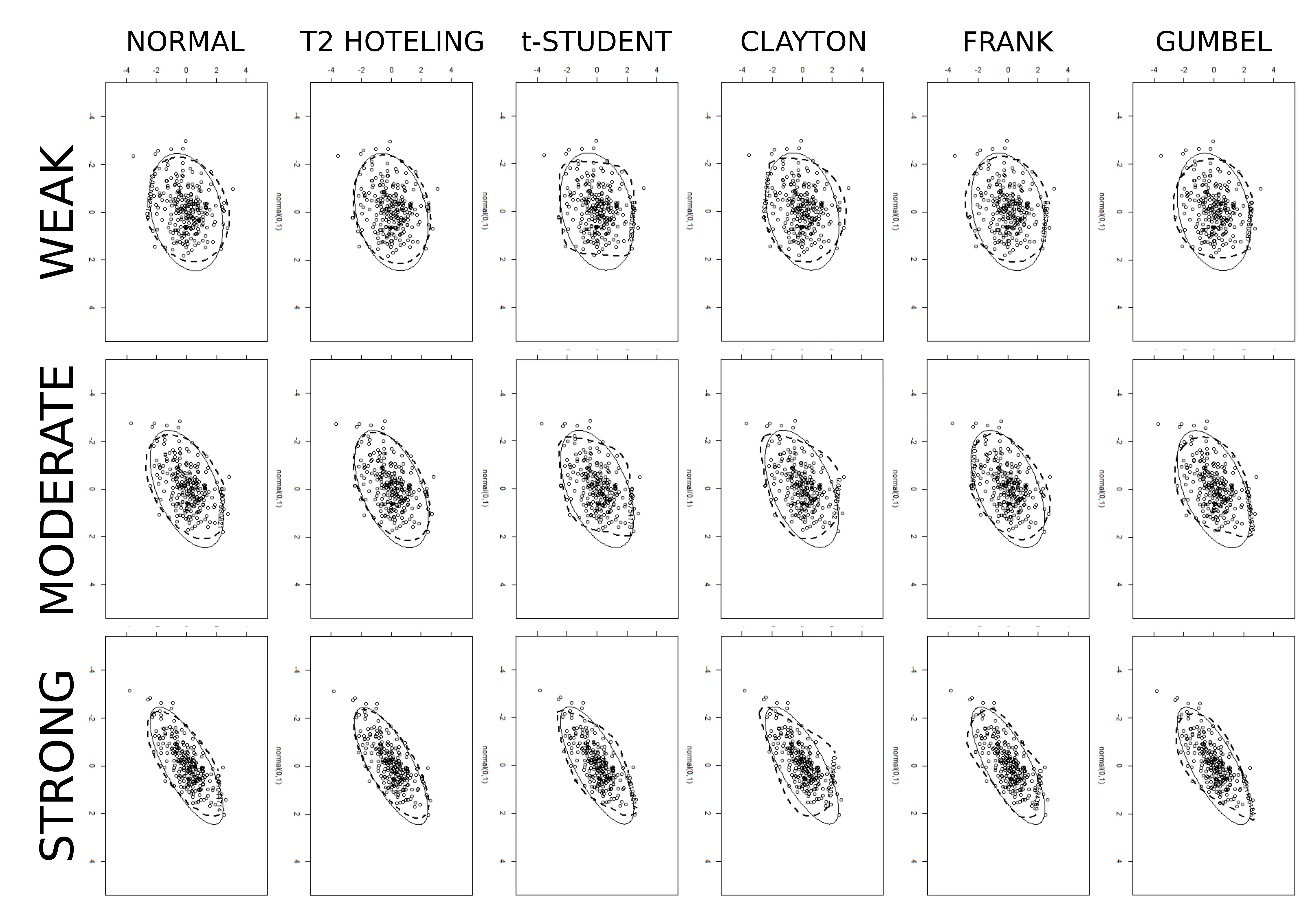}
\caption{Illustration of the (correlated) $\text{Normal}(0,1)$ marginal distributions, comparing the real tolerance region (continuous line) with the estimated one based on the parametric copula family (dashed line).}
\label{f_rt4_fr-mo-fo}
\end{center}
\end{figure}

Case \#4 portrays the traditional situation where the Hotelling's $T^2$ rule is indicated (bivariate normal distribution of the two characteristics studied). In addition to the theoretical tolerance region generated by this rule (ellipse), we also have the tolerance region estimated by it and by the parametric copula families. In Figure \ref{f_rt4_fr-mo-fo} (upper panels), all estimated tolerance regions are very similar. For moderate dependence (Figure \ref{f_rt4_fr-mo-fo}, middle panels), the copula families begin to present tolerance regions that are more distant from the traditional one, with the exception of the Gaussian copula, which maintains a behavior very similar to the Hotelling's $T^2$ rule even when the degree of dependence between the marginal distributions is strong (Figure \ref{f_rt4_fr-mo-fo}, lower panels).

%O \textbf{Caso 4} retrata a tradicional situa\c{c}\~ao em que a regra $T^2$ de Hotelling \'e indicada (distribuição normal bivariada das duas características estudadas). Al\'em da regi\~ao te\'orica gerada por esta regra (elipse), temos tamb\'em a regi\~ao estimada pela mesma e pelas fam\'ilias de c\'opulas. Na Figura \ref{f_rt4_fr} (associação fraca), todas as regi\~oes estimadas s\~ao muito similares. Para depend\^encia moderada, as fam\'ilias de c\'opulas passam a apresentar regi\~oes de toler\^ancia mais distantes da tradicional, com exce\c{c}\~ao da c\'opula Gaussiana (Figura \ref{f_rt4_mo}), que mant\'em o comportamento muito semelhante ao da regra $T^2$ de Hotelling mesmo quando o grau de depend\^encia entre as distribuições marginais \'e forte (Figura \ref{f_rt4_fo}).
 
% RT -CASO 4

%%%%%%%%%%%%%%%%%%%%%%%%%%%%%%%%%%%%%%%%%%%
The simulation results (obtained via a Monte Carlo approach with 1,000 realizations) corroborated those found by the graphical study of the tolerance region. In the bivariate parametric cases with non-normal marginal distributions, the Clayton family of copulas (which gave rise to the data) obtained the best performances, especially when there was a strong dependence between the variables. In the bivariate normality case, the Gaussian copula presented values of the in-control average run length ($\text{ARL}_0$) (Montgomery 2013) very close to those obtained by the Hotelling's $T^2$ rule, which fits very well regardless of the degree of association. Furthermore, the copula detected subtle changes more quickly, in which case the traditional $T^2$ is not indicated. These tests can be extended to the frameworks of semiparametric and nonparametric copulas, where the advantage of copula functions is expected to become evident when the degree of dependence between the margins is moderate or strong. \nocite{montgomery2013}

%Os resultados obtidos corroboraram aqueles encontrados pelo estudo gráfico da regi\~ao de toler\^ancia. Nos casos param\'etricos bivariados com distribuições marginais n\~ao-normais, a fam\'ilia de c\'opulas de Clayton (que deu origem aos dados) obteve os melhores desempenhos, especialmente quando havia forte depend\^encia entre as vari\'aveis. No caso da normalidade bivariada, a c\'opula Gaussiana apresentou valores de $\text{ARL}_0$ bem pr\'oximos dos obtidos pelo $T^2$, que se adequa muito bem independente do grau de associa\c{c}\~ao. Ademais, a c\'opula detectou mudan\c{c}as sutis mais rapidamente, caso em que o $T^2$ n\~ao \'e indicado. Tais testes poderão ser extendidos aos contextos de cópulas semi-param\'etricas e não param\'etricas, onde a supremacia das fun\c{c}\~oes c\'opula s\'o ficam evidentes quando o grau de depend\^encia \'e moderado ou forte.

%%%%%%%%%%%%%%%%%%%%%%%%%%%%%%%%%%%%%%%%%%%%%%%%%%%%%%%%%%%

\end{document}